\documentclass[journal]{IEEEtran}

\usepackage{float}
\usepackage{amssymb}
\usepackage{bm}
\usepackage{amsthm, amsmath,amsfonts}
\usepackage[ruled, vlined]{algorithm2e}
\usepackage{array}
\usepackage{textcomp}
\usepackage{stfloats}
\usepackage{url}
\usepackage{verbatim}
\usepackage{graphicx}
\usepackage{subfigure}
\usepackage{cite}
\usepackage{booktabs}    
\usepackage{mathrsfs} 
\usepackage{setspace}
\usepackage{geometry}
\usepackage{multirow}
\usepackage{amsmath} 
\allowdisplaybreaks[4] 
\usepackage{autobreak} 
\usepackage{xcolor}
\usepackage{hyperref} 
\usepackage{threeparttable}
\usepackage{diagbox}
\usepackage{pifont}
\theoremstyle{definition}
\makeatletter
\renewcommand{\maketag@@@}[1]{\hbox{\m@th\normalsize\normalfont#1}}%
\makeatother

\usepackage{xcolor} 

\makeatletter

\def\BibColor@get#1{%
	\@ifundefined{BibColor@#1}{}{%
		\csname BibColor@#1\endcsname}%
}

\AtBeginDocument{%
	\let\BC@orig@lbibitem\@lbibitem
	\let\BC@orig@bibitem \@bibitem
	
	\def\@bibitem#1{%
		\edef\BC@col{\BibColor@get{#1}}%
		\ifx\BC@col\@empty
		\color{black}%
		\else
		\color{\BC@col}%
		\fi
		\BC@orig@bibitem{#1}%
	}%
	
	\def\@lbibitem[#1]#2{%
		\edef\BC@col{\BibColor@get{#2}}%
		\ifx\BC@col\@empty
		\color{black}%
		\else
		\color{\BC@col}%
		\fi
		\BC@orig@lbibitem[#1]{#2}%
	}%
}
\makeatother

\begin{document}

\title{
Radiation Pattern Reconfigurable FAS-Empowered Interference-Resilient UAV Communication
}
\author{Zhuoran Li, Zhen Gao, \textit{Member}, \textit{IEEE}, Boyu Ning, \textit{Member}, \textit{IEEE}, and Zhaocheng Wang, \textit{Fellow}, \textit{IEEE}
	
	\thanks{
	The work of Zhen Gao was supported by the National Key Research and Development Program of China under Grant 2023YFC3341102, the Natural Science Foundation of China (NSFC) under Grant U2233216 and Grant 62471036, Shandong Province Natural Science Foundation under Grant ZR2025QA30 and ZR2022YQ62, the Beijing Natural Science Foundation under Grant L242011, and Beijing Nova Program 20220484025.
		
	Z. Li is with the School of Information and Electronics, Beijing Institute of Technology, Beijing 100081, China, and also with the Advanced Research Institute of Multidisciplinary Sciences, Beijing Institute of Technology, Beijing 100081, China (e-mail: \mbox{lizhuoran2000@qq.com}).
	
	Zhen Gao is with the State Key Laboratory of CNS/ATM and the MIIT Key Laboratory of Complex-Field Intelligent Sensing, Beijing 100081, China, also with Beijing Institute of Technology (BIT), BIT, Zhuhai 519088, China, also with the Advanced Technology Research Institute, BIT (Jinan), Jinan 250307, China, and also with the Yangtze Delta Region Academy, BIT (Jiaxing), Jiaxing 314019, China (e-mail: \mbox{gaozhen16@bit.edu.cn}).\textit{(Corresponding author: Zhen Gao.)}
		
	Boyu Ning is with the National Key Laboratory of Wireless Communications, University of Electronic Science and Technology of China, Chengdu 611731, China (e-mail: 	\mbox{boydning@outlook.com}).
	
	Z. Wang is with Beijing National Research Center for Information Science and Technology, Department of Electronic Engineering, Tsinghua University, Beijing 100084, China, and Z. Wang is also with Tsinghua Shenzhen International Graduate School, Shenzhen 518055, China (e-mail: \mbox{zcwang@tsinghua.edu.cn}).
	}
}


\maketitle
\begin{abstract}	
	The widespread use of uncrewed aerial vehicles (UAVs) has propelled the development of advanced techniques on countering unauthorized UAV flights. 
	However, the resistance of legal UAVs to illegal interference remains under-addressed. 
	This paper proposes radiation pattern reconfigurable fluid antenna systems (RPR-FAS)-empowered interference-resilient UAV communication scheme.
	This scheme integrates the reconfigurable pixel antenna technology, which provides each antenna with an adjustable radiation pattern and electromagnetic domain precoding capability.
	Therefore, RPR-FAS can enhance the angular resolution of a UAV with a limited number of antennas, thereby improving spectral efficiency (SE) and interference resilience.
	Specifically, we first design dedicated radiation pattern adapted from 3GPP-TR-38.901, where the beam direction and half power beamwidth are tailored for UAV communications.
	Furthermore, we propose a low-storage-overhead orthogonal matching pursuit multiple measurement vectors algorithm, which accurately estimates the angle-of-arrival (AoA) of the communication link, even in the single antenna case. 
	Particularly, by utilizing the Fourier transform to the radiation pattern gain matrix, we design a dimension-reduction technique to achieve 1--2 order-of-magnitude reduction in storage requirements.
	Meanwhile, we propose a maximum likelihood interference AoA estimation method based on the law of large numbers, so that the SE can be further improved.
	Finally, alternating optimization is employed to obtain the optimal uplink radiation pattern and combiner, while an exhaustive search is applied to determine the optimal downlink pattern, complemented by the water-filling algorithm for beamforming.
	Comprehensive simulations demonstrate that the proposed schemes outperform traditional methods in terms of angular sensing precision and spectral efficiency.
	The proposed scheme protects legal UAVs against unauthorized interference while maintaining stable data transmission, and effectively supports the security of the booming low-altitude economy.
	Simulation codes are provided to reproduce the results in this paper: \href{https://github.com/LiZhuoRan0/2025-JSAC-RadiationPatternReconfigurableAntenna}{https://github.com/LiZhuoRan0}.
\end{abstract}


\section{Introduction}\label{Sec_Introduction}
\IEEEPARstart{T}{he} emergence of the low-altitude economy and its reliance on dedicated low-altitude wireless networks has spurred the widespread adoption of uncrewed aerial vehicles (UAVs). However, this proliferation has also given rise to significant security concerns regarding unauthorized flights.
To address this, counter-UAV technologies are rapidly evolving~\cite{ref_2021_AESM_C_UAS,ref_2021_Access_AntiUAV,ref_UAV_com_1,ref_GenAI}.
For instance, passive sensing methods have been proposed to obtain UAV model parameters~\cite{ref_2021_TIM_FHS_Model,ref_DroneRFa,ref_ISAC_1}, while other approaches suggest jamming unauthorized UAVs by leveraging base station communication signals on the same frequency~\cite{ref_IoTJ}.
Furthermore, various monitoring technologies are applied to bolster the effectiveness of counter-UAV interference measures\cite{ref_2018_CM_AntiUAV,ref_UAV_com_2}.

With the development of counter-UAV technologies, new issues have emerged, where legitimate UAVs may be mistakenly or maliciously interfered with.
This necessitates enhancing the sensing abilities of UAVs to detect and counteract potential interference.
Unfortunately, civilian UAVs often face inherent limitations in sensing due to constraints related to their size, flight time, cost, and practicality. 
On the other hand, UAVs generally operate in the industrial, scientific, and medical (ISM) bands, which are characterized by long wavelength. 
This presents a challenge because the limited physical space on UAVs makes it impractical to install large antenna arrays, which are essential for effective angular sensing.

Recent research indicates that fluid antenna system (FAS) can enhance the sensing capabilities of limited-aperture arrays~\cite{ref_2025_CST}.
Specifically, reconfigurable antenna system is capable of altering the shape, position, and radio frequency characteristics of the radiation pattern~\cite{ref_2024_TCOM_KekeYing,ref_2024_WCM_KekeYing}, thereby enabling spatial diversity and flexibility, and it can be realized by the FAS~\cite{ref_2024_TWC_Outage_Diversity}.
By adjusting the radiation pattern of FAS, it becomes possible to achieve precise detection of electromagnetic signals in space even with a small number of antennas or a single antenna.
Therefore, deploying FAS on UAVs can significantly enhance their sensing capabilities, which is a vital prerequisite to enhance spectral efficiency (SE).

\subsection{Prior Works}
Due to the high spatial degrees of freedom of FAS, traditional sensing methods are difficult to apply, and new approaches are required to improve estimation accuracy and reduce overhead~\cite{ref_2024_arXiv_Maga}.
A method proposed in~\cite{ref_2025_TWC_ChannelEst} changes channel characteristics by adjusting the antenna position and radiation pattern, and reconstructs the channel using oversampling.
The maximum likelihood estimation method and Nyquist sampling method used in this study can theoretically improve estimation accuracy, yet they may still encounter practical challenges when implementing the required sampling rates in real-world systems.
A channel sensing method based on dynamic antenna position switching was proposed in~\cite{ref_2024_CL_ChannelEst}, which can greatly reduce pilot overhead.
J. Zou \textit{et al.} in~\cite{ref_2024_arXiv_ISAC} explored integrated sensing and communication (ISAC) under dynamic antenna position switching. 
By employing flexible port selection and beamforming optimization~\cite{ref_lsc}, FAS can achieve a better balance between sensing and communication capacities.
An innovative fluid antenna-assisted ISAC system was proposed in~\cite{ref_2024_WCL_ISAC}, where the dynamic optimization of antenna positions for base stations and terminal users can effectively enhance both communication rates and sensing performance.
However, the three studies~\cite{ref_2024_CL_ChannelEst, ref_2024_arXiv_ISAC, ref_2024_WCL_ISAC} on FAS-enabled ISAC mentioned above are not applicable to systems with reconfigurable radiation patterns.

With ongoing research on the hardware of radiation pattern reconfigurable FAS (RPR-FAS), the manufacturing process of RPR-FAS has been steadily progressing.
In~\cite{ref_2012_TAP_Hardware}, a reconfigurable antenna system based on parasitic layers was developed using a multi-objective optimization approach.
The study in~\cite{ref_2014_TAP_Hardware} designed a reconfigurable antenna with parasitic pixel layers, capable of independently adjusting operating frequency, radiation pattern, and polarization.
An antenna array capable of simultaneously controlling beamforming and adjusting beamwidth was introduced in~\cite{ref_2018_TAP_Hardware}.
This antenna enhanced network flexibility and efficiency, especially in hotspot and ultra-dense deployment scenarios. 
In~\cite{ref_2022_TAP_Hardware}, a horizontally polarized antenna was introduced, which utilizes Alford loops and four compact parasitic pixel loops to achieve highly reconfigurable radiation patterns. 
This antenna does not require phase modulator and can generate various radiation patterns, including omni-directional, single-directional, and multi-beam modes. 
In~\cite{ref_BoyuNing_Move}, B. Ning \textit{et al.} provided a comprehensive overview of antennas with reconfigurable radiation patterns, including both mechanically driven and electronically driven types. 
These antennas can change their radiation direction by adjusting their position through motors, a slide rail, liquid, or micro-electro-mechanical systems, and they can also alter their radiation pattern by exciting different transverse magnetic modes, using coding sequences, or employing pin-diodes.
In addition to radiation-pattern reconfiguration, recent studies have shown that dynamically relocating antenna elements—the so-called movable-antenna paradigm—can further enhance channel conditions and link performance. For example, \cite{refMAone,refMAtwo,refMAthree} reviewed system architectures, performance modeling, and practical challenges for movable-antenna systems, offering a complementary form of spatial adaptation alongside the pattern-reconfigurable FAS techniques examined here.

Benefiting from the advancements in hardware, algorithms for enabling efficient sensing in RPR-FAS have been progressing concurrently.
The significant potential of RPR-FAS in wireless communications has been revealed in studies~\cite{ref_2025_OJAP_ReconfigurableRadiationPattern,ref_2025_CM_ReconfigurableRadiationPattern, ref_2024_arXiv__ReconfigurableRadiationPattern}.
K. Ying \textit{et al.} in~\cite{ref_2024_WCM_KekeYing} innovatively explored on the architecture, theory, and applications of reconfigurable large-scale multiple-input multiple-output (MIMO) systems. 
It conceived the concept of electromagnetic domain precoding for the first time, i.e., leveraging the electromagnetic radiation field to enhance both SE and energy efficiency for enhanced information transmission.
H. Wang \textit{et al.} in~\cite{ref_2023_TWC_ContinuousOpt} designed continuously variable antenna radiation patterns by maximizing the rate optimization objective.
However, implementing such variable antenna radiation pattern in practice proved to be challenging.
K. Ying \textit{et al.} in~\cite{ref_2024_TCOM_KekeYing} proposed precoding design and channel sensing methods for reconfigurable large-scale MIMO systems, which introduced a new dimension of electromagnetic domain precoding on the basis of conventional fully-digital array or hybrid analog-digital array.
This research innovatively simplified channel sensing by decomposing antenna radiation patterns using spherical harmonic functions.
The iterative selection method used in~\cite{ref_2018_TWC_ModeSelection} circumvents the high complexity associated with antenna pattern selection and improves throughput in multi-user MIMO systems.
Moreover, A. C. Gurbuz \textit{et al.} in~\cite{ref_2020_AESM_CognitiveTrack} optimized direction estimation and signal-to-noise ratio (SNR) by iteratively selecting the radiation pattern for each antenna, enhancing the target-detection performance in cognitive radar systems.
To avoid the complexity of reconfigurable radiation patterns selection, T. Zhao \textit{et al.} in~\cite{ref_2021_TWC_ModeSelection} employed an online learning method to avoid high storage and computational overhead.
However, the above-mentioned studies are idealized and do not account for interference, rendering them unsuitable for practical interference-resilient UAV communication.

Meanwhile, existing UAV anti-interference research primarily focuses on spectrum sensing and resource coordination, which is inadequate for scenarios where UAVs are mistakenly or maliciously interfered with~\cite{ref_UAV_1,ref_UAV_2,ref_UAV_3,ref_UAV_4,ref_UAV_5}.
Although the studies in \cite{ref_UAV_2,ref_UAV_4} considered UAVs with multiple antennas, the limited number of antennas is insufficient to suppress interference effectively in the angular domain.
Therefore, further research is needed on the RPR-FAS based ISAC to equip UAVs with the capability to detect and resist interference.

\subsection{Our Contributions}
Building on prior hardware advancements detailed in~\cite{ref_2018_TAP_Hardware,ref_2022_TAP_Hardware}, we investigate a scenario in which a legitimate UAV communicates with its controller and is capable of detecting and counteracting illegal jamming sources.
By varying the half-power beam width (HPBW) of the reference radiation pattern in the 3rd generation partnership project (3GPP) technical report (TR) 38.901\cite{ref_38901}, the paper reveals the significant role of RPR-FAS in enhancing the angular sensing precision and spectral efficiency of UAVs under the severe interference.
The specific contributions of this paper are summarized as follows.
\begin{itemize}
    \item To address the failure of traditional angle estimation methods in single-antenna and two-antenna scenarios, a novel angle estimation model based on the RPR-FAS is established.
    The estimation of the UAV controller's angle of arrival (AoA) is formulated as a compressed sensing problem, where the radiation pattern gain matrix serves as the sensing matrix.
    To reduce the storage overhead of the radiation pattern gain matrix, a fast construction method is proposed.
    The designed dimension-reduction construction method, based on Fourier transform, can reduce storage overhead by 1--2 orders of magnitude.
    Leveraging the multiple measurement vector (MMV) characteristics enabled by the UAV's possible multi-antenna configuration, the low-storage-overhead orthogonal matching pursuit MMV (LSO-OMP-MMV) algorithm is utilized to estimate the AoA of the UAV controller.

    \item To save sensing time overhead, the UAV senses both the jammer and the UAV controller using the same received signal.
    After estimating the AoA of the UAV controller at the UAV, the impact of the UAV controller on the received signal is eliminated.
    The jammer's AoA is then estimated using a maximum likelihood estimation method based on the law of large numbers.

    \item After estimating the AoAs of the UAV controller and jammer at the UAV, an alternating optimization method is proposed for uplink transmission to jointly select the optimal radiation pattern and design the combiner for the UAV. For downlink transmission, a one-dimensional exhaustive search method is employed to determine the optimal radiation pattern, while the water-filling algorithm is applied to maximize SE.

    \item The proposed RPR-FAS-empowered interference-resilient UAV communication scheme is generalizable, requiring only substitution of the radiation pattern gain matrix with the actual one.
    Extensive simulations demonstrate that super-resolution algorithms fail to estimate AoAs with few receive antennas and under severe jamming conditions.
    In contrast, the proposed LSO-OMP-MMV algorithm based on reconfigurable radiation patterns achieves high-precision angle estimation performance. 
    At the same time, by selecting the optimal radiation pattern, the interference from the jammer can be significantly suppressed, thereby improving system SE. Furthermore, the sensing matrix design achieves low storage overhead with acceptable performance degradation.
\end{itemize}

The remainder of this paper is organized as follows.
Section \ref{sec_system_model} details the system model and the reconfigurable radiation pattern modes.
In Section \ref{sec_controller_angle}, we present the proposed UAV controller's angle estimation scheme.
In Section \ref{sec_jammer_angle}, we introduce the proposed jammer's angle estimation scheme.
After sensing the UAV controller and the jammer, an environment-aware optimal radiation pattern selection scheme is proposed in Section \ref{sec_pattern_selection} to maximize SE while decreasing the probability of the UAV being detected by the jammer.
Simulation results are presented in Section \ref{sec_simulation}.
Finally, we draw
our conclusions in Section \ref{sec_conclusion}.

\subsection{Notation}
 Unless otherwise stated, normal-face letters, boldface lower-case letters, boldface uppercase letters and calligraphic letters denote scalar variables, column vectors, matrices and tensors, respectively.
The transpose and conjugate transpose operators are denoted by $(\cdot)^{\text T}$ and $(\cdot)^{\text H}$, respectively.
$j=\sqrt{-1}$ is the imaginary unit, $\mathbb{R}$ and $\mathbb{C}$ are the sets of real-valued and complex-valued numbers, respectively.
For vector $\mathbf{x}\! \in\! \mathbb{C}^{N}$, $\mathbf{x}[n]$ denotes the $n$-th element of $\mathbf{x}$, and ${\bf{x}}(\theta)[n]$ means extracting the element of ${\bf{x}}$ indexed by $[n]$, where $\theta$ is the argument.
For matrix $\mathbf{X}\! \in\! \mathbb{C}^{N\times M}$, $\mathbf{X}[n,m]$ denotes the $(n,m)$-th element of $\mathbf{X}$.
${\bf{X}}^\dag$ denotes the pseudo-inverse of ${\bf{X}}$.
For tensor $\mathcal{Y}\! \in\! \mathbb{C}^{N\times M\times K}$, $\mathcal{Y}[n,m,k]$ denotes its $(n,m,k)$-th element.
$\mathcal{CN}(\mu, \Sigma)$ denotes the complex Gaussian distribution with mean $\mu$ and covariance $\Sigma$.
The magnitude of a scalar, or the element-wise magnitude of a vector, matrix, or tensor, is denoted by $|\cdot|$.
$\left\|  \cdot  \right\|_{0}$ and ${\left\|  \cdot  \right\|_{\text{F}}}$ denote zero norm and Frobenius norm, respectively.
For real number $s$, $\left\lfloor s \right\rfloor$ represents the largest integer less than or equal to $s$, and $\left[\kern-0.15em\left[ {s} \right]\kern-0.15em\right]$ is the integer closest to $s$.
$\mathbf{1}_n$ denotes the vector of size $n\times 1$ with all the elements being 1.
$\mathbf{I}_n$ denotes the identity matrix with dimension $n$.
$\odot$ represents the Hadamard product and $\oslash$ is the element-wise division, while $\bmod(x,y)$ denotes the remainder when $x$ is divided $y$.
The operator $\Leftarrow$ assigns the value of the variable on the right to the left. The speed of light is denoted as $c$, and $\Pi (t)$ is the unit rectangular window function, which equals to $1$ in the interval $[0,1)$ and $0$ otherwise.

\begin{figure*}[!t]
		\vspace{-9mm}
	\centering
	\color{black}
	\includegraphics[width=7in]{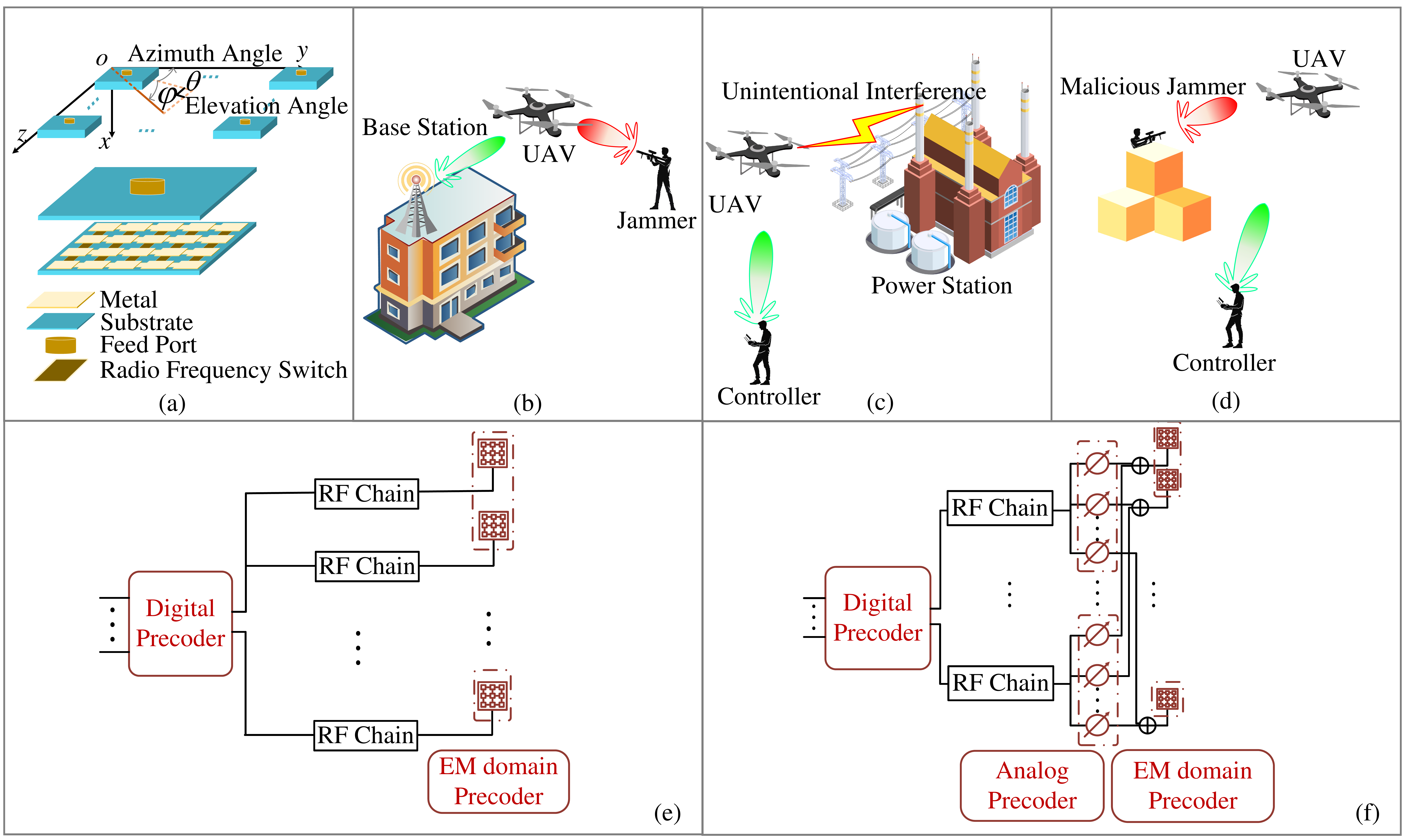}
	\caption{RPR-FAS and its application in enhancing interference-resilient UAV communications across various scenarios:
	(a) RPR-FAS with reconfigurable pixel antennas\cite{ref_2024_arXiv__ReconfigurableRadiationPattern};
	(b) networked UAVs jammed by illegal jammer;
	(c) remotely piloted UAVs interfered by unintentional interference;
	(d) remotely piloted UAVs jammed by illegal jammer;
	(e) reconfigurable fully digital array;
	(f) reconfigurable hybrid array.}
	\label{fig_scene}
\end{figure*}

\section{System Model}\label{sec_system_model}
In this section, we detail the UAV uplink transmission model, the corresponding channel model, and the radiation pattern model for the RPR-FAS.
\subsection{Transmission Model}
As shown in Fig.~\ref{fig_scene}(a), the UAV in this paper deploys RPR-FAS with reconfigurable pixel antennas\cite{ref_2024_TCOM_KekeYing,ref_2024_WCM_KekeYing,ref_2024_arXiv__ReconfigurableRadiationPattern}.
Figs.~\ref{fig_scene}(e) and (f) illustrate brand-new transceiver architectures based on RPR-FAS.
RPR-FAS can introduce a new dimension in the electromagnetic domain, namely electromagnetic-domain precoding, thereby enhancing information transmission on the basis of traditional all-digital arrays or hybrid analog-digital arrays (the latter being the emerging three-stage precoding MIMO architecture)\cite{ref_2024_TCOM_KekeYing,ref_2024_WCM_KekeYing}.
In Fig.~\ref{fig_scene}(b)--\ref{fig_scene}(d), three scenarios demonstrate how RPR-FAS can be employed to enhance interference-resilient UAV communications.
Since the UAV side generally has fewer antennas, the reconfigurable fully digital array shown in Fig.~\ref{fig_scene}(e) is used in this paper for the UAV side. However, for transceivers with a larger number of antennas, the reconfigurable hybrid array shown in Fig.~\ref{fig_scene}(f) can be adopted to reduce hardware complexity. 
Without loss of generality, we focus on the scenario depicted in Fig.~\ref{fig_scene}(d).
In this scenario, we consider a narrowband frequency-hopping transmission where the UAV controller transmits uplink pilot signals to the UAV, after which both the UAV and UAV controller engage in uplink and downlink communications.
The UAV estimates the AoA of the UAV controller using the RPR-FAS.
While receiving the UAV controller's signal, the UAV is also subject to interference from a jammer.
The jammer uses a single directional antenna to perform directional interference.
The controller is a legitimate ground station transmitting control and data signals to the UAV. Its channel can be estimated via pilot symbols.
Jammer is a malicious adversary whose sole goal is to disrupt the controller–UAV link. The jammer's channel is unknown and must be identified by the UAV.
We define the key symbols of system parameters and their associated definitions in Table~\ref{table_SensingBound}.

\begin{figure}[!t]
	\vspace{-7mm}
	\centering
	\color{black}
	\includegraphics[width=2.5in]{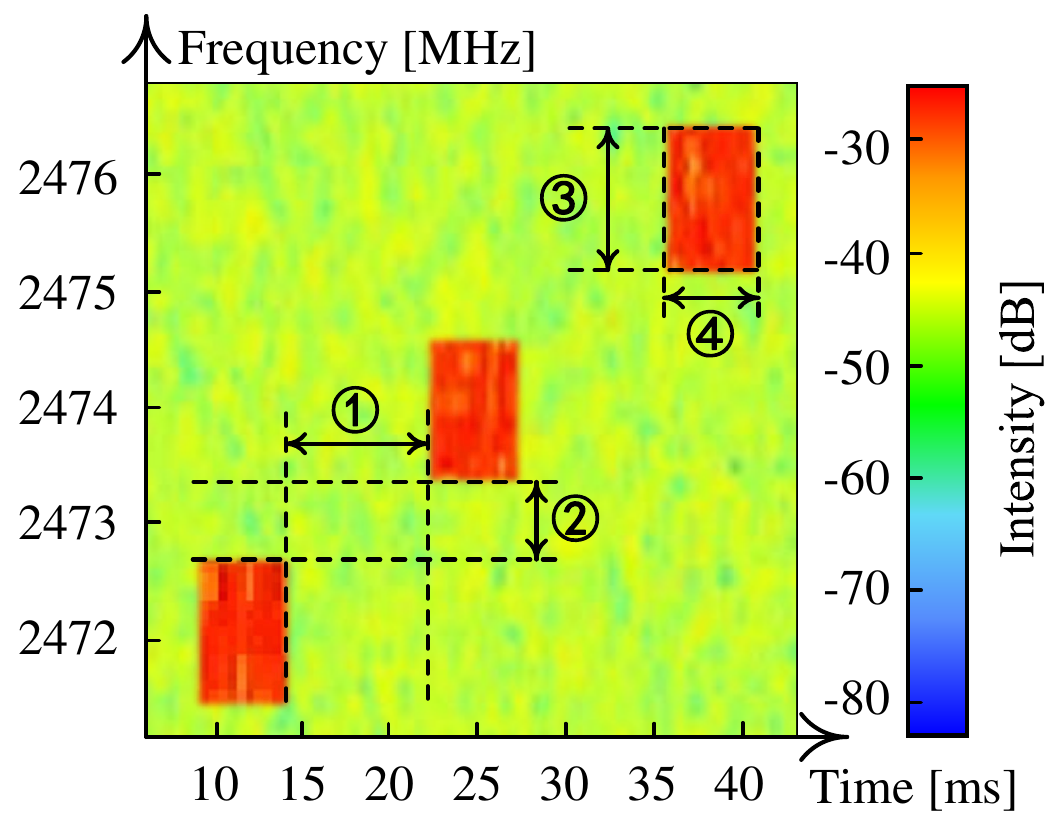}
	\caption{Time-frequency diagram of FHSS signal\cite{ref_DroneRFa}.
		\textcircled{1} $T_I$, time interval of adjacent FHSS blocks.
		\textcircled{2} $B_I$, bandwidth of adjacent FHSS blocks.
		\textcircled{3} $B_h$, FHSS signal bandwidth.
		\textcircled{4} $T_h$, FHSS signal time interval.}
	\label{fig_signal}
	\vspace{-4mm}
\end{figure}

Time-frequency diagram of frequency hopping spread spectrum (FHSS) signal transmitted by the UAV and the UAV controller is shown in Fig.~\ref{fig_signal}.
The FHSS signal can be modeled as 
\begin{align}
    x(t) = b(t)f(t),
\end{align}
where $t$ is the time, $b(t)$ is the baseband signal, and $f(t)$ is the carrier signal that undergoes frequency hopping.
$f(t)$ is modeled as 
\begin{align}
    f(t) = A\sum\limits_{n_h = 0}^{N_h - 1} \Pi\left(\frac{t - n_h{T_h}}{{T_h}}\right){e^{j\left( {2\pi {f_n}(t - n_h{T_h}) + {\phi _n}} \right)}},
\end{align}
where $n_h$ is the index of the FHSS pilot signals, $A$ is the amplitude,
$f_n$ is the hopping frequency,
and $\phi_n$ is the initial phase\cite{ref_2021_TIM_FHS_Model}.
${\mathcal{X}_{\text{C}}}[n_{\text{tx}}, \bar n_s, n_h]$ is the  $\bar n_s$-th sampling point of the $n_h$-th FHSS signal transmitted by the UAV controller at the $n_{\text{tx}}$-th transmit antenna.
Similarly, the FHSS signal transmitted by the UAV is given as ${\mathcal X_{\text{U}}}[n_{\text{tx}}, \bar n_s, n_h]$.
${{\bf  X_{\text{J}}}[{\bar n_s},{n_h}]}
\sim
\mathcal{CN}\left(
0, \sigma^2_{\text{J}}
\right)$ is the jamming signal transmitted by the jammer.
$\sigma^2_{\text{J}}$ is the variance of ${{\bf  X_{\text{J}}}[{\bar n_s},{n_h}]}$ and we define $\sigma^2_{\text{J}}=P_\text{J}$, where $P_\text{J}$ is the transmit power of the jammer.
Their dimensions are ${\mathcal{X}_{\text{C}}} \in {\mathbb{C}^{{N_{{\text{C}}}}\times {\bar N_s} \times {N_h}}},\;{\mathcal{X}_{\text{U}}} \in {\mathbb{C}^{{N_{{\text{U}}}} \times {\bar N_s} \times {N_h}}}, \text{and}\;{{\mathbf{X}}_{\text{J}}} \in {\mathbb{C}^{{\bar N_s} \times {N_h}}}$, respectively.

In the uplink payload data signal transmission stage, ${\mathcal{Y}_{\text{U}}} \in {\mathbb{C}^{{N_{{\text{U}}}} \times {\bar N_s} \times {N_h}}}$ is the uplink signal received by the UAV.
If the UAV controller activates all transmit antennas, the $\bar n_s$-th sampling point of the $n_h$-th FHSS signal received by the UAV at the $ n_{\text{rx}}$-th receive antenna is given as 
\begin{align}
\label{eq_up_tensor}
    {{\cal Y}_{\text{U}}}[{n_{{\text{rx}}}},{\bar n_s},{n_h}] &= {{\bf{H}}_{\text{J}}}[{n_{{\text{rx}}}},{n_h}]{{\bf{X}}_{\text{J}}}[{\bar n_s},{n_h}] + {{\cal W}_{\text{U}}}[{n_{{\text{rx}}}},{\bar n_s},{n_h}]\nonumber\\& + \sum\limits_{{n_{{\text{tx}}}} = 0}^{{N_{{\text{C}}}} - 1} {{{\cal H}_{\text{C}}}[{n_{{\text{rx}}}},{n_{{\text{tx}}}},{n_h}]{{\cal X}_{\text{C}}}[{n_{{\text{tx}}}},{\bar n_s},{n_h}]},
\end{align}
where ${\mathcal{W}_{\text{U}}}[{n_{{\text{rx}}}},{\bar n_s},{n_h}]\sim\mathcal{CN}\left( {0,\sigma _{\text{U}}^2} \right)$ is the noise, ${\mathcal{W}_{\text{U}}} \in {\mathbb{C}^{{N_{{\text{U}}}} \times {\bar N_s} \times {N_h}}}$, and $\sigma _{\text{U}}^2$ is the associated noise variance.
${\bf{H}}_{\text{J}}\in\mathbb{C}^{N_{\text{U}}\times N_h}$ is the channel from the jammer to the UAV.
${{\cal H}_{\text{C}}}\in\mathbb{C}^{N_{\text{U}}\times N_{\text{C}}\times N_h}$ is the channel from the UAV controller to the UAV.

In the uplink pilot signal transmission stage, the UAV controller activates only one antenna to achieve omni-directional transmission, the signal received by the UAV is given as
\begin{align}
    {{\cal Y}_{\text{U}}}[{n_{{\text{rx}}}},{\bar n_s},{n_h}]& = {{\bf{H}}_{\text{J}}}[{n_{{\text{rx}}}},{n_h}]{{\bf{X}}_{\text{J}}}[{\bar n_s},{n_h}] + {{\cal W}_{\text{U}}}[{n_{{\text{rx}}}},{\bar n_s},{n_h}] \nonumber\\&+ {{\bf{H}}_{\text{C}}}[{n_{{\text{rx}}}},{n_h}]{{\bf{X}}_{\text{C}}}[{\bar n_s},{n_h}],
\end{align}
where ${\mathbf{X}_{\text{C}}} \in \mathbb{C}^{ {\bar N_s} \times {N_h}}$ is the FHSS signal transmitted by the UAV controller and ${{\bf H}_{\text{C}}}\in\mathbb{C}^{N_{\text{U}}\times N_h}$ is the channel from the
UAV controller's first antenna to the UAV.
We define the power of ${\mathbf{X}_{\text{C}}}$ as $P_{\text{C}}$.
The number of sampling points for pilot signals can be obtained by $\bar N_s=T_uB_p$.

In the downlink data transmission stage, ${\mathcal{Y}_{\text{C}}} \in {\mathbb{C}^{{N_{{\text{C}}}} \times {\bar N_s} \times {N_h}}}$ is the signal received by the UAV controller.
The $\bar n_s$-th sampling point of the $n_h$-th FHSS signal received by the UAV controller at the $ n_{\text{rx}}$-th receive antenna is given as 
\begin{align}
\label{eq_dn_tensor}
    {{\cal Y}_{\text{C}}}[{n_{{\text{rx}}}},{\bar n_s},{n_h}] &= \sum\limits_{{n_{{\text{tx}}}} = 0}^{{N_{{\text{U}}}} - 1} {{{\cal H}_{\text{U}}}[{n_{{\text{rx}}}},{n_{{\text{tx}}}},{n_h}]{{\cal X}_{\text{U}}}[{n_{{\text{tx}}}},{\bar n_s},{n_h}]}\nonumber\\&+ {{\cal W}_{\text{C}}}[{n_{{\text{rx}}}},{\bar n_s},{n_h}] ,
\end{align}
where ${\mathcal{W}_{\text{C}}}[{n_{{\text{rx}}}},{\bar n_s},{n_h}]\sim\mathcal{CN}\left( {0,\sigma _{\text{C}}^2} \right)$ is the noise, ${\mathcal{W}_{\text{C}}} \in {\mathbb{C}^{{N_{{\text{C}}}} \times {\bar N_s} \times {N_h}}}$, and $\sigma _{\text{C}}^2$ is the associated noise variance.
${{\cal H}_{\text{U}}}\in\mathbb{C}^{N_{\text{C}}\times N_{\text{U}}\times N_h}$ is the channel from the UAV to the UAV controller and $\mathcal{X}_{\text{U}}\in\mathbb{C}^{N_{\text{U}}\times \bar N_{s}\times N_h}$ is the data signal transmitted by the UAV.

We assume that within the duration of a FHSS block, the UAV can adjust the reconfigurable radiation pattern multiple times.
Therefore, the aforementioned variables can be reshaped as follows: $\mathcal{X}_{\text{C}} \in \mathbb{C}^{N_{\text{C}}\times N_s \times N_p}$, $\mathcal{X}_{\text{U}}\in \mathbb{C}^{N_{\text{U}}\times N_s \times N_p}$, $\mathbf{X}_{\text{J}}\in \mathbb{C}^{N_s \times N_p}$, $\mathcal{Y}_{\text{C}}\in \mathbb{C}^{N_{\text{C}}\times N_s \times N_p}$, $\mathcal{Y}_{\text{U}}\in \mathbb{C}^{N_{\text{U}}\times N_s \times N_p}$,
$\mathbf{H}_{\text{C}}\in \mathbb{C}^{N_{\text{U}} \times N_p}$, $\mathcal{H}_{\text{U}}\in \mathbb{C}^{N_{\text{C}} \times N_{\text{U}} \times N_p}$, $\mathbf{H}_{\text{J}}\in \mathbb{C}^{N_{\text{U}} \times N_p}$,
$\mathcal{W}_{\text{C}}\in \mathbb{C}^{N_{\text{C}} \times N_s \times N_p}$, and $\mathcal{W}_{\text{U}}\in \mathbb{C}^{N_{\text{U}} \times N_s \times N_p}$, where $N_s=\bar N_s N_h / N_p$ and we assume $\bar N_s N_h$ can be divided by $N_p$ without loss of generality.
\begin{table}[!t]
	\vspace{-7mm}
	\centering
	\begin{threeparttable}
		\caption{Key Symbol Definitions}
		\label{table_SensingBound}
		\begin{small}
			\begin{tabular}{l@{\vrule width 1.5pt}l}
				\noalign{\hrule height 0.5mm}  
				\multicolumn{1}{l@{\vrule width 1.5pt}}{\hspace{-1mm}\textbf{Symbol  }  }  &\multicolumn{1}{c}{{\textbf{Definition}}} 
				\\ 
				\noalign{\hrule height 0.5mm}
				$f_c$ & \hspace{0.5mm}Carrier frequency\\  \hline
				$B$ & \hspace{0.5mm}Total work bandwidth\\ \hline
				$B_I$ & \hspace{0.5mm}Bandwidth of adjacent FHSS blocks\\ \hline
				$B_p$ & \hspace{0.5mm}Bandwidth for receiving hopping pilot signals\\ \hline
				$B_h$ & \hspace{0.5mm}FHSS signal bandwidth\\ \hline
				$T_h$ & \hspace{0.5mm}FHSS signal time interval\\  \hline
				$T_I$ & \hspace{0.5mm}Time interval of adjacent FHSS blocks\\ \hline
				$T_u$ & \hspace{0.5mm}{\begin{tabular}[l]{@{}l@{}} Time duration for one adjustment of the radiation\\  pattern\end{tabular}}\\ \hline
				$N_p$ & \hspace{0.5mm}{\begin{tabular}[l]{@{}l@{}} Number of reconfigurable radiation patterns,\\$N_p=N_{p,\text{azi}}N_{p,\text{ele}}$\end{tabular}} \\ \hline
				$N_{p,\text{azi}}$ & \hspace{0.5mm}{\begin{tabular}[l]{@{}l@{}}Number of reconfigurable radiation patterns in\\  azimuth dimension\end{tabular}} \\\hline
				$N_{p,\text{ele}}$ & \hspace{0.5mm}{\begin{tabular}[l]{@{}l@{}}Number of reconfigurable radiation patterns in\\  elevation dimension\end{tabular}} \\\hline
				$N_a$ & \hspace{0.5mm}Number of angle sampling points, $N_a=N_{a,\text{azi}}N_{a,\text{ele}}$\\ \hline
				$N_{a,\text{azi}}$ & \hspace{0.5mm}{\begin{tabular}[l]{@{}l@{}}Number of angle sampling points in azimuth\\  dimension\end{tabular}}\\ \hline
				$N_{a,\text{ele}}$ & \hspace{0.5mm}{\begin{tabular}[l]{@{}l@{}}Number of angle sampling points in elevation\\  dimension\end{tabular}} \\ \hline
				$N_h$ & \hspace{0.5mm}Number of hopping pilot signals \\ \hline
				$\bar N_s$ & \hspace{0.5mm}{\begin{tabular}[l]{@{}l@{}}Number of sampling points within each FHSS \\signal   \end{tabular}} \\ \hline
				$N_s$ & \hspace{0.5mm}{\begin{tabular}[l]{@{}l@{}}Number of sampling points within each mode of \\  radiation pattern, $N_s=\bar N_s N_h / N_p$ \end{tabular}} \\ \hline
				$N_{\text{C}}$ & \hspace{0.5mm}Number of controller's antennas, $N_{\text{C}}=N_{\text{C},\text{azi}}N_{\text{C},\text{ele}}$ \\ \hline
				$N_{\text{C},\text{azi}}$ & \hspace{0.5mm}Number of antennas in controller's azimuth dimension \\ \hline
				$N_{\text{C},\text{ele}}$ & \hspace{0.5mm}Number of antennas in controller's elevation dimension \\ \hline
				$N_{\text{U}}$ & \hspace{0.5mm}Number of UAV's antennas, $N_{\text{U}}=N_{\text{U},\text{azi}}N_{\text{U},\text{ele}}$ \\ \hline
				$N_{\text{U},\text{azi}}$ & \hspace{0.5mm}Number of antennas in UAV's azimuth dimension \\ \hline
				$N_{\text{U},\text{ele}}$ & \hspace{0.5mm}Number of antennas in UAV's elevation dimension \\
				\noalign{\hrule height 0.5mm} 
			\end{tabular}
		\end{small}
	\end{threeparttable}
	\vspace{-4mm}
\end{table}

\subsection{Channel Model}
Without loss of generality, we set the antenna spacing to be half the wavelength, although this is not a strict requirement in practice.
The UPA steering vector is given as $\mathbf{a}_{{N_{{\text{azi}}}},{N_{{\text{ele}}}}}\in{\mathbb{C}^{{N_{{\text{azi}}}}{N_{{\text{ele}}}}}}$ and it can be formulated as
\begin{align}
    {{\bf{a}}_{{N_{{\text{azi}}}},{N_{{\text{ele}}}}}}\left( {\varphi ,\theta } \right)\left[ n \right] = {e^{j\pi \left( {{n_{{\text{azi}}}}\cos {\varphi }\cos {\theta} + {n_{{\text{ele}}}}\sin {\varphi}} \right)}},
\end{align}
where $n = 0, \ldots ,{N_{{\text{azi}}}}{N_{{\text{ele}}}} - 1,\;
{n_{{\text{azi}}}} =  \bmod(n, {N_{{\text{azi}}}}) ,\;\text{and }{n_{{\text{ele}}}} = \left\lfloor {\frac{n}{{{N_{{\text{azi}}}}}}} \right\rfloor $\cite{ref_BoyuNing_OJCS}.
$\theta$ is the azimuth angle, $\varphi$ is the elevation angle.
As shown in Fig. \ref{fig_scene}(a), $N_{\text{azi}}$ is the number of antennas along the $y$-axis, and $N_{\text{ele}}$ is the number of antennas along the $z$-axis.

The channel from the jammer to the UAV, ${\bf{H}}_{\text{J}}\in\mathbb{C}^{N_{\text{U}}\times N_p}$, can be given as
\begin{align}
    {{\bf{H}}_{\text{J}}}\left[ {:,{n_p}} \right] = {\alpha _{\text{J}}}{e^{ - j2\pi {\varepsilon _{\text{J}}}}}{\mathbf{p}}\left( {{\varphi _{\text{J}}},{\theta _{\text{J}}}} \right)[n_p]{\bf{a}}_{{N_{{\text{U}},{\text{azi}}}},{N_{{\text{U}},{\text{ele}}}}}\left( {{\varphi _{\text{J}}},{\theta _{\text{J}}}} \right),
\end{align}
where ${\alpha _{\text{J}}}$ is the large-scale fading, $\varepsilon_\text{J}$ is the random phase,
$\varphi_{\text{J}}$ and $\theta_{\text{J}}$ are the elevation angle and azimuth angle of the jammer relative to the UAV, respectively.
$\mathbf{p}\in\mathbb{R}^{N_p}$ and $\mathbf{p}\left( {{\varphi _{\text{J}}},{\theta _{\text{J}}}} \right)[n_p]$ is the radiation pattern gain of the $n_p$-th radiation pattern mode when the elevation angle and azimuth angle are ${\varphi _{\text{J}}}$ and ${\theta _{\text{J}}}$, respectively.
The difference between RPR-FAS and traditional pattern-fixed antenna systems lies in $\mathbf{p}$. 
For pattern-fixed antenna systems, $\mathbf{p}\left( {{\varphi _{\text{J}}},{\theta _{\text{J}}}} \right)[n_p]$ remains constant, whereas for RPR-FAS, $\mathbf{p}\left( {{\varphi _{\text{J}}},{\theta _{\text{J}}}} \right)[n_p]$ changes with different antenna radiation patterns.

The channel from the UAV controller to the UAV, denoted as ${{\cal H}_{\text{C}}}\in\mathbb{C}^{N_{\text{U}}\times N_{\text{C}}\times N_p}$, is given as
\begin{align}
\label{eq_H_C}
    {{\cal H}_{\text{C}}}\left[ {:,:,{n_p}} \right] = {{\bf{A}}_{{\text{C}},{\text{rx}}}}{{\cal D}_{\text{C}}}\left[ {:,:,{n_p}} \right]{{\bf{A}}_{{\text{C}},{\text{tx}}}},
\end{align}
where  
\begin{align}
  &\mathbf{A}_{\text{C},\text{tx}} = \big[ \mathbf{a}_{N_{\text{C},\text{azi}} , N_{\text{C},\text{ele}}} \left( \boldsymbol\varphi_{\text{C},\text{tx}}[0], \boldsymbol\theta_{\text{C},\text{tx}}[0] \right), \ldots , \nonumber\\
  &\qquad\quad\ \ \ \mathbf{a}_{N_{\text{C},\text{azi}} , N_{\text{C},\text{ele}}} \left( \boldsymbol\varphi_{\text{C},\text{tx}}[, L_{\text{C}}-1], \boldsymbol\theta_{\text{C},\text{tx}}[L_{\text{C}}-1] \right) \big]^\text{T}, \\ 
  &\mathbf{A}_{\text{C},\text{rx}} = \big[ \mathbf{a}_{N_{\text{U},\text{azi}} , N_{\text{U},\text{ele}}} \left( \boldsymbol\varphi_{\text{C},\text{rx}}[0], \boldsymbol\theta_{\text{C},\text{rx}}[0] \right), \ldots , \nonumber\\
  &\qquad\quad\ \ \ \mathbf{a}_{N_{\text{U},\text{azi}} , N_{\text{U},\text{ele}}} \left( \boldsymbol\varphi_{\text{C},\text{rx}}[L_{\text{C}}-1], \boldsymbol\theta_{\text{C},\text{rx}}[L_{\text{C}}-1] \right) \big], \\ 
  &D_{\text{C}}[ : , : , n_p ] \hspace{-1mm}=\hspace{-1mm} \text{diag} \big( \boldsymbol\alpha_{\text{C}}[0] \mathbf{p} \left( \boldsymbol\varphi_{\text{C},\text{rx}}[0], \boldsymbol\theta_{\text{C},\text{rx}}[0] \right)[n_p]e^{-j 2 \pi \boldsymbol{\varepsilon}_{\text{C}}[0]},\nonumber\\
  &\qquad\quad\ldots , \boldsymbol\alpha_{\text{C}}[L_{\text{C}}-1] \mathbf{p}\big( \boldsymbol\varphi_{\text{C},\text{rx}}[L_{\text{C}}-1], \boldsymbol\theta_{\text{C},\text{rx}}[L_{\text{C}}-1] \big)[n_p]\nonumber\\
    & \quad\qquad e^{-j 2 \pi \boldsymbol{\varepsilon}_{\text{C}}[ L_{\text{C}}-1]} \big).
\end{align}
$\boldsymbol\varphi_{\text{C},\text{tx}}[l]$ and $\boldsymbol\theta_{\text{C},\text{tx}}[l]$ are the transmitted elevation angle and azimuth angle of the $l$-th path from the UAV controller to the UAV, respectively.
$\boldsymbol\varphi_{\text{C},\text{rx}}[l]$ and $\boldsymbol\theta_{\text{C},\text{rx}}[l]$ are the received elevation angle and azimuth angle of the $l$-th path from the UAV controller to the UAV, respectively.
$\boldsymbol\alpha_{\text{C}}[l]$ and $\boldsymbol\varepsilon_{\text{C}}[l]$ are the large-scale fading and random phase of the $l$-th path, respectively.
Therein, $l=0,\ldots,L_{\text{C}}-1$, and there are a total of $L_{\text{C}}$ paths.

If the UAV controller activates a single transmit antenna, the channel from the UAV controller to the UAV, denoted as ${{\bf H}_{\text{C}}}\in\mathbb{C}^{N_{\text{U}}\times N_p}$, is given by
\begin{align}
	\label{eq_H_C}
    {{\mathbf{H}}_{\text{C}}}\left[ {:,{n_p}} \right] = {{\mathbf{A}}_{{\text{C}},{\text{rx}}}}{\mathcal{D}_{\text{C}}}\left[ {:,:,{n_p}} \right]{{\mathbf{1}}_{{L_{\text{C}}}}}.
\end{align}
Similarly, the downlink channel from the UAV to the UAV controller, denoted as ${\mathcal{H}_{\text{U}}} \in {\mathbb{C}^{{N_{{\text{C}}}} \times {N_{{\text{U}}}} \times {N_p}}}$, is given as 
\begin{align}
    {{\cal H}_{\text{U}}}\left[ {:,:,{n_p}} \right] = {\cal H}_{\text{C}}^{\text{T}}\left[ {:,:,{n_p}} \right]
\end{align}
due to the reciprocity of the uplink and downlink channels.

\subsection{Radiation Pattern Model}
\begin{figure}[!t]
		\vspace{-7mm}
	\color{black}
	\subfigure[]{
		\begin{minipage}[t]{0.33\linewidth}
			\includegraphics[width=1.1in]{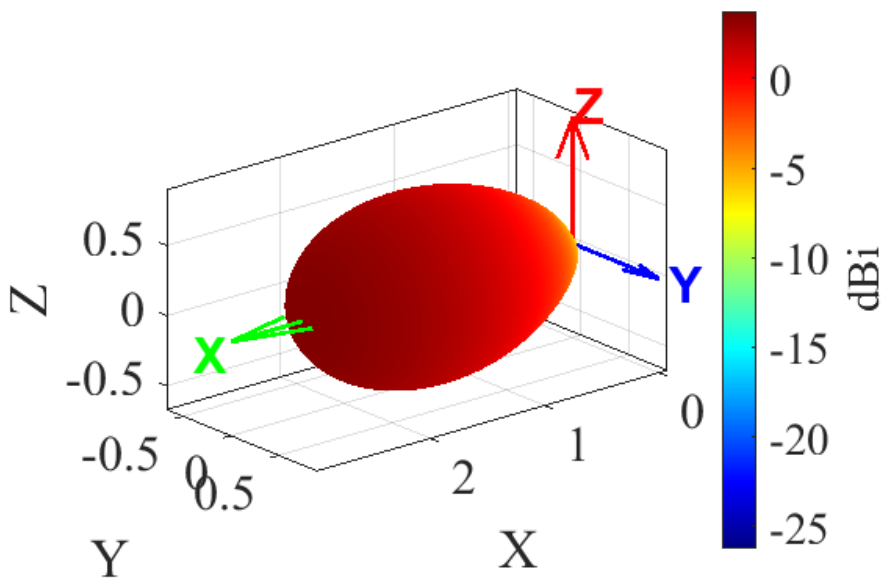}
			\label{fig_RadiationPattern_N_7_HPBW_65}
		\end{minipage}%
	}%
	\subfigure[]{
		\begin{minipage}[t]{0.33\linewidth}
			\includegraphics[width=1.1in]{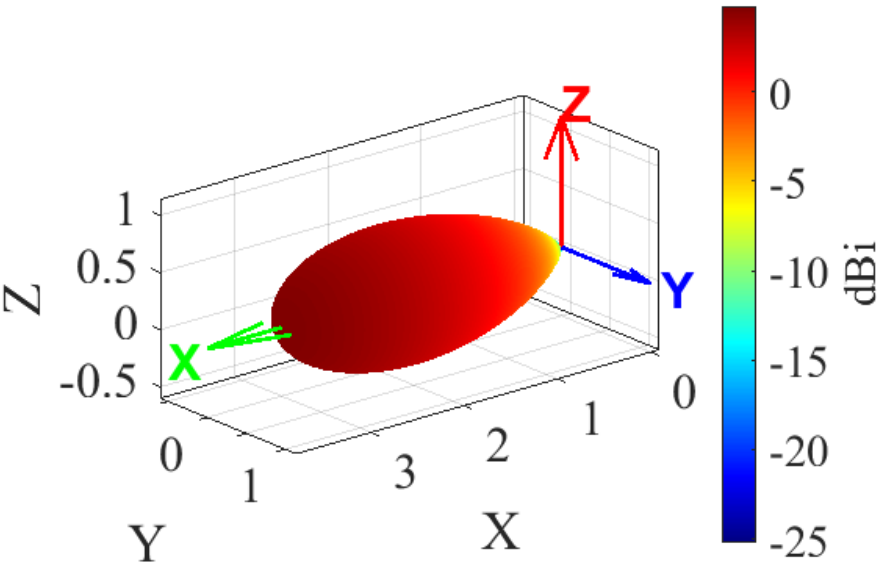}
			\label{fig_RadiationPattern_N_7_HPBW_45}
		\end{minipage}%
	}
	\subfigure[]{
		\begin{minipage}[t]{0.33\linewidth}
			\includegraphics[width=1.1in]{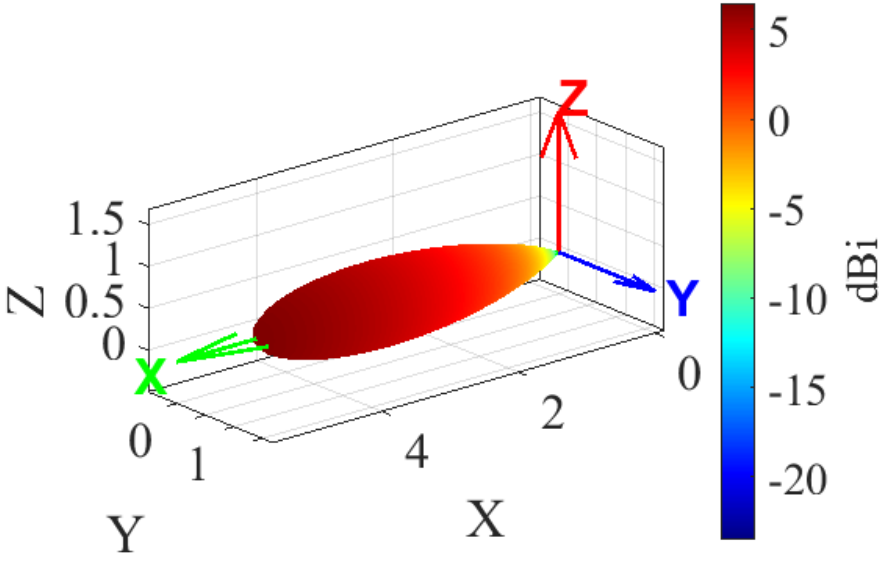}
			\label{fig_RadiationPattern_N_7_HPBW_25}
		\end{minipage}%
	}\\
	\subfigure[]{
		\begin{minipage}[t]{0.33\linewidth}
			\includegraphics[width=1.1in]{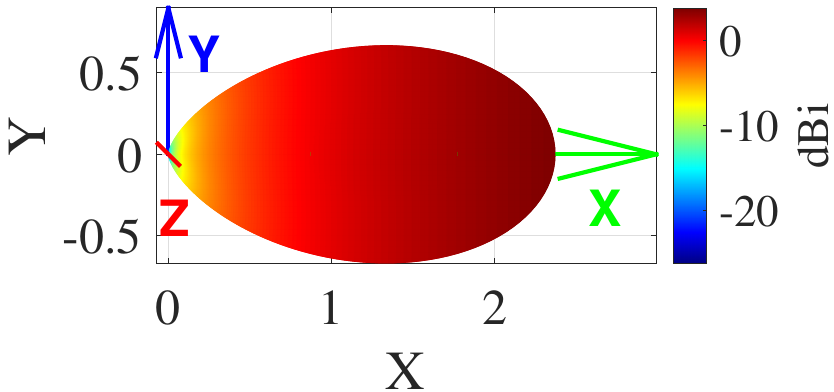}
			\label{fig_RadiationPattern_N_7_HPBW_65_plane}
		\end{minipage}%
	}%
	\subfigure[]{
		\begin{minipage}[t]{0.33\linewidth}
			\includegraphics[width=1.1in]{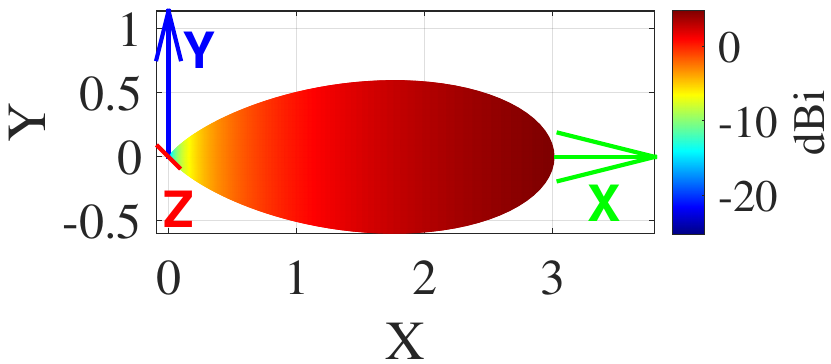}
			\label{fig_RadiationPattern_N_7_HPBW_45_plane}
		\end{minipage}%
	}
	\subfigure[]{
		\begin{minipage}[t]{0.33\linewidth}
			\includegraphics[width=1.1in]{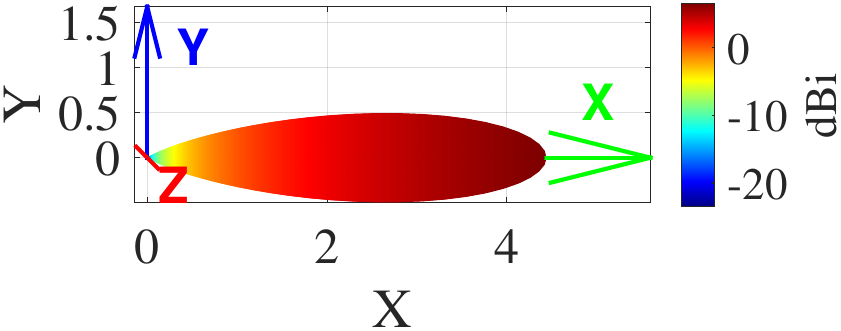}
			\label{fig_RadiationPattern_N_7_HPBW_25_plane}
		\end{minipage}%
	}\\
	\subfigure[]{
		\begin{minipage}[t]{0.33\linewidth}
			\includegraphics[width=0.8in]{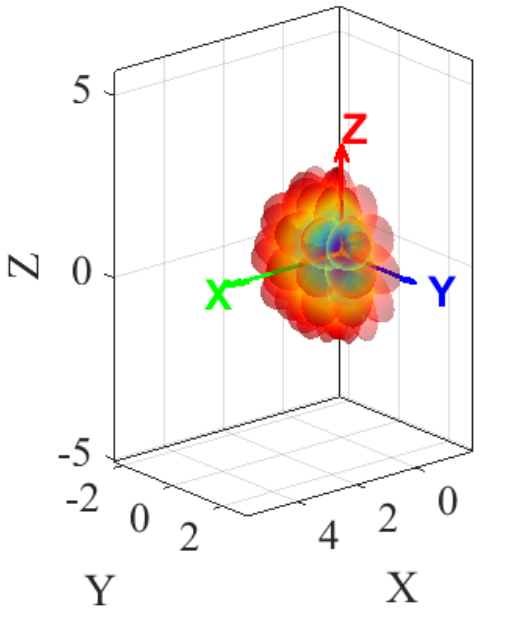}
			\label{fig_RadiationPattern_N_7_HPBW_65_All}
		\end{minipage}%
	}%
	\subfigure[]{
		\begin{minipage}[t]{0.33\linewidth}
			\includegraphics[width=0.8in]{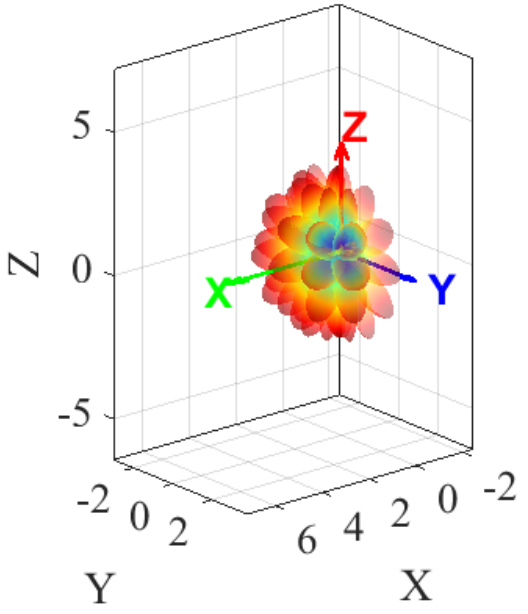}
			\label{fig_RadiationPattern_N_7_HPBW_45_All}
		\end{minipage}%
	}%
	\subfigure[]{
		\begin{minipage}[t]{0.33\linewidth}
			\includegraphics[width=0.8in]{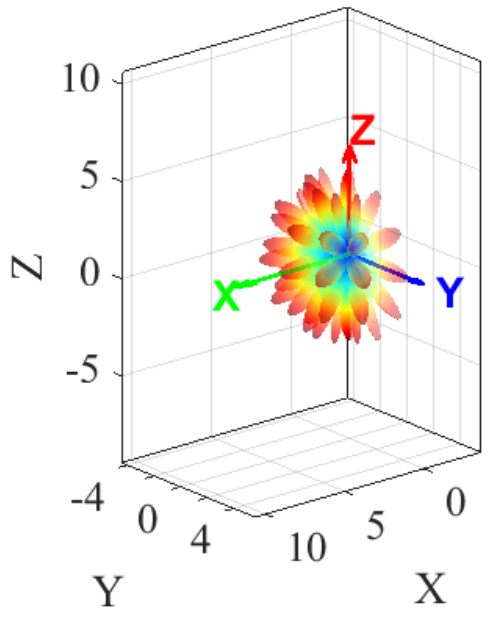}
			\label{fig_RadiationPattern_N_7_HPBW_25_All}
		\end{minipage}%
	}%
	\centering
	\caption{Visualization of different radiation patterns adopted in this paper.
	The power of antenna radiation patterns for different HPBW values has been normalized to 1 to ensure a fair comparison:
    (a) $\text{HPBW}=65^\circ$;
    (b) $\text{HPBW}=45^\circ$;
    (c) $\text{HPBW}=25^\circ$;
    (d) $\text{HPBW}=65^\circ$, X–Y plane view;
	(e) $\text{HPBW}=45^\circ$, X–Y plane view;
	(f) $\text{HPBW}=25^\circ$, X–Y plane view;  
    (g) $\text{HPBW}=65^\circ$, $N_{p,\text{azi}}=N_{p,\text{ele}}=7$;
    (h) $\text{HPBW}=45^\circ$, $N_{p,\text{azi}}=N_{p,\text{ele}}=7$;
    (i) $\text{HPBW}=25^\circ$, $N_{p,\text{azi}}=N_{p,\text{ele}}=7$.
    (a)--(f) show the benchmark radiation patterns, while (g)--(i) display all the radiation patterns simultaneously.}
	\label{fig_RadiationPattern}
\end{figure}
\begin{figure}[!t]
	\vspace{-7mm}
	\color{black}
	\subfigure[]{
		\begin{minipage}[t]{0.33\linewidth}
			\includegraphics[width=1.15in]{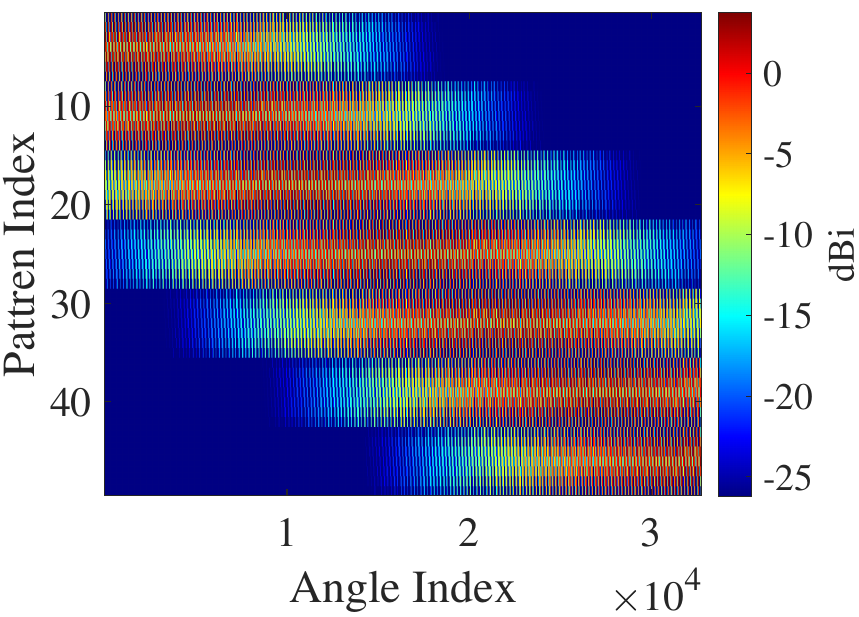}
			\label{fig_SensingMatrix_N_7_HPBW_65}
		\end{minipage}%
	}%
	\subfigure[]{
		\begin{minipage}[t]{0.33\linewidth}
			\includegraphics[width=1.15in]{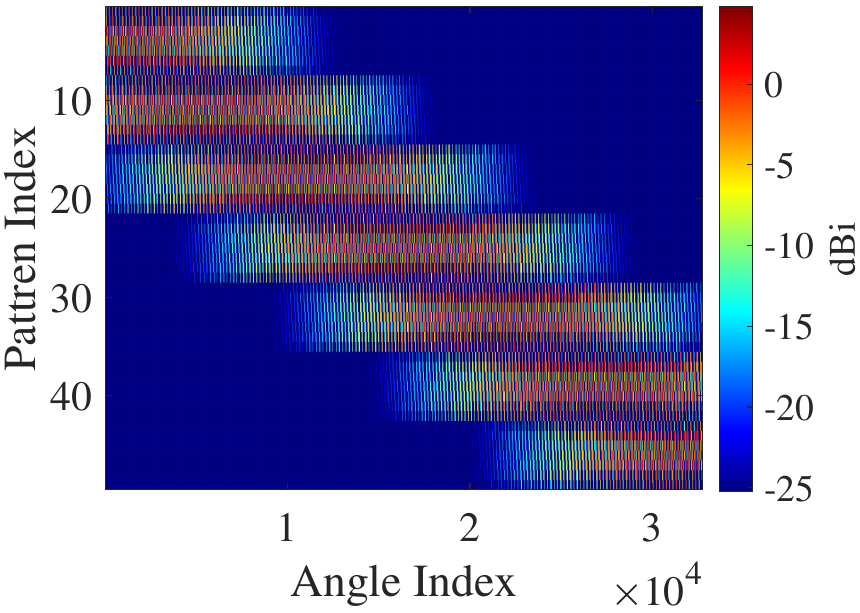}
			\label{fig_SensingMatrix_N_7_HPBW_45}
		\end{minipage}%
	}
	\subfigure[]{
		\begin{minipage}[t]{0.33\linewidth}
			\includegraphics[width=1.15in]{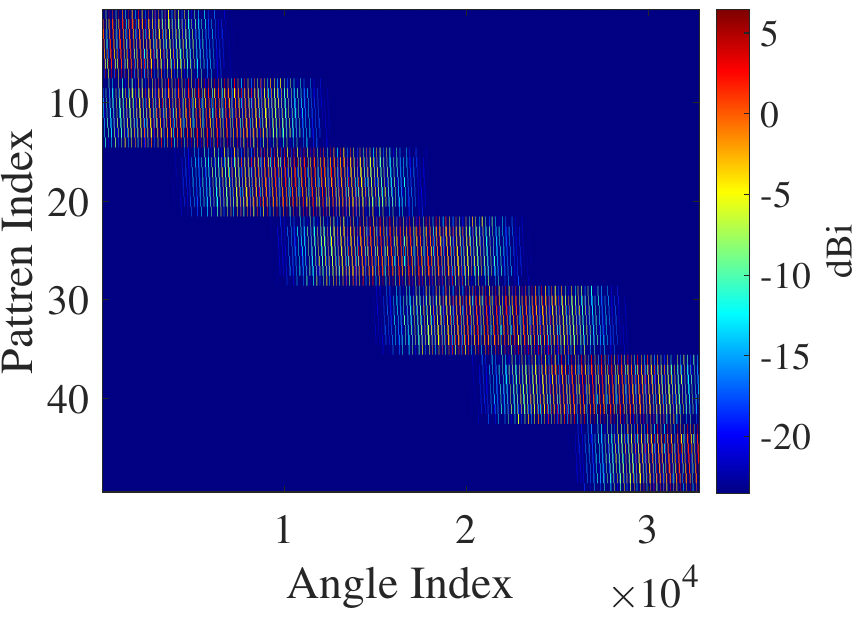}
			\label{fig_SensingMatrix_N_7_HPBW_25}
		\end{minipage}%
	}
	\centering
	\caption{Visualization of radiation pattern gain matrix $\mathbf{C}$:
		(a) $\text{HPBW}=65^\circ$, $N_{p,\text{ele}}=N_{p,\text{azi}}=7$;
		(b) $\text{HPBW}=45^\circ$, $N_{p,\text{ele}}=N_{p,\text{azi}}=7$;
		(c) $\text{HPBW}=25^\circ$, $N_{p,\text{ele}}=N_{p,\text{azi}}=7$.}
	\label{fig_SensingMatrix}
	\vspace{-3mm}
\end{figure}

Traditionally, the radiation pattern of an antenna is fixed.
Among the various radiation pattern models, the radiation pattern model in 3GPP-TR-38.901 is the widely used one~\cite{ref_38901}.
Therefore, the benchmark radiation pattern in this paper uses the 3GPP-TR-38.901 radiation pattern model to examine the angular resolution capability introduced by the variable radiation pattern.
The axis of the benchmark radiation pattern is set with elevation angle at 0$^\circ$ and azimuth angle at 90$^\circ$.
Different radiation patterns are realized by varying the axis angles.
The elevation and azimuth angles of the $n_p$-th radiation pattern are selected according to
\begin{equation}
\left\{
\begin{aligned}
\boldsymbol\varphi_{\text{axis}}[n_p]&=\left\lfloor {n_p/N_{p,\text{azi}}} \right\rfloor   \left[\kern-0.15em\left[ {180/(N_{p,\text{ele}}-1)} 
 \right]\kern-0.15em\right]-90,\\
\boldsymbol\theta_{\text{axis}}[n_p]&=\bmod\left(n_p, N_{p,\text{azi}}\right)\left[\kern-0.15em\left[ {180/(N_{p,\text{azi}}-1)} 
 \right]\kern-0.15em\right],
\end{aligned}
\right.
\end{equation}
where $\boldsymbol\varphi_{\text{axis}}\in\mathbb{C}^{N_p}$ and $\boldsymbol\theta_{\text{axis}}\in\mathbb{C}^{N_p}$.
The space is divided into 1$^\circ$ intervals for both the azimuth and elevation angles, with the range from 0$^\circ$ to 180$^\circ$ and $-90^\circ$ to 90$^\circ$ divided into 181 grid points.
That is, $N_{a,\text{ele}}=N_{a,\text{azi}}=181$ and $N_a=181^2$.
The radiation pattern gains at all angle grids together form the radiation pattern gain vector $\mathbf{c}_{n_p}\in\mathbb{R}^{N_a}$.
The elevation and azimuth angle of $\mathbf{c}_{n_p}$'s $n_a$-th element are
\begin{equation}
\left\{
\begin{aligned}
    \varphi  &= \left\lfloor {{n_a}/N_{a,\text{azi}}} \right\rfloor-90, \\
\theta  &= \bmod({n_a}, N_{a,\text{azi}}).
\end{aligned}
\right.
\end{equation}
All radiation pattern gain vectors form the radiation pattern gain matrix $\mathbf{C}\in\mathbb{C}^{N_p\times N_a}$, which is given as 
\begin{align}
    \mathbf{C}=[\mathbf{c}_0,\ldots,\mathbf{c}_{N_p-1}]^\text{T}.
\end{align}
To investigate the relationship among the characteristics of the reconfigurable radiation pattern, angular estimation capability, interference immunity, and communication performance, we modify the HPBW of the reference radiation pattern from 3GPP-TR-38.901.
During the modification, we ensure that the total radiation energy is normalized as 1 across different HPBW values.
We plot the reference radiation pattern and all radiation patterns for different HPBW and $N_p$ values in Fig.~\ref{fig_RadiationPattern}.
Additionally, we plot $\mathbf{C}$ in Fig.~\ref{fig_SensingMatrix}, where its subfigures (a), (b), and (c) have a one-to-one correspondence with subfigures (d), (e), and (f) of Fig.~\ref{fig_RadiationPattern}.

\section{Proposed Controller's Angle Estimation Scheme}\label{sec_controller_angle}
In this section, we formulate the controller's angle estimation as a compressed sensing problem and propose the LSO-OMP-MMV algorithm to solve it.
\subsection{Problem Formulation}
We assume ${{\bf{X}}_{\text{C}}}[{n_s},{n_p}]=1, \forall n_s,n_p$, without loss of generality.
The denoised and interference-reduced signal ${{\mathbf{S}}_{\text{U}}}\in\mathbb{C}^{N_{\text{U}}\times N_p}$ is obtained by arithmetically averaging ${\cal{Y}}_{\text{U}}$ along the sampling points dimension as
\begin{align}
    {{\mathbf{S}}_{\text{U}}}[{n_{{\text{rx}}}},{n_p}] = \sum\limits_{{n_s} = 0}^{{N_s} - 1} 
\frac{{{\cal{Y}_{\text{U}}}\left[ {{n_{{\text{rx}}}},{n_s},{n_p}} \right]} }{N_s}.
\end{align}
Assuming the channel remains constant during a single estimation period, the resulting optimization problem is formulated as 
\begin{align}
    \label{eq_problem_original}
    &\left\{ {{{{\mathbf{\hat {\boldsymbol{\alpha}} }}}_{\text{C}}},{{{\mathbf{\hat {\boldsymbol{\varphi}} }}}_{\text{C,rx}}},{{{\mathbf{\hat {\boldsymbol{\theta}} }}}_{\text{C,rx}}}} \right\} = \arg \mathop {\min }\limits_{\left\{ {{{{\mathbf{\hat {\boldsymbol{\alpha}} }}}_{\text{C}}}\left[ l \right],{{{\mathbf{\hat {\boldsymbol{\varphi}} }}}_{\text{C,rx}}}\left[ l \right],{{{\mathbf{\hat {\boldsymbol{\theta}} }}}_{\text{C,rx}}}\left[ l \right]} \right\},l = 0, \ldots ,{\hat{L}_{\text{C}}} - 1} \nonumber\\
    &\qquad\qquad\left\| {{{\mathbf{S}}_{\text{U}}} - \sum\limits_{l = 0}^{{\hat{L}_{\text{C}}} - 1} {{{{\mathbf{\hat {\boldsymbol\alpha} }}}_{\text{C}}}\left[ l \right]{{{\mathbf{\hat S}}}_{{\text{U}},l}}\left( {{{{\mathbf{\hat {\boldsymbol\varphi} }}}_{\text{C,rx}}}\left[ l \right],{{{\mathbf{\hat {\boldsymbol\theta} }}}_{\text{C,rx}}}\left[ l \right]} \right)} } \right\|_{\text{F}}^2,
\end{align}
where ${\hat {\boldsymbol{\alpha}} }_{\text{C}}\in\mathbb{R}^{\hat {L}_{\text{C}}}$, ${\hat {\boldsymbol{\varphi}} }_{\text{C,rx}}\in\mathbb{R}^{\hat {L}_{\text{C}}}$, and ${\hat {\boldsymbol{\theta}} }_{\text{C,rx}}\in\mathbb{R}^{\hat {L}_{\text{C}}}$ are the estimated gain vector, received elevation angle, and received azimuth angle, respectively.
${{{\mathbf{\hat S}}}_{{\text{U}},l}}\left( {{{{\mathbf{\hat {\boldsymbol\varphi} }}}_{\text{C,rx}}}\left[ l \right],{{{\mathbf{\hat {\boldsymbol\theta} }}}_{\text{C,rx}}}\left[ l \right]} \right)\in\mathbb{C}^{N_{\text{U}}\times N_p}$ is the estimated noise- and interference-free signal coming from the $l$-th path, and there are totally $\hat L_{\text{C}}$ estimated paths.
The parameter estimation problem in (\ref{eq_problem_original}) is a sparse recovery problem.
Since the number of reconfigurable radiation patterns $N_p$ is much smaller than $N_a$, the estimation of the angle from the UAV controller to the UAV can be formulated as a compressed sensing problem, with the sensing matrix being ${\mathbf{{ C}}}_{\text{N}}\in\mathbb{C}^{N_p\times N_a}$~\cite{ref_kml,ref_zxy,ref_lhs}.
${\mathbf{{ C}}}_{\text{N}}$ is the result of column-wise normalization of ${\mathbf{{ C}}}$.
The constructed compressed sensing problem can be formulated as 
\begin{equation}
    \label{eq_OMP_problem}
    \begin{aligned}
  &{\arg \mathop {\min }\limits_{\mathbf{G}_{\text{N}}} \left\| {{\mathbf{S}}_{\text{U}}^{\text{T}} - {\mathbf{ C}}}_{\text{N}}{\mathbf{G}}_{\text{N}} \right\|_{\text{F}}^2}, \\ 
  &{{\text{s}}.{\text{t}}.\ {{\left\| {{\mathbf{G}_{\text{N}}}\left[ {:,{n_{{\text{rx}}}}} \right]} \right\|}_0} = {{\hat L}_{\text{C}}}}, n_{\text{rx}}=0,\ldots,N_{\text{U}}-1,\\
  &\hspace{6mm}{\text{supp}}({{\mathbf{G}}_{\text{N}}}\left[ {:,{\mkern 1mu} m} \right]) = {\text{supp}}({{\mathbf{G}}_{\text{N}}}\left[ {:,{\mkern 1mu} n} \right]),\\ 
  &\hspace{6mm}m=0,\ldots,N_{\text{U}}-1; n=0,\ldots,N_{\text{U}}-1, 
\end{aligned}
\end{equation}
where ${\text{supp}}(\cdot)$ denotes the set of indices corresponding to the non-zero elements.

\begin{figure}[!t]
		\vspace{-7mm}
	\color{black}
	\subfigure[]{
		\begin{minipage}[t]{0.5\linewidth}
			\includegraphics[width=1.6in]{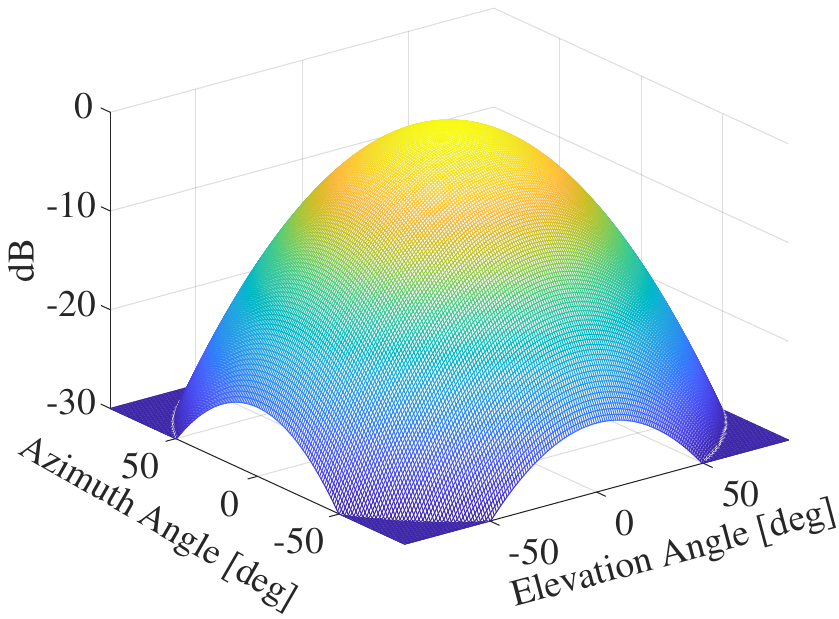}
			\label{fig_Original}
		\end{minipage}%
	}%
	\subfigure[]{
		\begin{minipage}[t]{0.5\linewidth}
			\includegraphics[width=1.7in]{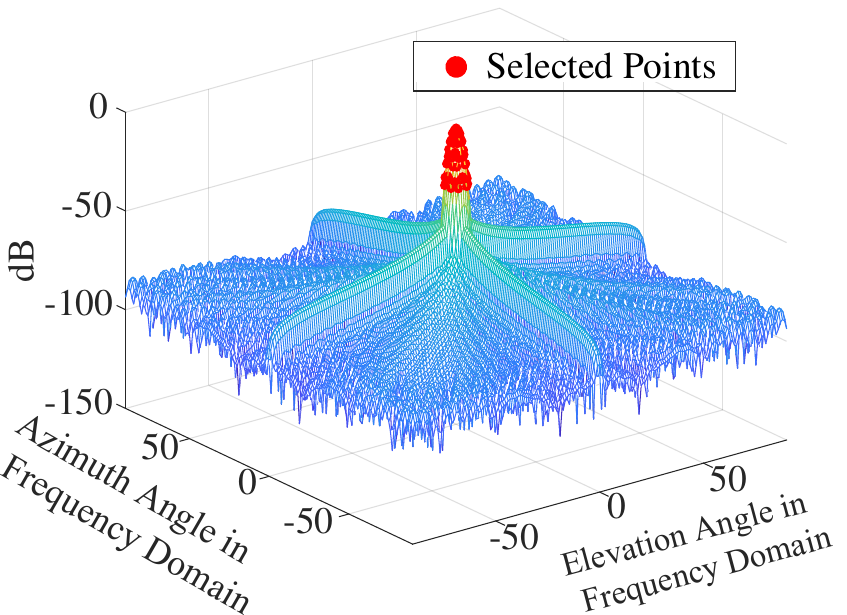}
			\label{fig_Low}
		\end{minipage}%
	}
	\centering
	\caption{
		Reduction of storage overhead for the radiation pattern gain matrix through Fourier transform ($\text{HPBW}=65^\circ$ and $N_{p,\text{ele}}=N_{p,\text{azi}}=7$):
		(a) $\mathbf{C}_{25}$, the matrix representation of the 25th row of the radiation pattern gain matrix;
		(b) $\mathbf{\tilde C}_{25}$, Fourier transform of $\mathbf{C}_{25}$.}
	\label{fig_LowSaveOverhead}
		\vspace{-3mm}
\end{figure}

\vspace{-2mm}
\subsection{Low Storage Overhead Sensing Matrix Generation Scheme}
$N_p$ is on the order of tens, while $N_a$ is $181^2$.
As a result, the number of elements in the radiation pattern gain matrix is considerably large, reaching an order of magnitude of $10^6$.
By visualizing the matrix representation of a representative row of the radiation pattern gain matrix, specifically the 25th row, in Fig.~\ref{fig_Original}, it can be observed that low-frequency components are more dominant.
Therefore, transforming the radiation pattern gain matrix to the frequency domain can reduce the computational cost.

We reshape the $n_p$-th row of $\mathbf{C}$, i.e., $\mathbf{c}_{n_p}^{\text{T}}$, into $\mathbf{C}_{n_p}\in\mathbb{C}^{N_{a,\text{ele}}\times N_{a,\text{azi}}}$, where the rows represent the elevation angles and the columns represent the azimuth angles.
We add hamming window on $\mathbf{C}_{n_p}$ to obtain
$\mathbf{\bar C}_{n_p}$ as 
\begin{align}
    \mathbf{\bar C}_{n_p} = \left(\mathbf{w}_{N_{a,\text{ele}}}\mathbf{w}_{N_{a,\text{azi}}}^\text{T}\right) \odot \mathbf{ C}_{n_p},
\end{align}
where $\mathbf{w}_{N_{a,\text{ele}}}\in\mathbb{R}^{_{N_{a,\text{ele}}}}$ and $\mathbf{w}_{N_{a,\text{azi}}}\in\mathbb{R}^{_{N_{a,\text{azi}}}}$ are the hamming windows.
We then apply two-dimensional Fourier transform on $\mathbf{\bar C}_{n_p}$ to obtain $\mathbf{\tilde C}_{n_p}$ as 
\begin{align}
    {{\mathbf{\tilde C}}_{{n_p}}} = {\mathbf{F}}{{\mathbf{\bar C}}_{{n_p}}}{{\mathbf{F}}^{\text{H}}},
\end{align}
where $\mathbf{F}$ is the Fourier transform matrix.
Visualization of $\mathbf{C}_{n_p}$ and $\mathbf{\tilde C}_{n_p}$ is shown in Fig.~\ref{fig_LowSaveOverhead}, where $n_p$ is specifically set to 25.
The red selected points in Fig.~\ref{fig_Low} indicate that their normalized amplitudes are greater than $-30$ dB.
We can see that the ``frequency-domain" version of $\mathbf{C}_{n_p}$ exhibits an energy concentration effect.
We only store the indices whose normalized magnitudes are greater than $-30$ dB along with their corresponding magnitudes.
Magnitudes below $-30$ dB are considered to be zero.
The ``frequency-domain" representation of this radiation pattern matrix is denoted as $\mathbf{\tilde C}_{n_p,\text{L}}$.
We can then obtain the reconstructed radiation pattern matrix as 
\begin{align}
    \label{eq_radiation_pattern_matrix}
    \mathbf{ C}_{n_p,\text{L}} = \text{Re}\left(\left(\mathbf{F}^{\text{H}}\mathbf{\tilde C}_{n_p,\text{L}}\mathbf{F}\right) \oslash \left(\mathbf{w}_{N_{a,\text{ele}}}\mathbf{w}_{N_{a,\text{azi}}}^\text{T}\right)\right),
\end{align}
where $\text{Re}(\cdot)$ means taking the real part.
Once $\mathbf{ C}_{n_p,\text{L}}$ is obtained, we can reshape it into $\mathbf{ c}_{n_p,\text{L}}\in\mathbb{R}^{N_{a,\text{ele}}N_{a,\text{azi}}}$ and obtain the radiation pattern gain matrix with low storage overhead $\mathbf{C}_{\text{L}}\in\mathbb{C}^{N_p\times N_a}$, which is given as 
\begin{align}
\label{eq_radiation_pattern_gain_matrix}
    \mathbf{C}_{\text{L}}=[\mathbf{c}_{0,\text{L}},\ldots,\mathbf{c}_{N_p-1,\text{L}}]^\text{T}.
\end{align}
The detailed normalized mean square error (NMSE) of the radiation pattern gain matrix with low storage overhead and the associated storage overhead ratio compared to the original matrix are shown in Table~\ref{table_NMSE}.
In Table~\ref{table_NMSE}, NMSE is calculated as
\begin{align}
    \text{NMSE} = 10\log_{10}\left(\left\| {{{\mathbf{C}}_{\text{L}}} - {\mathbf{C}}} \right\|_{\text{F}}^2/\left\| {\mathbf{C}} \right\|_{\text{F}}^2\right),
\end{align}
and the percentage is the storage overhead of $\mathbf{C}_{\text{L}}$ compared with $\mathbf{C}$.
\begin{table}[!t]
	\vspace{-7mm}
	\caption{NMSE and Storage Overhead Ratio}
	\label{table_NMSE}
	\centering
	\small
	\renewcommand{\arraystretch}{1} 
	\setlength{\tabcolsep}{8pt} 
	\begin{tabular}{|p{0.5cm}|p{1.8cm}|p{1cm}|p{1cm}|p{1cm}|}
		\hline
		\multicolumn{2}{|c|}{\diagbox{HPBW}{$N_p$}} & \hspace{3mm}11    & \hspace{4mm}9     & \hspace{4mm}7 \\ 
		\hline
		\multirow{2}{*}{\centering \hspace{1mm}65}  & NMSE (dB)  & \hspace{1mm}-25.5  & \hspace{1mm}-26.1  & \hspace{1mm}-24.9  \\ 
		\cline{2-5}
		& Ratio & \hspace{1mm}2.1\% & \hspace{1mm}2.4\% & \hspace{1mm}2.3\% \\ \hline
		\multirow{2}{*}{\centering \hspace{1mm}45}  & NMSE (dB)  & \hspace{1mm}-31.0  & \hspace{1mm}-29.9  & \hspace{1mm}-29.3  \\ 
		\cline{2-5}
		& Ratio & \hspace{1mm}3.1\% & \hspace{1mm}3.3\% & \hspace{1mm}3.4\% \\ \hline
		\multirow{2}{*}{\centering \hspace{1mm}25}  & NMSE (dB)  & \hspace{1mm}-36.4  & \hspace{1mm}-36.5  & \hspace{1mm}-37.7  \\ 
		\cline{2-5}
		& Ratio & \hspace{1mm}5.1\% & \hspace{1mm}4.9\% & \hspace{1mm}5.9\% \\ \hline
		\multirow{2}{*}{\centering \hspace{1mm}15}  & NMSE (dB)  & \hspace{1mm}-39.2  & \hspace{1mm}-37.8  & \hspace{1mm}-35.6  \\ 
		\cline{2-5}
		& Ratio & \hspace{1mm}6.8\% & \hspace{1mm}7.8\% & \hspace{1mm}9.4\% \\ \hline
	\end{tabular}
	\vspace{-3mm}
\end{table}
\vspace{-2mm}
\subsection{LSO-OMP-MMV Algorithm}
The optimization problem described in (\ref{eq_OMP_problem}) can be solved through OMP algorithm\cite{ref_LZR_JSTSP}, with the low storage overhead gain matrix $\mathbf{C}_{\text{L}}$ serving as sensing matrix.
$\mathbf{C}_{\text{N}}$ is also denoted as the result of column-wise normalization of $\mathbf{C}_{\text{L}}$ to avoid too many symbol definitions.
First, we define the residual as ${\mathbf{\Phi }} = {\mathbf{S}}_{\text{U}}^{\text{T}}$ and we denote $\eta_{0}=\left\|{\mathbf{S}}_{\text{U}}\right\|_{\text{F}}^{2}$.
The initial index list $\zeta$ is an empty list.
Then, the inner product of the residual and the sensing matrix is denoted as ${\mathbf{\Gamma }}\in\mathbb{C}^{N_a\times N_{\text{U}}}$ and can be obtained as 
\begin{align}
    \label{eq_inner}
    {\mathbf{\Gamma }} = {{\mathbf{ C}}_{\text{N}}^{\text{T}}}{\mathbf{\Phi }}.
\end{align}
By leveraging the UAV's multiple antennas, we can improve the SNR through 
\begin{align}
    \label{eq_MMV}
     {\mathbf{\bar \Gamma }} = {\mathbf{\Gamma }} \odot {\mathbf{\bar A}},
\end{align}
where ${\mathbf{\bar A}}\in\mathbb{C}^{N_a\times N_{\text{U}}}$ is composed of the steering vectors in every angle grids and it can be given as 
\begin{align}
    {\mathbf{\bar A}}[n_a,:] = {{\bf{a}}_{{N_{{\text{U,azi}}}},{N_{{\text{U,ele}}}}}^{\text{T}}}\left( \left\lfloor n_a/N_{a,\text{azi}}\right\rfloor ,\bmod(n_a, N_{a,\text{azi}})  \right).
\end{align}
Subsequently, the gain vector ${\boldsymbol{\gamma }}\in\mathbb{C}^{N_a}$ composed all angle grids can be obtained as 
\begin{align}
    \label{eq_gain_vector}
    {\boldsymbol{\gamma }} = \sum\limits_{{n_{{\text{U}}}} = 0}^{{N_{{\text{U}}}} - 1} {{\mathbf{\bar \Gamma }}\left[{:},{{n_{{\text{U}}}}} \right]}. 
\end{align}
We select the index of the ${\boldsymbol{\gamma }}$'s maximum as 
\begin{align}
    \label{eq_index}
     n_a^* = \arg \mathop {\max }\limits_{{n_a}} \left| {{\boldsymbol{\gamma }}[{n_a}]} \right|.
\end{align}
Finally, we update the index list as 
\begin{align}
    \label{eq_update_list}
    \zeta  = \left [ {\zeta ,n_a^*} \right ]
\end{align}
and update the residual as
\begin{align}
    \label{eq_update_residual}
    {\mathbf{\Phi }} = {\mathbf{S}}_{\text{U}}^{\text{T}} - {\mathbf{ C}}_{\text{L}}\left[ {:,\zeta } \right]{{\mathbf{ C}}_{\text{L}}^\dag }\left[ {:,\zeta } \right]{\mathbf{S}}_{\text{U}}^{\text{T}}.
\end{align}
Given a predefined constant $\eta_{\text{th}}$, if $\eta_{\text{th}}<\left\| {{\mathbf{\Phi }}} \right\|_{\text{F}}^2/\eta_0$, we continue to update the index list according to (\ref{eq_inner})--(\ref{eq_update_residual}).
Otherwise, we can obtain the channel angles as
\begin{align}
\label{eq_controller_angle}
\left\{
\begin{aligned}
  \hat{\boldsymbol{\varphi}}_{\text{C,rx}}[l_{\text{C}}] &= \left\lfloor \frac{\hat{\mathbf i}_{\text{C}}[l_{\text{C}}]}{N_{a,\text{azi}}} \right\rfloor-90, \\[1mm]
  \hat{\boldsymbol{\theta}}_{\text{C,rx}}[l_{\text{C}}] &=\bmod\left( \hat{\mathbf i}_{\text{C}}[l_{\text{C}}], N_{a,\text{azi}}\right),
\end{aligned}
\right.
\end{align}
where ${\hat{\mathbf i}_{{\text{C}}}[{l_{\text{C}}}]}$ is the ${l_{\text{C}}}$-th element of $\zeta$, $l_{\text{C}}=0,\ldots,\hat L_{\text{C}}-1$, and $\hat L_{\text{C}}$ is the number of elements of $\zeta$.
The complex gain of each path on each receive antenna can be expressed as ${{\mathbf{G}}_{\text{N}}} \in {\mathbb{C}^{{{\hat L}_{\text{C}}} \times {N_{\text{U}}}}}$ and ${{\mathbf{G}}} \in {\mathbb{C}^{{{\hat L}_{\text{C}}} \times {N_{\text{U}}}}}$.
${{\mathbf{G}}_{\text{N}}} $ represents the gain corresponding to ${{{\mathbf{C}}_{\text{N}}}}$, and ${\mathbf{G}}$ represents the gain corresponding to ${{{\mathbf{C}}_{\text{L}}}}$.
They can be given as 
\begin{align}
    \label{eq_gain_1}
  {{\mathbf{G}}_{\text{N}}} &= {\left( {{{\mathbf{C}}_{\text{N}}}[:,\zeta ]} \right)^{ - 1}}{\mathbf{S}}_{\text{U}}^{\text{T}},  \\
  \label{eq_gain_controller}
  {\mathbf{G}} &= {\left( {{{\mathbf{C}}_{\text{L}}}[:,\zeta ]} \right)^{ - 1}}{\mathbf{S}}_{\text{U}}^{\text{T}}.
\end{align}
The complete process of the proposed LSO-OMP-MMV algorithm is summarized in Algorithm \ref{alg_ParaEst}.

\subsection{Complexity Analysis}
The computational complexity of Algorithm~\ref{alg_ParaEst} can be summarized as follows. The initialization (lines~\ref{line_1}–\ref{line_4}) requires
$\mathcal{O}\bigl(N_p N_a \log_2 N_a + 2N_p N_a + N_p N_\text{U} + N_s N_\text{U} N_p\bigr)$. Evaluating (\ref{eq_inner}) costs $\mathcal{O}\bigl(N_p N_a N_\text{U}\bigr)$, while (\ref{eq_MMV}) and (\ref{eq_gain_vector}) each costs $\mathcal{O}\bigl(N_a N_\text{U}\bigr)$, and (\ref{eq_index}) costs $\mathcal{O}(N_a)$. In the $n_{\rm it}$-th iteration, computing (\ref{eq_update_residual}) incurs $\mathcal{O}\bigl(n_{\rm it}^3 + 2n_{\rm it}^2 N_p + n_{\rm it}N_p^2 + N_p^2 N_\text{U}\bigr)$, and (\ref{eq_gain_1})–(\ref{eq_gain_controller}) together incur $\mathcal{O}\bigl(2(n_{\rm it}^3 + 2n_{\rm it}^2 N_p + n_{\rm it}N_p N_\text{U})\bigr)$. Hence, over a total of $N_{\rm it}$ iterations, the overall complexity is
$\mathcal{O}\Bigl(\tfrac{3}{4}N_{\rm it}^4 + 2N_p N_{\rm it}^3 + \bigl(N_p N_\text{U} + \tfrac12N_p^2\bigr)N_{\rm it}^2 + \bigl(N_p N_a N_\text{U} + N_p^2 N_\text{U}\bigr)N_{\rm it} + N_p N_a\log_2 N_a + N_s N_\text{U} N_p\Bigr)$.

\begin{algorithm}[!t]
	\begin{small}
		\label{alg_ParaEst}
		\caption{Proposed LSO-OMP-MMV Algorithm}
		\LinesNumbered 
		\KwIn{Uplink signal received by the UAV ${\mathcal{Y}_{\text{U}}} \in {\mathbb{C}^{{N_{{\text{U}}}} \times {N_s} \times {N_p}}}$, predefined threshold for terminating the loop $\eta_\text{th}$.
		}
		\KwOut{Number of paths $\hat L_{\text{C}}$, AoA of the controller $\hat{\boldsymbol{\theta}}_{\text{C,rx}} \in \mathbb{R}^{ \hat L_{\text{C}}}$ and $\hat{\boldsymbol{\varphi}}_{\text{C,rx}}  \in \mathbb{R}^{ \hat L_{\text{C}}}$, index list $\zeta \in \mathbb{R}^{ \hat L_{\text{C}}}$ in radiation pattern gain matrix 
			$\mathbf{C}_{\text{L}}$ corresponding to the received angle $\hat{\boldsymbol{\theta}}_{\text{C,rx}} $ and $\hat{\boldsymbol{\varphi}}_{\text{C,rx}} $, gain of each path on each receive antenna ${{\mathbf{G}}_{\text{N}}} \in {\mathbb{C}^{{{\hat L}_{\text{C}}} \times {N_{\text{U}}}}}$
			and ${{\mathbf{G}}} \in {\mathbb{C}^{{{\hat L}_{\text{C}}} \times {N_{\text{U}}}}}$.
			
		}
		Obtain the radiation pattern gain matrix $\mathbf{C}_{\text{L}}$ through the low storage overhead ``frequency-domain" representation as (\ref{eq_radiation_pattern_matrix}) and (\ref{eq_radiation_pattern_gain_matrix})\label{line_1}\;
		Obtain the result of column-wise normalization of $\mathbf{C}_{\text{L}}$ and denote it as $\mathbf{C}_{\text{N}}$\;
		Initialize the empty list $\zeta$\;	
		Initialize the residual as ${\mathbf{\Phi }} = {\mathbf{S}}_{\text{U}}^{\text{T}}$ and calculate $\eta_{0}=\left\|{\mathbf{S}}_{\text{U}}\right\|_{\text{F}}^{2}$\label{line_4}\;	
		\While{$\eta_{\text{th}}<\left\| {{\mathbf{\Phi }}} \right\|_{\text{F}}^2/\eta_0$}{
			Obtain the inner product of the residual and the sensing matrix as (\ref{eq_inner})\;
			Improve SNR by using MMV property as (\ref{eq_MMV})\;
			Obtain the gain vector as (\ref{eq_gain_vector})\;
			Select the index of the gain vector's maximum as (\ref{eq_index})\;
			Update the index list as (\ref{eq_update_list})\;
			Update the residual as (\ref{eq_update_residual})\;
		}
		Obtain the received angles $\big(\hat{\boldsymbol{\theta}}_{\text{C,rx}} \in \mathbb{C}^{ \hat L_{\text{C}}}
		\text{ and } \hat{\boldsymbol{\varphi}}_{\text{C,rx}}  \in \mathbb{C}^{ \hat L_{\text{C}}}\big)$ and gains $\big({{\mathbf{G}}_{\text{N}}} \in {\mathbb{C}^{{{\hat L}_{\text{C}}} \times {N_{\text{U}}}}}\text{ and }{{\mathbf{G}}} \in {\mathbb{C}^{{{\hat L}_{\text{C}}} \times {N_{\text{U}}}}}\big)$ associated with the channel $\mathbf{H}_{\text{C}}$ as (\ref{eq_controller_angle})--(\ref{eq_gain_controller}).
	\end{small}
\end{algorithm}
\section{Proposed Jammer's Angle Estimation Scheme}\label{sec_jammer_angle}
In this section, we propose a maximum likelihood estimation method, based on the law of large numbers, to estimate the jammer's AoA.
This method reuses the same received signal utilized in Section~\ref{sec_controller_angle}, thereby reducing sensing overhead.
\subsection{Problem Formulation}
Let ${{\bf{\hat H}}_{\text{J}}}$ be the estimate of $\left|{{\bf{H}}_{\text{J}}}\right|$, and we suppose that ${\mathbf{\hat H}}_{\text{J}}$ has been extracted from ${\cal{Y}_{\text{U}}}$.
Then, we can construct ${\mathbf{H}}_{\text{J}}$ based on the received angle to approximate ${\mathbf{\hat H}}_{\text{J}}$ and obtain the received angle.
The least square problem can be formulated as 
\begin{align}
	\label{eq_jammer_angle_problem}
    \{{{\hat \varphi }_{\text{J}}},{{\hat \theta }_{\text{J}}}\} = \arg \mathop {\min }\limits_{{\varphi _{\text{J}}},{\theta _{\text{J}}}} \left\| {\mathbf{\hat H}}_{\text{J}}^{} - \left|{\mathbf{H}}_{\text{J}}^{}\left( {{\varphi _{\text{J}}},{\theta _{\text{J}}}} \right)\right| \right\|_{\text{F}}^2,
\end{align}
where ${{\hat \varphi }_{\text{J}}}$ and ${{\hat \theta }_{\text{J}}}$ are the estimated received elevation angle and azimuth angle from the jammer, respectively.
Therefore, the jammer's angle estimation problem can be divided into two subproblems.
The first subproblem involves extracting  ${{\bf{\hat H}}_{\text{J}}}$ from ${\cal{Y}_{\text{U}}}$, while the second subproblem focuses on obtaining ${{\hat \varphi }_{\text{J}}}$ and ${{\hat \theta }_{\text{J}}}$ by solving (\ref{eq_jammer_angle_problem}).

Under different radiation‐pattern configurations, the random interference from the jammer is multiplied by different gain coefficients. The only variation in the received signal across those configurations therefore come from the envelope of the random signal. By taking the magnitude of $\mathbf{H}_\text{J}$, we directly extract the directional gain that each pattern provides toward the jammer, rendering any phase diversity of the random interference immaterial.

\subsection{Law of Large Numbers based Maximum Likelihood Estimation Method}
Initially, subtract the controller's impact from ${{\cal Y}_{\text{U}}}$ as 
\begin{align}
	\label{eq_Y_U}
    {{\cal Y}_{\text{U}}}[{n_{{\text{rx}}}},{n_s},{n_p}] \Leftarrow {{\cal Y}_{\text{U}}}[{n_{{\text{rx}}}},{n_s},{n_p}] - {{\bf{\hat H}}_{\text{C}}}[{n_{{\text{rx}}}},{n_p}], 
\end{align}
since we can assume ${{\bf{X}}_{\text{C}}}[{n_s},{n_p}]=1, \forall n_s,n_p$, without loss of generality.
${{\bf{\hat H}}_{\text{C}}}$ is the estimate of ${{\bf{H}}_{\text{C}}}$, and it can be constructed like (\ref{eq_H_C}) by using estimated ${\hat {\boldsymbol{\alpha}} }_{\text{C}}$, ${\hat {\boldsymbol{\varphi}} }_{\text{C,rx}}$, and ${\hat {\boldsymbol{\theta}} }_{\text{C,rx}}$ in Algorithm \ref{alg_ParaEst}.
The sampled points associated with each radiation pattern mode are accumulated to obtain ${\mathbf{S}}_{\text{J}} \in \mathbb{C}^{N_{\text{U}}\times N_p}$ as
\begin{align}
\label{eq_rationale}
    {\mathbf{S}}_{\text{J}}[{n_{{\text{rx}}}},{n_p}] = \sum\limits_{{n_s} = 0}^{{N_s} - 1} 
\frac{{\left| {{\mathcal{Y}_{\text{U}}}[{n_{{\text{rx}}}},{n_s},{n_p}]} \right|}}{N_s}.
\end{align}
The rationale behind (\ref{eq_rationale}) is based on the assumption that ${{\mathbf{X}}_{\text{J}}}[{n_s},{n_p}]\sim\mathcal{CN}\left( {0,\sigma _{\text{J}}^2} \right)$ and ${\mathcal{W}_{\text{U}}}[{n_{{\text{rx}}}},{n_s},{n_p}]\sim\mathcal{CN}\left( {0,\sigma _{\text{U}}^2} \right)$.
Therefore, ${\mathcal{Y}_{\text{U}}}[{n_{{\text{rx}}}},{n_s},{n_p}]\sim\mathcal{CN}\left( {0,\left|{\mathbf{H}}_{\text{J}}[{n_{{\text{rx}}}},{n_p}]\right|^2\sigma _{\text{J}}^2 + \sigma _{\text{U}}^2} \right)$ and $\left| {{\mathcal{Y}_{\text{U}}}[{n_{{\text{rx}}}},{n_s},{n_p}]} \right|$ follows a folded normal distribution.
Let $\tilde{\boldsymbol{\Sigma}}^2[{n_{{\text{rx}}}},{n_p}] = \left|{\mathbf{H}}_{\text{J}}[{n_{{\text{rx}}}},{n_p}]\right|^2\sigma _{\text{J}}^2 + \sigma _{\text{U}}^2$, the expectation and variance of $\left| {{\mathcal{Y}_{\text{U}}}[{n_{{\text{rx}}}},{n_s},{n_p}]} \right|$ can be obtained as 
\begin{equation}
\left\{    
\begin{aligned}
  \mathbb{E}\left( {\left| {{\mathcal{Y}_{\text{U}}}[{n_{{\text{rx}}}},{n_s},{n_p}]} \right|} \right) &= {\tilde{\boldsymbol{\Sigma}}[{n_{{\text{rx}}}},{n_p}]}\sqrt {\frac{2}{\pi }} ,  \\
  {\text{Var}}\left( {\left| {{\mathcal{Y}_{\text{U}}}[{n_{{\text{rx}}}},{n_s},{n_p}]} \right|} \right)& = \tilde{\boldsymbol{\Sigma}}^2[{n_{{\text{rx}}}},{n_p}]\frac{{\pi  - 2}}{\pi },
\end{aligned}
\right.
\end{equation}
where $\mathbb{E}(\cdot)$ and $\text{Var}(\cdot)$ represent expectation operator and variance operator, respectively.
Therefore, the expectation and variance of ${\mathbf{S}}_{\text{J}}[{n_{{\text{rx}}}},{n_p}]$ can be obtained as 
\begin{equation}
\left\{    
\begin{aligned}
  \mathbb{E}\left( {{\mathbf{S}}_{\text{J}}[{n_{{\text{rx}}}},{n_p}]} \right) &= {\tilde{\boldsymbol{\Sigma}}[{n_{{\text{rx}}}},{n_p}]}\sqrt {\frac{2}{\pi }}  , \\
  {\text{Var}}\left( {{\mathbf{S}}_{\text{J}}[{n_{{\text{rx}}}},{n_p}]} \right) &= \tilde{\boldsymbol{\Sigma}}^2[{n_{{\text{rx}}}},{n_p}]\frac{{\pi  - 2}}{{\pi {N_s}}}.
\end{aligned}
\right.    
\end{equation}
As long as $N_s$ is sufficiently large, the variance of ${\mathbf{S}}_{\text{J}}[{n_{{\text{rx}}}},{n_p}]$ becomes sufficiently small, and ${{\mathbf{S}}_{\text{J}}}[{n_{{\text{rx}}}},{n_p}]$ can be approximated as 
\begin{align}
    \mathop {\lim }\limits_{{N_s} \to \infty } {{\mathbf{S}}_{\text{J}}}[{n_{{\text{rx}}}},{n_p}] = {\tilde{\boldsymbol{\Sigma}}[{n_{{\text{rx}}}},{n_p}]}\sqrt {\frac{2}{\pi }}. 
\end{align}
Therefore, 
\begin{align}
    \mathop {\lim }\limits_{{N_s} \to \infty }\left|{\mathbf{H}}_{\text{J}}[{n_{{\text{rx}}}},{n_p}]\right| = \sqrt {\left( {{{\mathbf{S}}_{\text{J}}^2}[{n_{{\text{rx}}}},{n_p}]\frac{\pi }{2} - \sigma _{\text{U}}^2} \right)\frac{1}{{\sigma _{\text{J}}^2}}}, 
\end{align}
and we define the estimation of $\left|{\mathbf{H}}_{\text{J}}[{n_{{\text{rx}}}},{n_p}]\right|$ as
\begin{align}
	\label{eq_H_J}
    {\mathbf{\hat H}}_{\text{J}}[{n_{{\text{rx}}}},{n_p}] = \sqrt {\left( {{{\mathbf{S}}_{\text{J}}^2}[{n_{{\text{rx}}}},{n_p}]\frac{\pi }{2} - \sigma _{\text{U}}^2} \right)\frac{1}{{\sigma _{\text{J}}^2}}}. 
\end{align}
After we have obtained ${\mathbf{\hat H}}_{\text{J}}[{n_{{\text{rx}}}},{n_p}]$, we now try to solve the problem mentioned in (\ref{eq_jammer_angle_problem}).
Since we assume that $\mathbf{H}_{\text{J}}$ is a line-of-sight (LoS) path dominant channel, as otherwise the jamming energy would be too low to cause effective jamming due to path loss, we employ the maximum likelihood estimation method to obtain the received angle of the jammer as
\begin{align}
    \left\{
    \begin{aligned}
      \hat{\varphi}_{\text{J}} &= \left\lfloor \frac{\hat{i}_{\text{J}}}{N_{a,\text{azi}}} \right\rfloor-90, \\ 
      \hat{\theta}_{\text{J}} &= \bmod\left( \hat{i}_{\text{J}}, N_{a,\text{azi}} \right) ,
    \end{aligned}
    \right.
\end{align}
where 
\begin{equation}
\begin{aligned}
	\label{eq_MLE}
  {{\hat i}_{\text{J}}} &= \arg \mathop {\max }\limits_{i} {\mathbf{a}}[i], \\
  {\text{s}}.{\text{t}}.\quad {\mathbf{a}} &= {\left( {{\mathbf{\hat H}}_{\text{J}}^{}{\mathbf{ C}}_{\text{N}}} \right)^{\text{T}}{{\mathbf{1}}_{{N_{{\text{U}}}}}}}.
\end{aligned}
\end{equation}

\subsection{Complexity Analysis}
The computational complexity of the law of large numbers based maximum likelihood estimation method is as follows: updating $\mathcal{Y}_{\text{U}}$ in (\ref{eq_Y_U}) requires $\mathcal{O}(N_{\text{U}}N_{s}N_{p})$, obtaining $\mathbf{S}_{\text{J}}$ in (\ref{eq_rationale}) requires $\mathcal{O}(N_{\text{U}}N_{s}N_{p})$, obtaining $\mathbf{H}_{\text{J}}$ in (\ref{eq_H_J}) requires $\mathcal{O}(N_{\text{U}}N_{p})$, and solving the problem in (\ref{eq_MLE}) requires $\mathcal{O}(N_{\text{U}}N_{p}N_{a})$. Consequently, the overall complexity is $\mathcal{O}\bigl(N_{\text{U}}N_{p}(N_{a} + 2N_{s})\bigr)$.

\section{Proposed Environment-Aware Optimal Radiation Pattern Selection Scheme}\label{sec_pattern_selection}
In this section, building on the estimated AoAs of the controller and jammer, we propose an environment-aware optimal radiation pattern selection scheme.
By leveraging an alternating optimization approach, this scheme maximizes the uplink SE.
\subsection{Problem Formulation}
Assuming there is only one uplink data stream, if the radiation pattern is kept fixed, the uplink data transmission model will change from (\ref{eq_up_tensor}) to
\begin{align}
	\label{eq_data_tx}
    {y_{\text{U}}} = \mathbf{b}^\text{H}{{\mathbf{h}}_{\text{J}}}{{x}}_{\text{J}} + \mathbf{b}^\text{H}{{\mathbf{w}}_{\text{U}}} + \mathbf{b}^\text{H}{{\mathbf{H}}_{\text{C}}}\mathbf{f}{{{x}}_{\text{C}}},
\end{align}
where $\mathbf{b}\in\mathbb{C}^{N_{\text{U}}}$ is the combiner,
$\mathbf{f}\in\mathbb{C}^{N_{\text{C}}}$ is the beamformer, ${{\mathbf{h}}_{\text{J}}}\in\mathbb{C}^{N_{\text{U}}}$, ${{\mathbf{w}}_{\text{U}}}\in\mathbb{C}^{N_{\text{U}}}$, and ${{\mathbf{H}}_{\text{C}}}\in\mathbb{C}^{N_{\text{U}}\times {N_{\text{C}}}}$.
The meanings of the variables in (\ref{eq_data_tx}) can be analogized to those in (\ref{eq_up_tensor})
The signal model of (\ref{eq_data_tx}) can be simplified as
\begin{align}
	{y_{\text{U}}} = \mathbf{b}^\text{H}{{\mathbf{H}}_{\text{C}}}\mathbf{f}{{{x}}_{\text{C}}} + {{\tilde w}_{\text{U}}},
\end{align}
where ${{\tilde w}_{\text{U}}} = {{\mathbf{b}}^{\text{H}}}{{\mathbf{h}}_{\text{J}}}{x_{\text{J}}} + {{\mathbf{b}}^{\text{H}}}{{\mathbf{w}}_{\text{U}}}$.
We assume $\mathbb{E}\left(x_{\text{C}}x_{\text{C}}^* \right) = P_{\text{C}}$ and $P_{\text{C}}$ is the controller's transmit power.
Then, the uplink SE can be given as
\begin{align}
	\label{eq_SE}
	\text{SE}_{\text{up}}={\log _2}\left( {1 + P_{\text{C}}\frac{{{{\mathbf{b}}^{\text{H}}}{{\mathbf{H}}_{\text{C}}}{\mathbf{f}}{{\mathbf{f}}^{\text{H}}}{\mathbf{H}}_{\text{C}}^{\text{H}}{\mathbf{b}}}}{{\tilde \sigma _{\text{U}}^2}}} \right),
\end{align}
where $\tilde \sigma _{\text{U}}^2$ is the variance of ${{\tilde w}_{\text{U}}}$ and ${{\tilde w}_{\text{U}}}\sim\mathcal{C}\mathcal{N}\left( {0,\tilde \sigma _{\text{U}}^2} \right)$.
However, the UAV only knows the received angle, meaning it only has partial information about $\mathbf{H}_{\text{C}}$.
Therefore, by assuming
\begin{align}
	\label{eq_omni}
	|{\mathbf{a}}_{{N_{{\text{C,azi}}}},{N_{{\text{C,ele}}}}}^{\text{H}}\left( {{\boldsymbol\varphi _{\text{C,tx}}}\left[ {{l_{\text{C}}}} \right],{\boldsymbol\theta _{\text{C,tx}}}\left[ {{l_{\text{C}}}} \right]} \right){\mathbf{f}}|=\sqrt{P_{\text{tx}}}
\end{align}
for different $l_{\text{C}}$, maximizing the SE in (\ref{eq_SE}) is converted to selecting the optimal radiation pattern and designing the combiner to maximize the SE.
The rationale behind (\ref{eq_omni}) is the requirement for omni-directional transmission to ensure the quality of the signal received by UAVs.
The expression of the approximate SE can be formulated as
\begin{align}\label{eq_SE_up}
	\widehat{\text{SE}}_{\text{up}}={\log _2}\left( {1 + P_{\text{C}}P_{\text{tx}}\frac{{\sum\limits_{{l_{\text{C}}} = 0}^{{L_{\text{C}}} - 1} {{{\left| {{{\boldsymbol{\alpha }}_{\text{C}}}\left[ {{l_{\text{C}}}} \right]} \right|}^2}{{\mathbf{b}}^{\text{H}}}\mathcal{A}\left[ {:,:,{l_{\text{C}}}} \right]{\mathbf{b}}} }}{{\tilde \sigma _{\text{U}}^2}}} \right),
\end{align}
where $\mathcal{A}\left[ {:,:,{l_{\text{C}}}} \right] = {\mathbf{a}}_{{N_{{\text{U,azi}}}},{N_{{\text{U,ele}}}}}^{}\left( {{\boldsymbol\varphi _{\text{C,rx}}}\left[ {{l_{\text{C}}}} \right],{\boldsymbol\theta _{\text{C,rx}}}\left[ {{l_{\text{C}}}} \right]} \right)$
${{\mathbf{a}}_{{N_{{\text{U,azi}}}},{N_{{\text{U,ele}}}}}^{\text{H}}\left( {{\boldsymbol\varphi _{\text{C,rx}}}\left[ {{l_{\text{C}}}} \right],{\boldsymbol\theta _{\text{C,rx}}}\left[ {{l_{\text{C}}}} \right]} \right)}$ and $\boldsymbol{\alpha}_{\text{C}}$ in (\ref{eq_SE_up}) includes both large-scale fading and antenna radiation pattern gain for expression simplicity.
After obtaining the optimal radiation pattern and the combiner, $\mathbf{f}$ can be computed by the water-filling algorithm to obtain the ultimate SE\cite{ref_Tse}.
Since the power of the jamming signal is much larger than that of the noise, the uplink SE maximizing problem can be approximately formulated by maximizing the signal to interference ratio as 
\begin{align}
\label{eq_P_0}
  &{P_0}:\arg \mathop {\max }\limits_{n,{\mathbf{b}}} \quad\frac{{\sum\limits_{{l_{\text{C}}} = 0}^{{{\hat L}_{\text{C}}} - 1} {{\boldsymbol{\hat \alpha }}_{\text{C}}^2\left[ {{l_{\text{C}}}} \right]{{\mathbf{C}}^2}\left[ {n,{\hat{\mathbf i}_{{\text{C}}}[{l_{\text{C}}}]}} \right]{P_{\text{B}}}\left( {{\hat{\mathbf i}_{{\text{C}}}[{l_{\text{C}}}]},{\mathbf{b}}} \right)} }}{{{{\mathbf{C}}^2}\left[ {n,{{\hat i}_{\text{J}}}} \right]{P_{\text{B}}}\left( {{{\hat i}_{\text{J}}},{\mathbf{b}}} \right)}}, \nonumber\\
  &{\text{s}}.{\text{t}}.\ {P_{\text{B}}}\left( {{\hat{\mathbf i}_{{\text{C}}}[{l_{\text{C}}}]},{\mathbf{b}}} \right) = {{\mathbf{b}}^{\text{H}}}{\mathbf{a}}_{{N_{{\text{U,azi}}}},{N_{{\text{U,ele}}}}}^{}\left( {{\hat{\boldsymbol \varphi }_{{\text{C,rx}}}[{l_{\text{C}}}]},{\hat{\boldsymbol \theta }_{{\text{C,rx}}}[{l_{\text{C}}}]}} \right)\nonumber\\&\qquad\qquad\qquad\qquad\;\;\;\;{\mathbf{a}}_{{N_{{\text{U,azi}}}},{N_{{\text{U,ele}}}}}^{\text{H}}\left( {{\hat{\boldsymbol \varphi }_{{\text{C,rx}}}[{l_{\text{C}}}]},{\hat{\boldsymbol \theta }_{{\text{C,rx}}}[{l_{\text{C}}}]}} \right){\mathbf{b}},  \nonumber\\
  &\hspace{12mm} {P_{\text{B}}}\left( {{{\hat i}_{\text{J}}},{\mathbf{b}}} \right) = {{\mathbf{b}}^{\text{H}}}{\mathbf{a}}_{{N_{{\text{U,azi}}}},{N_{{\text{U,ele}}}}}^{}\left( {{{\hat \varphi }_{\text{J}}},{{\hat \theta }_{\text{J}}}} \right)\nonumber\\
  &\hspace{12mm}{\mathbf{a}}_{{N_{{\text{U,azi}}}},{N_{{\text{U,ele}}}}}^{\text{H}}\left( {{{\hat \varphi }_{\text{J}}},{{\hat \theta }_{\text{J}}}} \right){\mathbf{b}}, 
\end{align}
where ${{\boldsymbol{\hat \alpha }}_{\text{C}}} = \sum\nolimits_{{n_{\text{U}}}{\text{ = 0}}}^{{N_{\text{U}}}{{ - 1}}} {\left| {{\mathbf{G}}[:,{n_{\text{U}}}]} \right|} /{N_{\text{U}}}$, $n$ is the index of the best radiation pattern, and $\mathbf{b}\in\mathbb{C}^{N_{\text{U}}}$ is the best combiner.
The problem $P_0$ described in (\ref{eq_P_0}) will be addressed in Section \ref{sec_alternating_opt}.

If the radiation pattern is kept fixed, the downlink data transmission model will change from (\ref{eq_dn_tensor}) to 
\begin{align}
	{{\mathbf{Y}}_{\text{C}}} = {{\mathbf{W}}_{\text{C}}} + {{\mathbf{H}}_{\text{U}}}{{\mathbf{X}}_{\text{U}}},
\end{align}
where ${{\mathbf{Y}}_{\text{C}}}\in\mathbb{C}^{N_{\text{C}}\times N_s}$, ${{\mathbf{W}}_{\text{C}}}\in\mathbb{C}^{N_{\text{C}}\times N_s}$, ${{\mathbf{H}}_{\text{U}}}\in\mathbb{C}^{N_{\text{C}}\times {N_{\text{U}}}}$, and ${{\mathbf{X}}_{\text{U}}}\in\mathbb{C}^{N_{\text{U}}\times {N_s}}$.
The downlink SE can be achieved up to the channel capacity using the classic water-filling algorithm, where the channel capacity can be adjusted by modifying the radiation pattern.
Meanwhile, it is necessary to minimize the energy received by the jammer from the UAV as much as possible to increase the difficulty of detecting the UAV.
Therefore, by adjusting the radiation pattern, the optimization problem of maximizing SE while minimizing the energy received by the jammer can be modeled as 
\begin{align}
\label{eq_pattern_selection_dn}
    \arg \mathop {\max }\limits_n \frac{{\sum\limits_{{l_{\text{C}}} = 0}^{{{\hat L}_{\text{C}}} - 1} {{\boldsymbol{\hat \alpha }}_{\text{C}}^2\left[ {{l_{\text{C}}}} \right]{{\mathbf{C}}^2}\left[ {n,{\hat{\mathbf i}_{{\text{C}}}[{l_{\text{C}}}]}} \right]} }}{{{{\mathbf{C}}^2}\left[ {n,{{\hat i}_{\text{J}}}} \right]}}.
\end{align}
The problem formulated in (\ref{eq_pattern_selection_dn}) can be efficiently solved using the exhaustive one-dimensional search.

\subsection{Alternating Optimization of Radiation Pattern and Combiner for the Uplink Transmission}\label{sec_alternating_opt}
Problem $P_0$ described in (\ref{eq_P_0}) is non-convex.
It is difficult to optimize $n$ and $\mathbf{b}$ simultaneously.
Therefore, we adopt alternating optimization method to optimize $n$ and $\mathbf{b}$.
Let the selected radiation mode and the estimated combiner in the previous step be $n_{k-1}$ and $\mathbf{b}_{k-1}$, respectively.
The selection of $n_k$ can then be obtained using the exhaustive one-dimensional search, and the optimization problem is formulated as 
\begin{align}
  & {P_1}\hspace{-1mm}:\hspace{-1mm}n_k\hspace{-1mm}=\hspace{-1mm}\arg \mathop {\max }\limits_n \hspace{-1mm}\frac{{\sum\limits_{{l_{\text{C}}} = 0}^{{{\hat L}_{\text{C}}} - 1} {{\boldsymbol{\hat \alpha }}_{\text{C}}^2\left[ {{l_{\text{C}}}} \right]{{\mathbf{C}}^2}\left[ {n,{\hat{\mathbf i}_{{\text{C}}}[{l_{\text{C}}}]}} \right]{P_{\text{B}}}\left( {{\hat{\mathbf i}_{{\text{C}}}[{l_{\text{C}}}]},{{\mathbf{b}}_{k - 1}}} \right)} }}{{{{\mathbf{C}}^2}\left[ {n,{{\hat i}_{\text{J}}}} \right]{P_{\text{B}}}\left( {{{\hat i}_{\text{J}}},{{\mathbf{b}}_{k - 1}}} \right)}}, \nonumber \\
  &{\text{s}}.{\text{t}}.\; {P_{\text{B}}}\left( {{\hat{\mathbf i}_{{\text{C}}}[{l_{\text{C}}}]},{\mathbf{b}}_{k - 1}^{}} \right) = {\mathbf{b}}_{k - 1}^{\text{H}}{\mathbf{a}}_{{N_{{\text{U,azi}}}},{N_{{\text{U,ele}}}}}^{}\left( {{\hat{\boldsymbol \varphi }_{{\text{C,rx}}}[{l_{\text{C}}}]},{\hat{\boldsymbol \theta }_{{\text{C,rx}}}[{l_{\text{C}}}]}} \right)\nonumber\\
  &\hspace{37mm}{\mathbf{a}}_{{N_{{\text{U,azi}}}},{N_{{\text{U,ele}}}}}^{\text{H}}\left( {{\hat{\boldsymbol \varphi }_{{\text{C,rx}}}[{l_{\text{C}}}]},{\hat{\boldsymbol \theta }_{{\text{C,rx}}}[{l_{\text{C}}}]}} \right){\mathbf{b}}_{k - 1}^{}, \nonumber \\
  &\hspace{11.75mm}{P_{\text{B}}}\left( {{{\hat i}_{\text{J}}},{\mathbf{b}}_{k - 1}^{}} \right) = {\mathbf{b}}_{k - 1}^{\text{H}}{\mathbf{a}}_{{N_{{\text{U,azi}}}},{N_{{\text{U,ele}}}}}^{}\left( {{{\hat \varphi }_{\text{J}}},{{\hat \theta }_{\text{J}}}} \right)\nonumber \\&\hspace{37mm}{\mathbf{a}}_{{N_{{\text{U,azi}}}},{N_{{\text{U,ele}}}}}^{\text{H}}\left( {{{\hat \varphi }_{\text{J}}},{{\hat \theta }_{\text{J}}}} \right){\mathbf{b}}_{k - 1}^{}.
\end{align}
After obtaining $n_k$, the optimization problem for estimating $\mathbf{b}_{k}$ is formulated as 
\begin{align}
  &{P_2}\hspace{-1mm}:\hspace{-1mm}{{\mathbf{b}}_k}\hspace{-1mm}=\hspace{-1mm}\arg \mathop {\max }\limits_{{{\mathbf{b}}_{}}} \frac{{\sum\limits_{{l_{\text{C}}} = 0}^{{{\hat L}_{\text{C}}} - 1} {{\boldsymbol{\hat \alpha }}_{\text{C}}^2\left[ {{l_{\text{C}}}} \right]{{\mathbf{C}}^2}\left[ {{n_k},{\hat{\mathbf i}_{{\text{C}}}[{l_{\text{C}}}]}} \right]{P_{\text{B}}}\left( {{\hat{\mathbf i}_{{\text{C}}}[{l_{\text{C}}}]},{\mathbf{b}}} \right)} }}{{{{\mathbf{C}}^2}\left[ {{n_k},{{\hat i}_{\text{J}}}} \right]{P_{\text{B}}}\left( {{{\hat i}_{\text{J}}},{{\mathbf{b}}_{}}} \right)}}, \nonumber\\
  &{\text{s}}.{\text{t}}.\; {P_{\text{B}}}\left( {{\hat{\mathbf i}_{{\text{C}}}[{l_{\text{C}}}]},{\mathbf{b}}_{}^{}} \right) = {\mathbf{b}}_{}^{\text{H}}{\mathbf{a}}_{{N_{{\text{U,azi}}}},{N_{{\text{U,ele}}}}}^{}\left( {{\hat{\boldsymbol \varphi }_{{\text{C,rx}}}[{l_{\text{C}}}]},{\hat{\boldsymbol \theta }_{{\text{C,rx}}}[{l_{\text{C}}}}]} \right)\nonumber\\
  &\hspace{32mm} {\mathbf{a}}_{{N_{{\text{U,azi}}}},{N_{{\text{U,ele}}}}}^{\text{H}}\left( {{\hat{\boldsymbol \varphi }_{{\text{C,rx}}}[{l_{\text{C}}}]},{\hat{\boldsymbol \theta }_{{\text{C,rx}}}[{l_{\text{C}}}]}} \right){\mathbf{b}}_{}^{}, \nonumber\\
  &\hspace{11.75mm}{P_{\text{B}}}\left( {{{\hat i}_{\text{J}}},{\mathbf{b}}_{}^{}} \right) = {\mathbf{b}}_{}^{\text{H}}{\mathbf{a}}_{{N_{{\text{U,azi}}}},{N_{{\text{U,ele}}}}}^{}\left( {{{\hat \varphi }_{\text{J}}},{{\hat \theta }_{\text{J}}}} \right)\nonumber\\
  &\hspace{32mm}{\mathbf{a}}_{{N_{{\text{U,azi}}}},{N_{{\text{U,ele}}}}}^{\text{H}}\left( {{{\hat \varphi }_{\text{J}}},{{\hat \theta }_{\text{J}}}} \right){\mathbf{b}}_{}^{}.
\end{align}
$P_2$ can be reformulated to $P_2'$ as
\begin{align}
  {P_2'}:{{\mathbf{b}}_k}&=\arg \mathop {\max }\limits_{{{\mathbf{b}}_{}}} \frac{{{\mathbf{b}}_{}^{\text{H}}{\mathbf{Ab}}_{}^{}}}{{{\mathbf{b}}_{}^{\text{H}}{\mathbf{Bb}}_{}^{}}}, \nonumber \\
  {\text{s}}.{\text{t}}.\  {\mathbf{A}} &= \sum\limits_{{l_{\text{C}}} = 0}^{{{\hat L}_{\text{C}}} - 1} {\boldsymbol{\hat \alpha }}_{\text{C}}^2\left[ {{l_{\text{C}}}} \right]{{\mathbf{C}}^2}\left[ {{n_k},{\hat{\mathbf i}_{{\text{C}}}[{l_{\text{C}}}]}} \right]\nonumber\\
  &\hspace{5mm}{\mathbf{a}}_{{N_{{\text{U,azi}}}},{N_{{\text{U,ele}}}}}^{}\left( {{\hat{\boldsymbol \varphi }_{{\text{C,rx}}}[{l_{\text{C}}}]},{\hat{\boldsymbol \theta }_{{\text{C,rx}}}[{l_{\text{C}}}]}} \right)\nonumber\\
  &\hspace{5mm}{\mathbf{a}}_{{N_{{\text{U,azi}}}},{N_{{\text{U,ele}}}}}^{\text{H}}\left( {{\hat{\boldsymbol \varphi }_{{\text{C,rx}}}[{l_{\text{C}}}]},{\hat{\boldsymbol \theta }_{{\text{C,rx}}}[{l_{\text{C}}}]}} \right) , \nonumber \\
  {\mathbf{B}} &= {{\mathbf{C}}^2}\left[ {{n_k},{{\hat i}_{\text{J}}}} \right]{\mathbf{a}}_{{N_{{\text{U,azi}}}},{N_{{\text{U,ele}}}}}^{}\left( {{{\hat \varphi }_{\text{J}}},{{\hat \theta }_{\text{J}}}} \right)\nonumber \\
  &\hspace{5mm}{\mathbf{a}}_{{N_{{\text{U,azi}}}},{N_{{\text{U,ele}}}}}^{\text{H}}\left( {{{\hat \varphi }_{\text{J}}},{{\hat \theta }_{\text{J}}}} \right).
\end{align}
Problem $P_2'$ is a generalized Rayleigh quotient problem, which aims to find the generalized eigenvector ${\mathbf{b}}$ corresponding to the largest generalized eigenvalue $\lambda$ that satisfies the following equation:
\begin{align}
    {\mathbf{Ab}} = \lambda {\mathbf{Bb}}.
\end{align}
Since $\mathbf{B}$ is a non-invertible positive semi-definite matrix, we apply diagonal loading to make it invertible, yielding $\mathbf{B}_{\epsilon} = \mathbf{B} + \epsilon \mathbf{I}$, where $\epsilon=10^{-8}$ in the simulations. 
We then perform Cholesky decomposition on $\mathbf{B}_{\epsilon}$ as $\mathbf{B}_{\epsilon} = \mathbf{L} \mathbf{L}^{\text{H}}$. In this way, the solution to the original problem can be obtained by solving the following equation:
\begin{align}
\label{eq_alg_begin}
    {\mathbf{Ab}} = \lambda {\mathbf{B}}_{\epsilon}{\mathbf{b}}.
\end{align}
By moving $\mathbf{L}$ to the left-hand side of the equation, we obtain
${{\mathbf{L}}^{-1}} \mathbf{A} \mathbf{b} = \lambda {{\mathbf{L}}^{\text{H}}} \mathbf{b}$. 
Letting ${{\mathbf{L}}^{\text{H}}} \mathbf{b} = \tilde{\mathbf{b}}$, we have
\begin{align}
    {{\mathbf{L}}^{-1}} \mathbf{A} \left( {{\mathbf{L}}^{\text{H}}} \right)^{-1} \tilde{\mathbf{b}} = \lambda \tilde{\mathbf{b}}.    
\end{align}
Therefore, we can obtain $\mathbf{b}_k$ through the eigenvalue decomposition of $\mathbf{\tilde A} = \mathbf{L}^{-1} \mathbf{A} \left( \mathbf{L}^{\text{H}} \right)^{-1}$.
Let $\mathbf{\tilde b}_{\max}$ be the eigenvector corresponding to the largest eigenvalue of $\mathbf{\tilde A}$, then 
\begin{align}
\label{eq_alg_end}
    \mathbf{b}_k = \left( \mathbf{L}^{\text{H}} \right)^{-1} \mathbf{\tilde b}_{\max}.
\end{align}
Iterate $P_1$ and $P_2$ alternately until the increase in the objective function between two consecutive iterations is less than $\varepsilon=10^{-8}$.

The complete process of the proposed alternating optimization of the radiation pattern and the combiner for the uplink transmission is summarized in Algorithm \ref{alg_AltOpt}.

\subsection{Convergence and Computational Complexity Analysis}

In subproblem $P_1$, the index $n$ is obtained via maximum likelihood estimation method. Since both the numerator and denominator of the objective function are finite, the resulting objective value is also finite. Moreover, the update rule in the maximum likelihood estimation method guarantees that the objective does not decrease.
In subproblem $P_2$, one maximizes a Rayleigh quotient, and the update—being the principal eigenvector of the associated Hermitian matrix—yields the global optimum of this subproblem\cite{refRayleighQuotient}.
Therefore, under the alternating-optimization scheme, the overall objective sequence is non-decreasing and upper-bounded, and hence converges.

Algorithm~\ref{alg_AltOpt} can be decomposed into two subproblems: a maximum‐likelihood selection of the radiation pattern (corresponding to $P_1$) and a Rayleigh‐quotient maximization for the optimal combiner (corresponding to $P_2$). The complexity of solving $P_1$ is $\mathcal{O}\bigl(N_{p}^2 + \hat L_{\text{C}}\bigr) $, 
and the complexity of solving $P_2$ is $\mathcal{O}\bigl((\hat L_{\text{C}} + 2)\,N_{\text{U}}^2 + 6\,N_{\text{U}}^3\bigr) $. 
Therefore, the overall complexity of Algorithm~\ref{alg_AltOpt} is $\mathcal{O}\!\Bigl(N_{\text{it}}\bigl((\hat L_{\text{C}} + 2)\,N_{\text{U}}^2 \;+\; 6\,N_{\text{U}}^3 \;+\; N_{p}^2\bigr)\Bigr) $, where $N_{\text{it}}$ is the total number of iterations, which in our simulations typically ranges from two to three.

\begin{algorithm}[!t]
	\begin{small}
		\label{alg_AltOpt}
		\caption{Proposed Alternating Optimization Scheme of Radiation Pattern and Combiner}
		\LinesNumbered 
		\KwIn{Received angle from the controller $\hat{\boldsymbol{\theta}}_{\text{C,rx}} \in \mathbb{C}^{ \hat L_{\text{C}}}$ and $\hat{\boldsymbol{\varphi}}_{\text{C,rx}}  \in \mathbb{C}^{ \hat L_{\text{C}}}$, index in radiation pattern gain matrix 
			$\mathbf{C}_{\text{L}}$ corresponding to the received angle $\hat{\boldsymbol{\theta}}_{\text{C,rx}} $ and $\hat{\boldsymbol{\varphi}}_{\text{C,rx}} $, gain of each path on each receive antenna ${{\mathbf{G}}} \in {\mathbb{C}^{{{\hat L}_{\text{C}}} \times {N_{\text{U}}}}}$.
		}
		\KwOut{Index of the
			best radiation pattern $n$, the best combiner $\mathbf{b}\in\mathbb{C}^{N_{\text{U}}}$.
		}
		Decompose problem $P_0$ into $P_1$ and $P_2$\;
		Transform $P_2$ into $P_2'$\;
		Solve the generalized eigenvectors of $\mathbf{A}$ and $\mathbf{B}$ through equations (\ref{eq_alg_begin})–(\ref{eq_alg_end}) to obtain the solution of $P_2'$\;
		Alternately iterate between $P_1$ and $P_2'$ until the algorithm converges, i.e., the objective function of $P_0$ between two consecutive iterations is less than $\varepsilon=10^{-8}$.
	\end{small}
\end{algorithm}

\section{Simulation Results}\label{sec_simulation}
In this section, we validate the effectiveness of the proposed RPR-FAS-empowered interference-resilient UAV communication system by comparing it to fixed-pattern antenna systems.
Additionally, we utilize the MUSIC algorithm as a benchmark to demonstrate the superiority of the proposed LSO-OMP-MMV algorithm in estimating the angles of both the communication link and the jammer.
\subsection{System Parameters and Performance Metrics}
\begin{figure*}[!t]
		\vspace{-9mm}
	\color{black}
	\subfigure[]{
		\begin{minipage}[t]{0.33\linewidth}
			\includegraphics[width=2.2in]{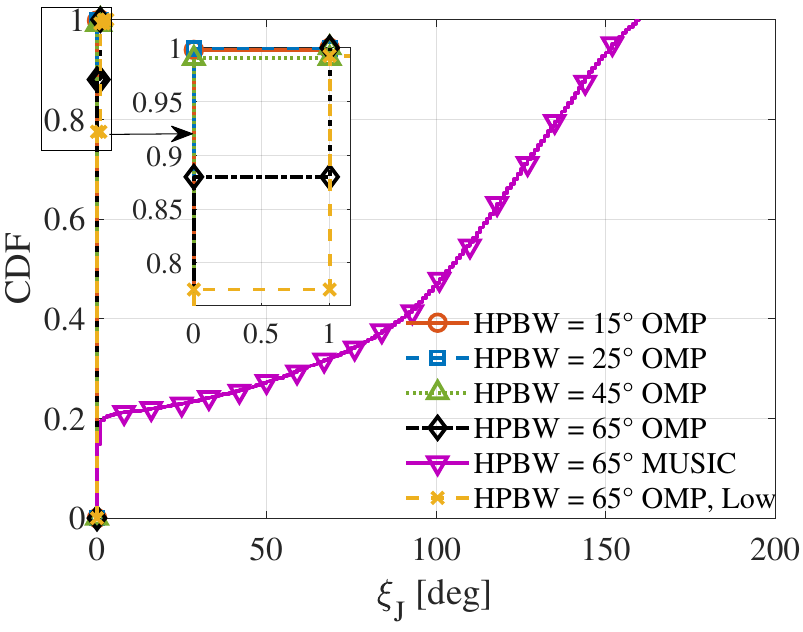}
			\label{fig_Jammer_HPBW}
		\end{minipage}%
	}%
	\subfigure[]{
		\begin{minipage}[t]{0.33\linewidth}
			\includegraphics[width=2.2in]{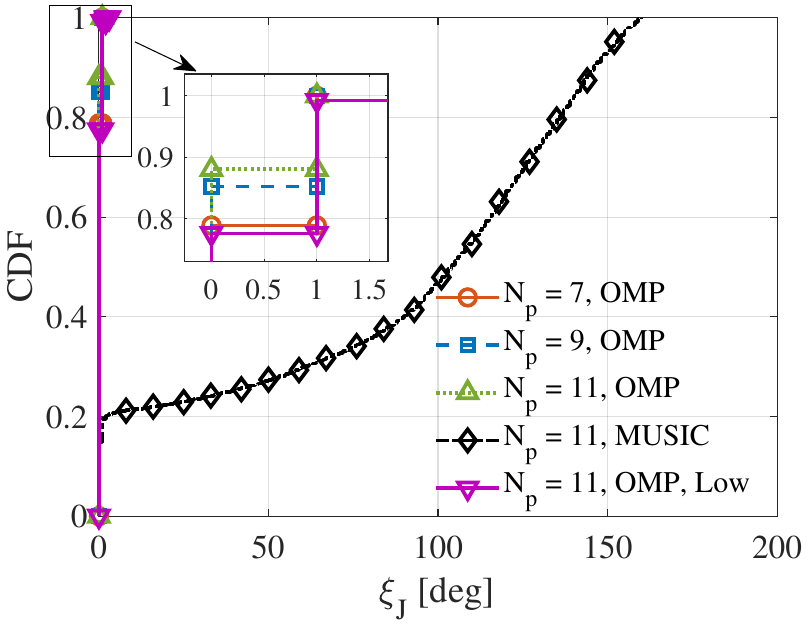}
			\label{fig_Jammer_PatternNum}
		\end{minipage}%
	}
	\subfigure[]{
		\begin{minipage}[t]{0.33\linewidth}
			\includegraphics[width=2.2in]{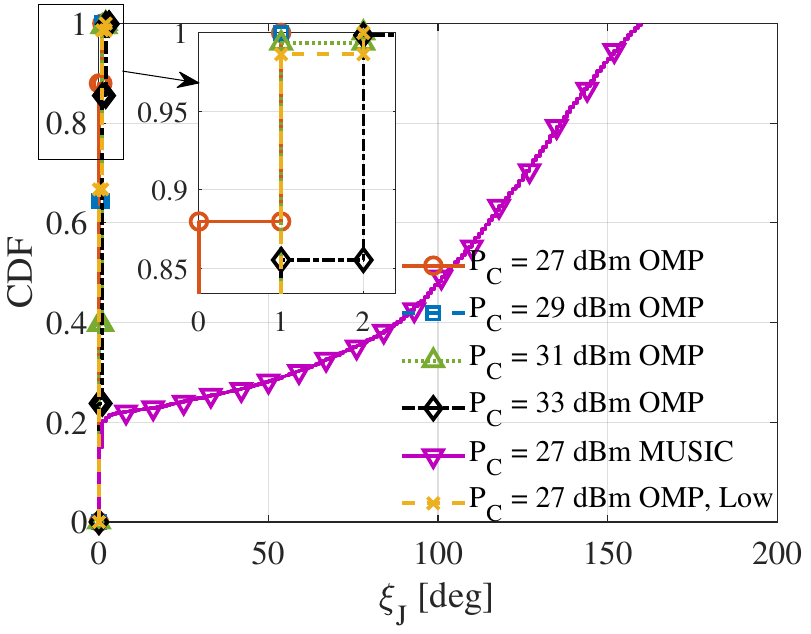}
			\label{fig_Jammer_Pt}
		\end{minipage}%
	}\\
	\subfigure[]{
		\begin{minipage}[t]{0.33\linewidth}
			\includegraphics[width=2.2in]{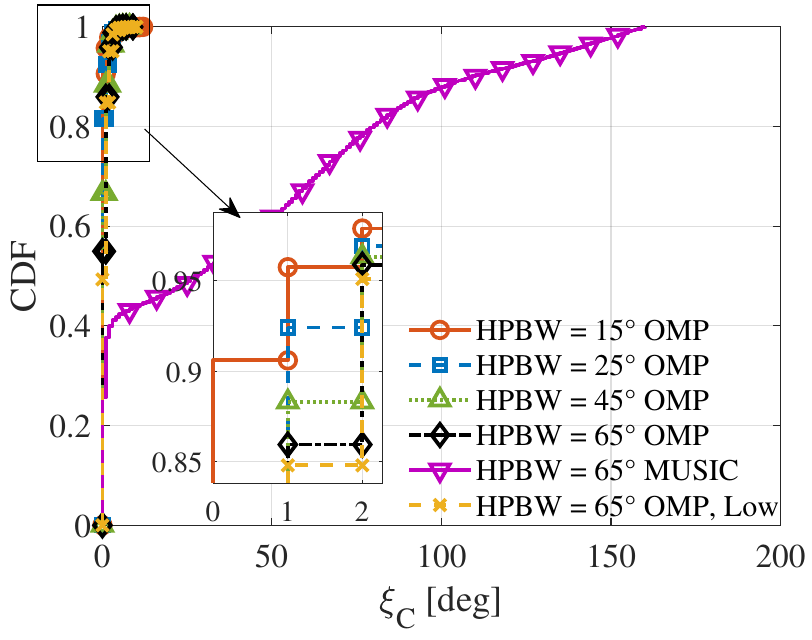}
			\label{fig_Control_HPBW}
		\end{minipage}%
	}%
	\subfigure[]{
		\begin{minipage}[t]{0.33\linewidth}
			\includegraphics[width=2.2in]{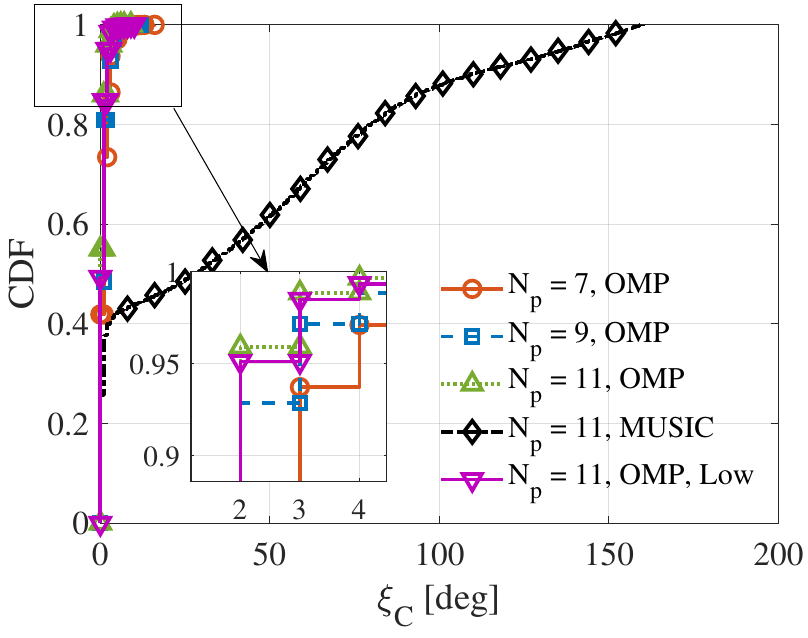}
			\label{fig_Control_PatternNum}
		\end{minipage}%
	}%
	\subfigure[]{
		\begin{minipage}[t]{0.33\linewidth}
			\includegraphics[width=2.2in]{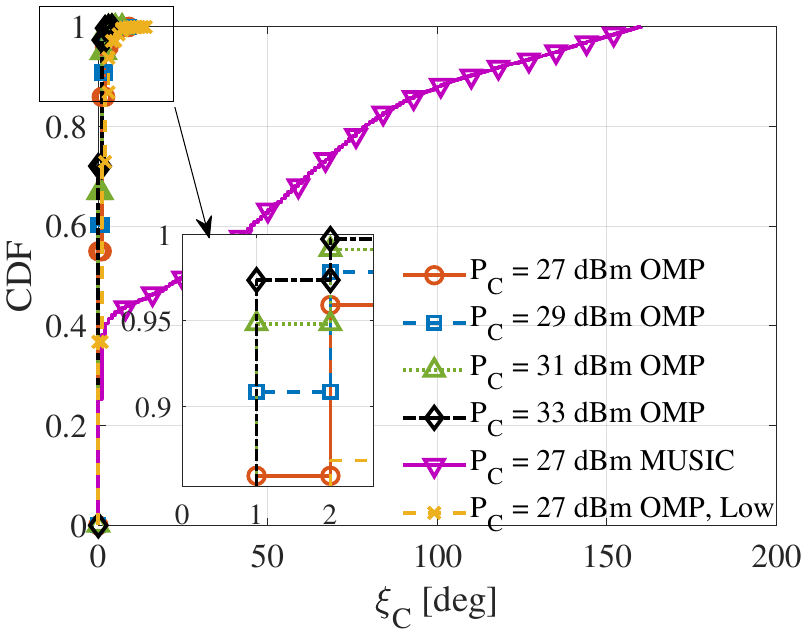}
			\label{fig_Control_Pt}
		\end{minipage}%
	}%
	\centering
		\vspace{-2mm}	
	\caption{
		Relationship between the AoA estimation errors of the uplink signals from the jammer as well as the UAV controller at the UAV and the system parameters:
		(a) AoA estimation errors of the uplink signals from the jammer at the UAV with different HPBWs;
		(b) AoA estimation errors of the uplink signals from the jammer at the UAV with different number of reconfigurable radiation patterns $N_p$s;
		(c) AoA estimation errors of the uplink signals from the jammer at the UAV with different transmit power $P_{\text{C}}$s;
		(d) AoA estimation errors of the uplink signals from the UAV controller at the UAV with different HPBWs;
		(e) AoA estimation errors of the uplink signals from the UAV with different number of reconfigurable radiation patterns $N_p$s;
		(f) AoA estimation errors of the uplink signals from the UAV controller at the UAV with different transmit powers $P_{\text{C}}$s.
	}
	\label{fig_Angle_Estimation}
		\vspace{-3mm}
\end{figure*}

\begin{figure*}[!t]
			\vspace{-7mm}
	\color{black}
	\subfigure[]{
		\begin{minipage}[t]{0.25\linewidth}
			\includegraphics[width=1.6in]{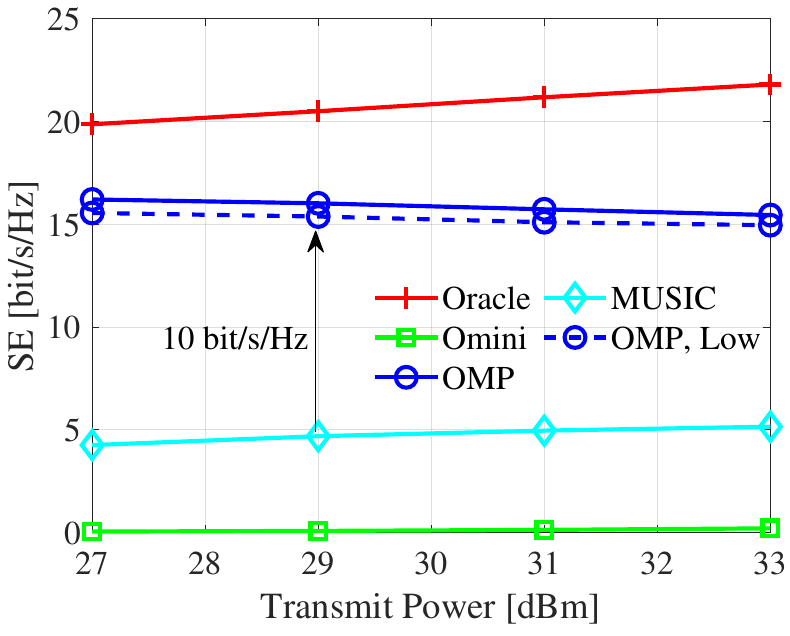}
			\label{fig_SE_algorithm_Up}
		\end{minipage}%
	}%
	\subfigure[]{
		\begin{minipage}[t]{0.25\linewidth}
			\includegraphics[width=1.6in]{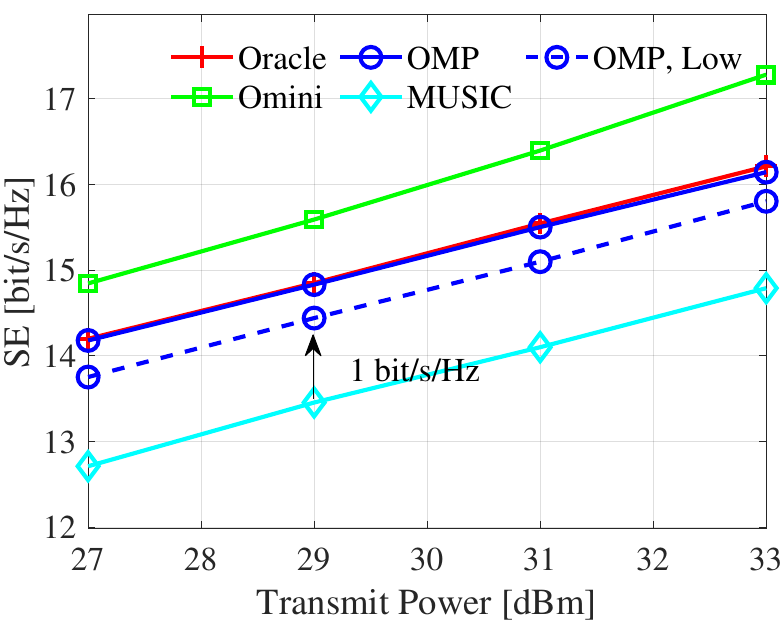}
			\label{fig_SE_algorithm_Dn}
		\end{minipage}%
	}
	\subfigure[]{
		\begin{minipage}[t]{0.25\linewidth}
			\includegraphics[width=1.6in]{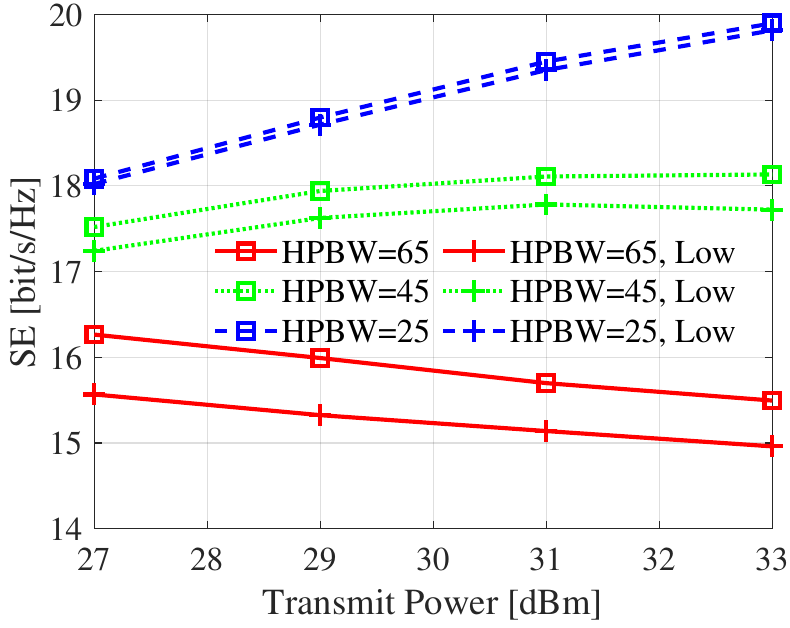}
			\label{fig_SE_HPBW_Up}
		\end{minipage}%
	}%
	\subfigure[]{
		\begin{minipage}[t]{0.25\linewidth}
			\includegraphics[width=1.6in]{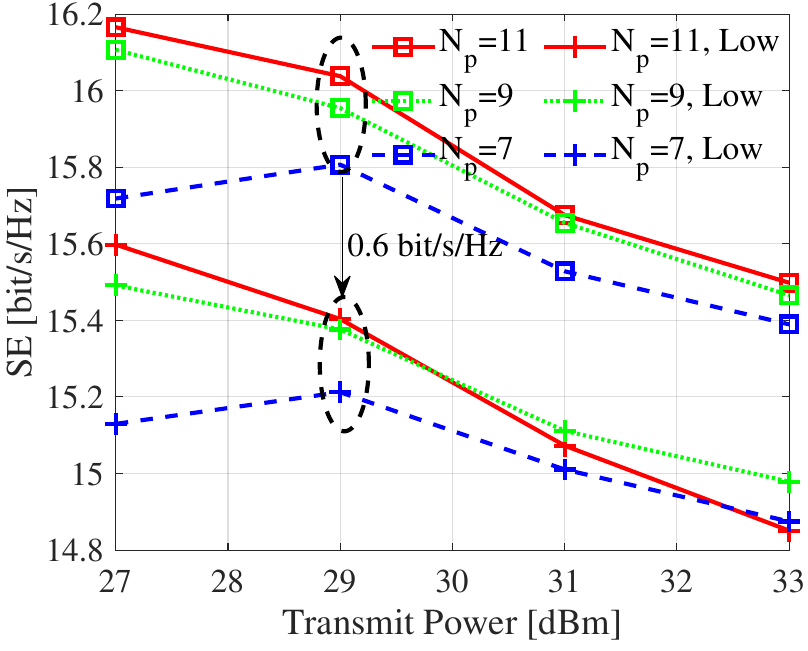}
			\label{fig_SE_PatternNum_Up}
		\end{minipage}%
	}
	\centering
	\caption{
			Relationship between uplink/downlink SE and system parameters:
			(a) uplink SE obtained using different algorithms;
			(b) downlink SE obtained using different algorithms; 
			(c) uplink SE under Algorithm \ref{alg_ParaEst} with different HPBW values;
			(d) uplink SE under Algorithm \ref{alg_ParaEst} with different $N_p$ values.
	}
	\label{fig_SE}
	\vspace{-3mm}
\end{figure*}

\begin{figure*}[!t]
	\vspace{-2mm}
	\color{black}
	\subfigure[]{
		\begin{minipage}[t]{0.33\linewidth}
			\includegraphics[width=2.2in]{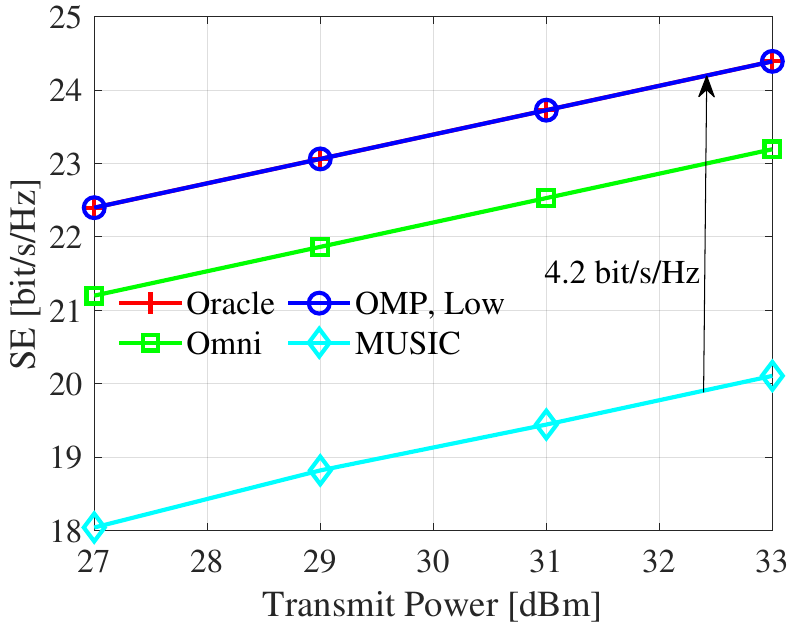}
			\label{fig_NoJammer_Power}
		\end{minipage}%
	}%
	\subfigure[]{
		\begin{minipage}[t]{0.33\linewidth}
			\includegraphics[width=2.2in]{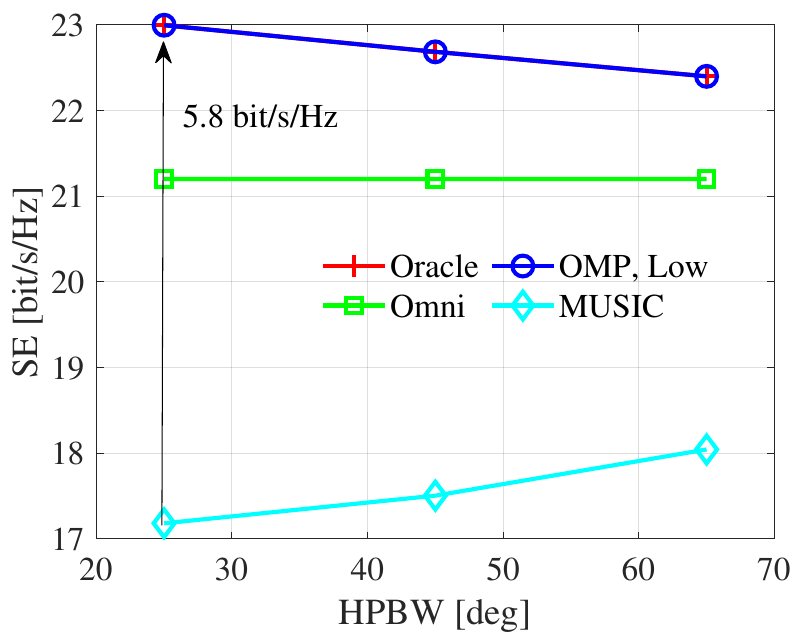}
			\label{fig_NoJammer_HPBW}
		\end{minipage}%
	}
	\subfigure[]{
		\begin{minipage}[t]{0.33\linewidth}
			\includegraphics[width=2.2in]{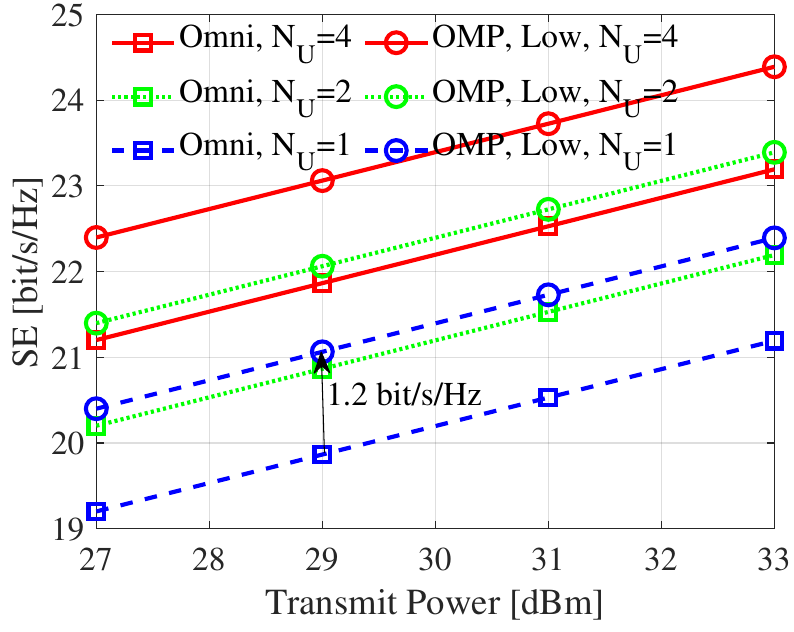}
			\label{fig_NoJammer_Nt}
		\end{minipage}%
	}
	\centering
	\vspace{-4mm}	
	\caption{
			Uplink SE in the absence of jamming:
			(a) uplink SE with varying transmit power;
			(b) uplink SE with different HPBW values;
			(c) uplink SE with different $N_\text{U}$ values.
	}
	\label{fig_SE_NoJammer}
	\vspace{-3mm}
\end{figure*}

\begin{table}[!b]
	\vspace{-7mm}
	\centering
	\small
	\begin{threeparttable}
		\caption{Parameter Settings}
		\label{table_para_value}
		\begin{tabular}{l  l@{\vrule width 1.5pt} l  l}
			\noalign{\hrule height 0.5mm}
			\textbf{Parameter} & \textbf{Value} & \hspace{1.5mm}\textbf{Parameter} & \textbf{Value}
			\\ \noalign{\hrule height 0.5mm}
			$f_c$ & 2.44 GHz &\hspace{1.5mm} $N_a$ & $32761=181^2$\\ \hline
			$B$ & 84 MHz &\hspace{1.5mm} $N_{a,\text{azi}}$ & 181\\ \hline
			$B_I$ & 20 MHz &\hspace{1.5mm} $N_{a,\text{ele}}$ & 181 \\ \hline
			$B_p$ & 10 MHz &\hspace{1.5mm} $N_s$ & 1000 \\ \hline
			$B_h$ & 1 MHz &\hspace{1.5mm} $N_p$ & 121,81,49\\ \hline
			$T_h$ & 1 ms &\hspace{1.5mm} $N_{p,\text{azi}}$ & 11,9,7\\ \hline
			$T_I$ & 12 ms&\hspace{1.5mm} $N_{p,\text{ele}}$ & 11,9,7\\ \hline
			$T_u$ &  0.1 ms &\hspace{1.5mm} $N_{\text{C}}$ & 2\\ \hline
			$P_{\text{J}}$ & 50 dBm &\hspace{1.5mm} $N_{\text{U}}$ & 1,2,4 \\ \hline
			$P_{\text{C}}$ & 27--33 dBm \hspace{0.5mm}&\hspace{1.5mm} $N_{\text{U},\text{azi}}$ & 1,2\\ \hline
			$N_{\text{C},\text{azi}}$ & 2 &\hspace{1.5mm} $N_{\text{U},\text{ele}}$ & 1,2 \\ \hline
			$N_{\text{C},\text{ele}}$ & 1 &  & \\
			\noalign{\hrule height 0.5mm}
		\end{tabular}
	\end{threeparttable}
	\vspace{-4mm}
\end{table}
The default system parameters are shown in Table~\ref{table_para_value}\cite{ref_DroneRFa}.
The power of antenna radiation patterns for different HPBW values has been normalized to 1 to ensure a fair comparison.
The noise power spectrum density is set to $-174$ dBm/Hz.
$\eta_{\text{th}}=10^{-3}$ is the predefined threshold for terminating the loop in Algorithm \ref{alg_ParaEst}.
The large-scale fading of the one-way path can be calculated as $\! \lambda_c/4\pi R$, where $R$ is the distance from the transmitter to the receiver and $\lambda_c=c/f_c$ is the wavelength.
The large-scale fading of the two-way path can be calculated as $\sqrt {{S_{{\text{eff}}}}{\lambda_c ^2}} /\sqrt {{{\left( {4\pi } \right)}^3}{{({R_1})}^2}{{({R_2})}^2}} $, where $S_{\mathrm{eff}}$ is the effective reflection area of the scatterers and is set to 10~$\text{m}^2$ in the simulations, $R_1$ and $R_2$ are the distance from the transmitter to the scatterer and the distance from the scatterer to the receiver, respectively.
The distance of the LoS path from the jammer (or UAV controller) to the UAV and the distances of the two-hop non-LoS (NLoS) paths from the jammer (or UAV controller) to the UAV are set to 200 meters, 200 meters, and 100 meters, respectively.
The transmitted and received azimuth and elevation angles of each path from the jammer and UAV controller relative to the UAV are uniformly distributed between  $-70^{\circ}$ and $70^{\circ}$.
The angle estimation errors in Fig.~\ref{fig_Angle_Estimation}, Table~\ref{table_HPBW_65}, and Table~\ref{table_HPBW_15} are defined as 
\begin{align}
  \xi_\text{J}\hspace{-1mm}&=\hspace{-1mm}\max \left( \left| \left[\kern-0.15em\left[ {{\theta _{\text{J}}}} 
 \right]\kern-0.15em\right] - {{\hat \theta }_{\text{J}}} \right|,\left|{\left[\kern-0.15em\left[ {{\varphi _{\text{J}}}} 
 \right]\kern-0.15em\right] - {{\hat \varphi }_{\text{J}}}}\right| \right), \\
  \xi_\text{C}\hspace{-1mm}&=\hspace{-1mm}\max \Big( \left| {\left[\kern-0.3em\left[ {{{\boldsymbol{\theta }}_{\text{C,rx}}}[{{\bar l}_{\text{C}}}]} 
 \right]\kern-0.3em\right] \hspace{-1mm}-\hspace{-1mm} {{{\boldsymbol{\hat \theta }}}_{\text{C,rx}}}[{{\tilde l}_{\text{C}}}]} \right|,\left| {\left[\kern-0.3em\left[ {{{\boldsymbol{\varphi }}_{\text{C,rx}}}[{{\bar l}_{\text{C}}}]} 
 	\right]\kern-0.3em\right] \hspace{-1mm}-\hspace{-1mm} {{{\boldsymbol{\hat \varphi }}}_{\text{C,rx}}}[{{\tilde l}_{\text{C}}}]} \right| \Big) ,
\end{align}
where $\xi_{\text{J}}$ and $\xi_{\text{C}}$ are the uplink AoA estimation errors of the jammer and the UAV controller at the UAV, respectively, $\bar l_{\text{C}}$ and $\tilde l_{\text{C}}$ are the indices corresponding to the maximum values in $\boldsymbol \alpha_{\text{C}}$ and $\sum\nolimits_{{n_{\text{U}}} = 0}^{{N_{\text{U}}} - 1} {\left| {{\mathbf{G}}[:,{n_{\text{U}}}]} \right|} $, respectively.
Multiple signal classification (MUSIC)~\cite{ref_LZR_JSTSP} algorithm is used as the benchmark and can be applied when $N_\text{U}>1$.
We assume that MUSIC has complete knowledge of the number of paths, meaning that MUSIC does not use Akaike information criterion or minimum description length criteria to estimate the number of incoming waves.
Since $N_{\text{U}}$ is too small, the use of these two criteria would degrade the performance of MUSIC, making it worse than the performance presented in the paper.
The angular search interval of MUSIC is $1^\circ$.
After estimating the AoAs of the jammer and the controller through the MUSIC algorithm, Algorithm \ref{alg_AltOpt} is also employed for radiation pattern selection and combiner design.
``Low" in Figs.~\ref{fig_Angle_Estimation}, \ref{fig_SE}, and \ref{fig_SE_NoJammer} indicates the use of the low-storage-overhead reconfigurable radiation pattern gain matrix $\mathbf{C}_{\text{L}}$.
In Figs.~\ref{fig_Angle_Estimation}, \ref{fig_SE}, and \ref{fig_SE_NoJammer}, the channel from the UAV controller to the UAV includes one LoS path and one NLoS path, while the channel from the jammer to the UAV includes only one LoS path.
In Table~\ref{table_HPBW_65} and Table~\ref{table_HPBW_15}, the channel from the UAV controller to the UAV includes one LoS path and one NLoS path or two NLoS paths, while the channel from the jammer to the UAV includes one LoS path or one NLoS path.

\subsection{Angle Estimation Performance}
Unless otherwise specified, the parameters in Fig.~\ref{fig_Angle_Estimation} are set to $N_{\text{U}}=4$, $\text{HPBW}=65^{\circ}$, $N_p=11$, and $P_{\text{C}}=27$ dBm.
CDF in Fig.~\ref{fig_Angle_Estimation} is the abbreviation of ``cumulative distribution function".
The super-resolution algorithm requires the number of antennas to exceed the number of targets for its applicability and is not suitable for multipath scenarios.
Therefore, MUSIC will fail when $N_{\text{U}}$ is 1 or 2.
In Fig.~\ref{fig_Angle_Estimation}, $N_{\text{U}}=4$ and
MUSIC algorithm still fails, demonstrating that the necessity of our proposed RPR-FAS-empowered interference-resilient UAV communication scheme.
The smaller the HPBW and the larger the $N_p$, the higher the AoA estimation accuracy.
The low-storage-overhead reconfigurable radiation pattern gain matrix $\mathbf{C}_{\text{L}}$ reduces storage overhead while maintaining AoA estimation performance.
As the transmit power of the controller increases, the accuracy of the jammer's AoA estimation decreases.
This is because the impact of the controller’s transmitted signal needs to be mitigated when estimating the jammer's AoA. 
Although a higher transmit power of the controller improves the accuracy of the controller's AoA estimation, it also increases the residual interference from the controller's signal, thereby degrading the jammer's AoA estimation performance.

\begin{figure*}[!t]
	\color{black}
	\subfigure[]{
		\begin{minipage}[t]{0.25\linewidth}
			\includegraphics[width=1.6in]{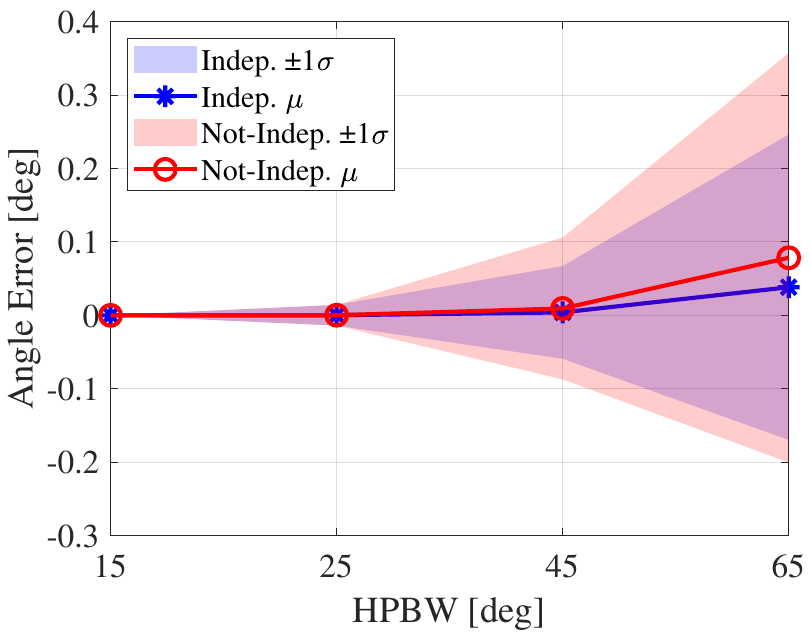}
			\label{fig_AngleEst_Jammer_HPBW}
		\end{minipage}%
	}%
	\subfigure[]{
		\begin{minipage}[t]{0.25\linewidth}
			\includegraphics[width=1.6in]{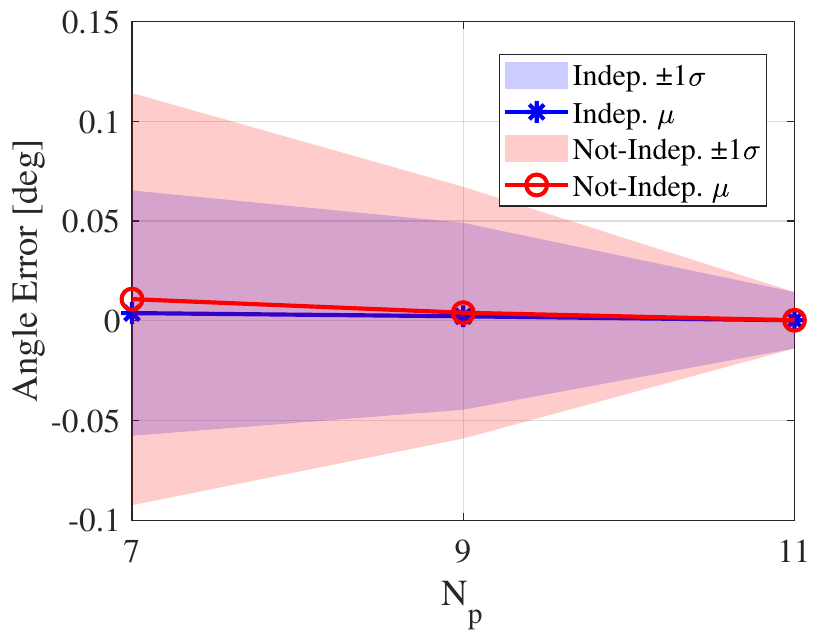}
			\label{fig_AngleEst_Jammer_Np}
		\end{minipage}%
	}
	\subfigure[]{
		\begin{minipage}[t]{0.25\linewidth}
			\includegraphics[width=1.6in]{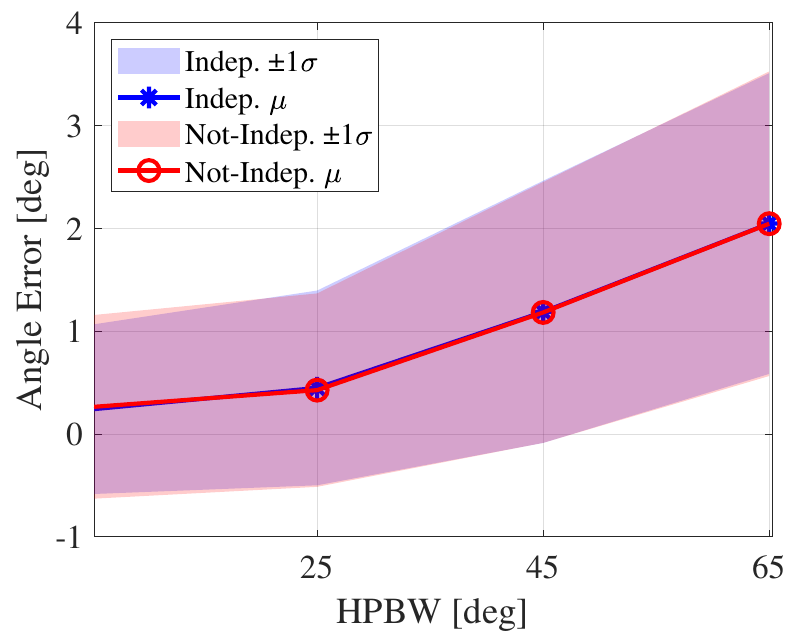}
			\label{fig_AngleEst_Controller_HPBW}
		\end{minipage}%
	}%
	\subfigure[]{
		\begin{minipage}[t]{0.25\linewidth}
			\includegraphics[width=1.6in]{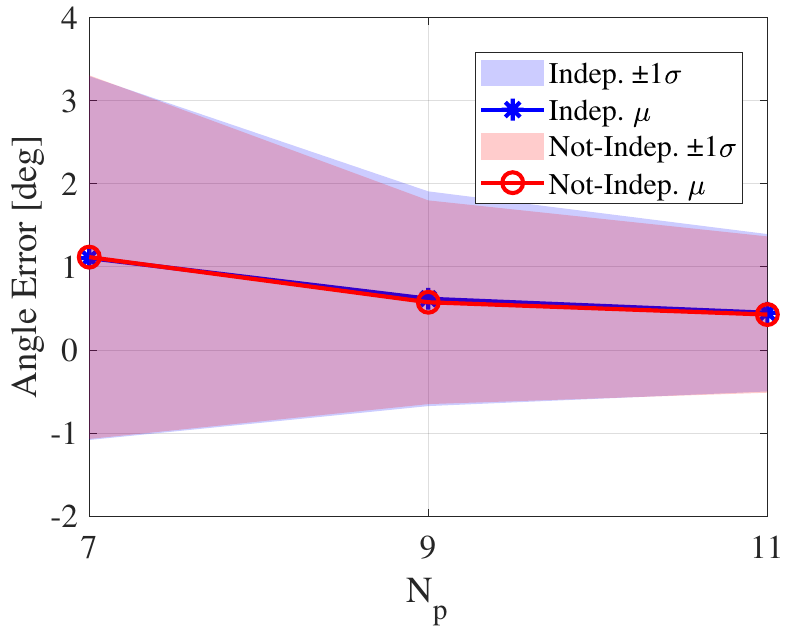}
			\label{fig_AngleEst_Controller_Np}
		\end{minipage}%
	}
	\centering
	\caption{
			Uplink AoA estimation performance for independent (Indep.) and proposed joint (Not-Indep.) cases with $P_{\text{C}}=27$~dBm, $N_{\text{U}}=1$, and $N_p=11$.
			Indep.: jammer angle estimation is performed using received signals containing only the jammer, while controller angle estimation employs the method proposed in this paper—jammer angle and controller angle estimations are based on independent signals; Not-Indep.: both jammer and controller angle estimations employ the method proposed in this paper—jammer angle and controller angle estimations are based on non-independent signals.
			(a) AoA estimation errors of the uplink signals from the jammer at the UAV with different HPBWs;
			(b) AoA estimation errors of the uplink signals from the jammer at the UAV with different $N_p$s;
			(c) AoA estimation errors of the uplink signals from the UAV controller at the UAV with different HPBWs;
			(d) AoA estimation errors of the uplink signals from the UAV controller at the UAV with different $N_p$s.
	}
	\label{fig_AngleEst_IndepOrNot}
	\vspace{-3mm}
\end{figure*}
\begin{figure}[!t]
	\color{black}
	\subfigure[]{
		\begin{minipage}[t]{0.5\linewidth}
			\includegraphics[width=1.6in]{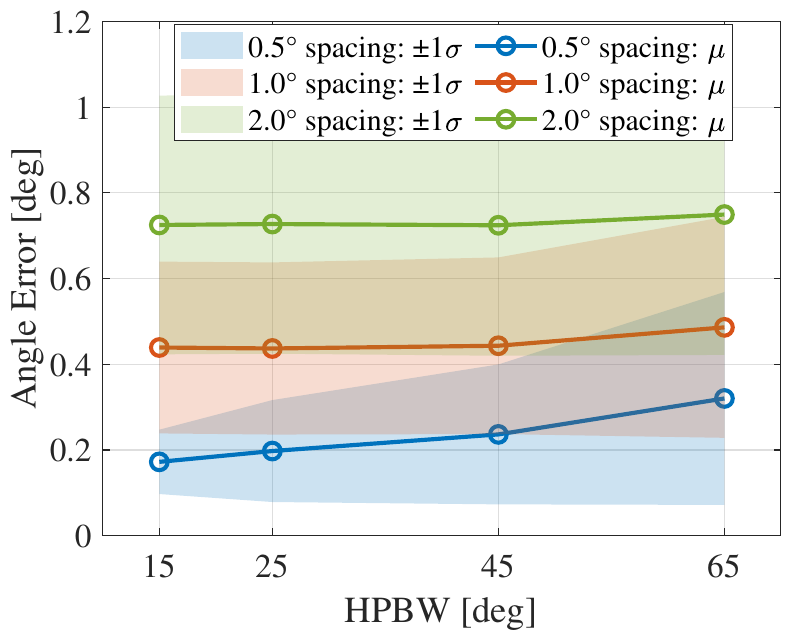}
			\label{fig_AngleEst_Jammer_DictScale}
		\end{minipage}%
	}%
	\subfigure[]{
		\begin{minipage}[t]{0.5\linewidth}
			\includegraphics[width=1.6in]{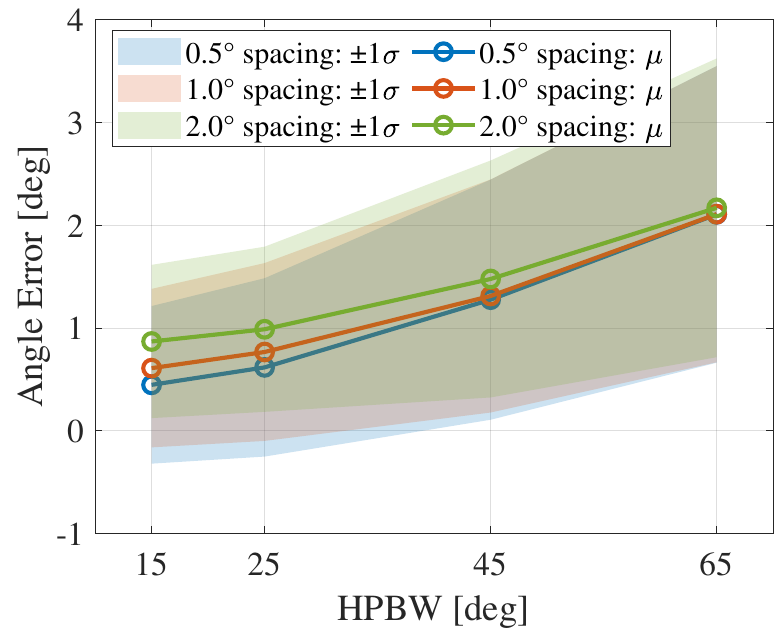}
			\label{fig_AngleEst_Controller_DictScale}
		\end{minipage}%
	}
	\centering
	\caption{
			Uplink AoA estimation performance with different scales of dictionary used in $\mathbf{C}_{\text{L}}$ with $P_{\text{C}}=27$~dBm, $N_{\text{U}}=1$, and $N_p=11$:
			(a) AoA estimation errors of the uplink signals from the jammer at the UAV with different HPBWs;
			(b) AoA estimation errors of the uplink signals from the UAV controller at the UAV with different HPBWs.
	}
	\label{fig_AngleEst_DictScale}
	\vspace{-3mm}
\end{figure}

Fig.~\ref{fig_AngleEst_IndepOrNot} presents the mean ($\mu$) and standard deviation ($\sigma$) of angle estimates when using the same signal versus different signals for estimation. Employing the scheme proposed in this paper—i.e., using the same signal for both estimates—reduces the time overhead to half that required when using separate signals. ``Indep." denotes ``Independent", indicating that the jammer's and controller's AoAs are estimated based on independent signals. ``Not Indep." denotes ``Not Independent", meaning that only one signal is used to estimate the AoAs of the jammer and the controller—jammer and controller angle estimations are based on non-independent signals.

Figs.~\ref{fig_AngleEst_Jammer_HPBW} and \ref{fig_AngleEst_Jammer_Np} show that, when estimating the jammer's AoA relative to the UAV, a smaller HPBW and a larger number of $N_p$ yield more accurate estimates. In addition, using an extra signal dedicated to jammer's AoA estimation achieves higher accuracy than relying on a single signal, because the superposition of controller's and jammer's signals introduces mutual interference. However, the improvement is limited, which is one reason why this paper adopts a unified approach to estimate both jammer's and controller's AoAs from the same signal.

Figs.~\ref{fig_AngleEst_Controller_HPBW} and \ref{fig_AngleEst_Controller_Np} lead to similar conclusions for the controller's AoA estimation. Nevertheless, by comparing Figs.~\ref{fig_AngleEst_Jammer_HPBW} and \ref{fig_AngleEst_Jammer_Np} with Figs.~\ref{fig_AngleEst_Controller_HPBW} and \ref{fig_AngleEst_Controller_Np}, it becomes apparent that the controller's AoA estimates are affected by the jammer regardless of whether one or two signals are used.

Fig.~\ref{fig_AngleEst_DictScale} shows the AoA estimation performance of Algorithm~\ref{alg_ParaEst} for different angular spacings in the dictionary. When the HPBW is small, the choice of dictionary angular spacing has a pronounced effect on estimation accuracy. As the HPBW increases, these differences shrink, and the AoA discrepancy for the UAV controller falls to zero at $\text{HPBW}=65^\circ$. This indicates that the UAV controller's angle estimation performance is particularly sensitive to HPBW variations.


\subsection{Spectral Efficiency Performance}
Unless otherwise specified, the parameters in Figs.~\ref{fig_SE} and \ref{fig_SE_NoJammer} are set to $N_{\text{U}}=4$, $\text{HPBW}=65^{\circ}$, $N_p=11$, and $P_{\text{C}}=27$ dBm.
``Omni" represents the use of an omni-directional radiation pattern; ``Oracle" assumes that the exact AoAs of the jammer and the UAV controller are perfectly known; ``OMP" indicates that the AoAs of the jammer and the UAV controller are obtained using Algorithm \ref{alg_ParaEst}.

In Fig.~\ref{fig_SE_algorithm_Up}, the omni-directional radiation pattern suffers from jammer's interference, resulting in a low SE.
Since the MUSIC algorithm fails in angle estimation, the SE obtained using the angles estimated by the MUSIC algorithm is also low.
As the transmit power increases, SE can be improved if the AoAs of the jammer and the UAV controller are perfectly estimated.
However, the SE of the proposed Algorithm \ref{alg_ParaEst} remains nearly unchanged with increasing transmit power for two reasons.
On the one hand, higher transmit power improves the accuracy of the UAV controller's AoA estimation but degrades the accuracy of the jammer's AoA estimation (as analyzed in Fig.~\ref{fig_Angle_Estimation}).
On the other hand, ``Oracle" assumes perfect knowledge of the actual AoAs of both the jammer and the controller , whereas Algorithm \ref{alg_ParaEst} estimates these angles with an accuracy of only 1$^\circ$.

In Fig.~\ref{fig_SE_algorithm_Dn}, for the downlink transmission scenario, the UAV controller is not affected by the jammer's interference.
When using the water-filling algorithm, the omni-directional radiation pattern can reduce the energy disparity among different paths compared to the directional radiation pattern, thereby enabling the utilization of NLoS paths for data transmission. 
The SE improvement gained by transmitting data streams through NLoS paths is greater than reallocating this energy to the LoS path.
This is because SE follows a logarithmic function, and once the logarithmic term reaches a sufficiently large value, further increases lead to diminishing returns in SE improvement.
In this case, adding a new data stream can effectively enhance SE.
Another reason why the omni-directional radiation pattern results in higher downlink SE is that it does not consider minimizing the signal energy received by the jammer from the UAV. 
Therefore, the omni-directional radiation pattern increases the likelihood of the UAV being detected by the jammer.
The angles estimated using Algorithm \ref{alg_ParaEst} achieve the optimal downlink SE that would be attained with the true angles.
Additionally, employing the low-storage-overhead reconfigurable radiation pattern gain matrix results in an SE loss of less than 0.5~bit/s/Hz.
Although the angle estimation of the MUSIC algorithm is highly inaccurate, the UAV controller, unaffected by the jammer's interference, can still achieve a non-zero SE even with arbitrarily selected reconfigurable radiation patterns.
However, compared to the proposed scheme, the scheme using the MUSIC algorithm still suffers an SE loss of approximately 1.0~bit/s/Hz.

In Fig.~\ref{fig_SE_HPBW_Up}, when the HPBW is small, the SE increases as the UAV controller's transmit power increases.
This is because a smaller HPBW leads to more accurate estimates of the AoAs of both the UAV controller and the jammer, enabling the design of more effective radiation patterns and combiners.
However, as the HPBW increases, the SE either saturates or decreases with the increasing transmit power of the UAV controller.
This is because a larger HPBW results in less accurate estimates of the UAV controller's and jammer's AoAs, which affects Algorithm \ref{alg_AltOpt}'s subsequent design of the radiation pattern and combiner.
When the radiation pattern and combiner are inadequate to mitigate jammer's interference, the SE is significantly degraded.
Additionally, the estimation accuracy of the UAV controller's and jammer's AoAs is interdependent.
On the one hand, accurate estimation of the UAV controller's AoA improves the accuracy of the signals used for the jammer's AoA estimation.
On the other hand, increasing the UAV controller's transmit power makes it more difficult to completely eliminate its signal when estimating the jammer's AoA, thereby degrading the accuracy of the jammer's AoA estimation. 
Consequently, the design of the radiation pattern and the combiner is affected by the previously inaccurate estimation of the jammer's AoA.
The combined effects of HPBW and UAV controller's transmit power result in the SE performance shown in Fig.~\ref{fig_SE_HPBW_Up}. 
Therefore, when selecting the HPBW of the radiation pattern, both coverage and angle estimation accuracy should be considered.
Similarly, the UAV controller's transmit power should be determined based on the actual SE requirements and the selected HPBW value.

In Fig.~\ref{fig_SE_PatternNum_Up}, as expected, a larger number of reconfigurable radiation patterns leads to higher SE, as it provides more observations.
When the UAV controller's transmit power increases, SE may either decrease or first increase and then decrease. 
This phenomenon is similar to the analysis in Fig.~\ref{fig_SE_HPBW_Up}, where the estimation performances of the UAV controller's and jammer's AoAs are interdependent.
Using the low-storage-overhead reconfigurable radiation pattern gain matrix results in an acceptable SE loss of 0.6~bit/s/Hz.
Furthermore, it can be observed that $N_p=7$ is adequate, as its performance degrades by only 0.2~bit/s/Hz compared to that of $N_p=11$. 
Therefore, excessively increasing $N_p$ to improve SE performance is unnecessary, as it would lead to an increase in pilot overhead.

\begin{figure}[!t]
	\color{black}
	\subfigure[]{
		\begin{minipage}[t]{0.5\linewidth}
			\includegraphics[width=1.6in]{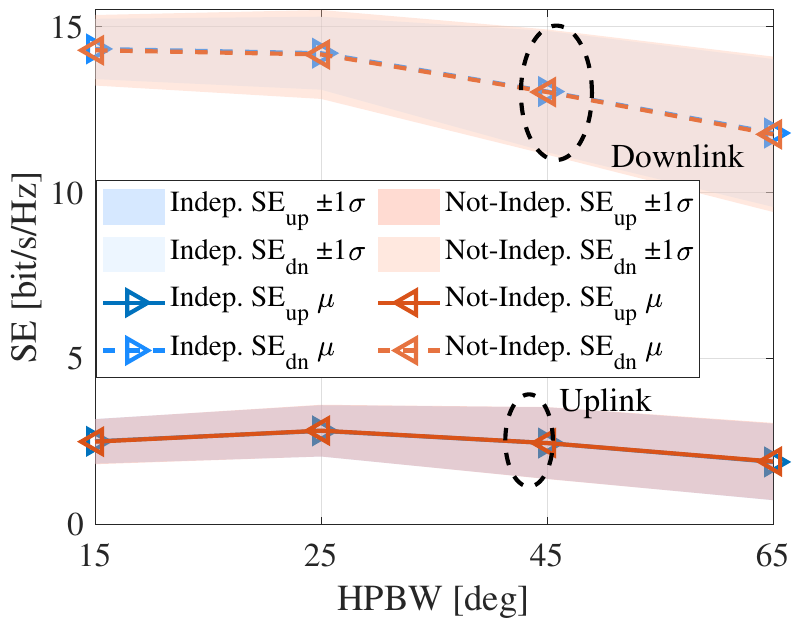}
			\label{fig_SE_IndepOrNot_HPBW}
		\end{minipage}%
	}%
	\subfigure[]{
		\begin{minipage}[t]{0.5\linewidth}
			\includegraphics[width=1.6in]{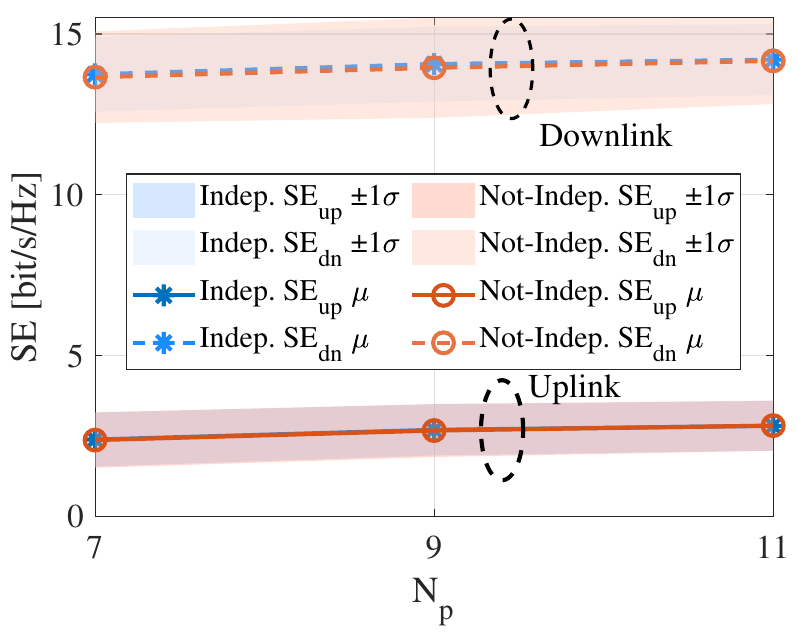}
			\label{fig_SE_IndepOrNot_Np}
		\end{minipage}%
	}
	\centering
	\caption{
			SE performance for independent (Indep.) and proposed joint (Not-Indep.) cases with $P_\text{C}=27$~dBm and $N_{\text{U}}=1$:
			(a) SE performance for different HPBWs with $N_p=11$;
			(b) SE performance for different $N_p$s with HPBW$=25^\circ$.
	}
	\label{fig_SE_IndepOrNot}
	\vspace{-3mm}
\end{figure}
\begin{figure}[!t]
	\color{black}
	\subfigure[]{
		\begin{minipage}[t]{0.5\linewidth}
			\includegraphics[width=1.6in]{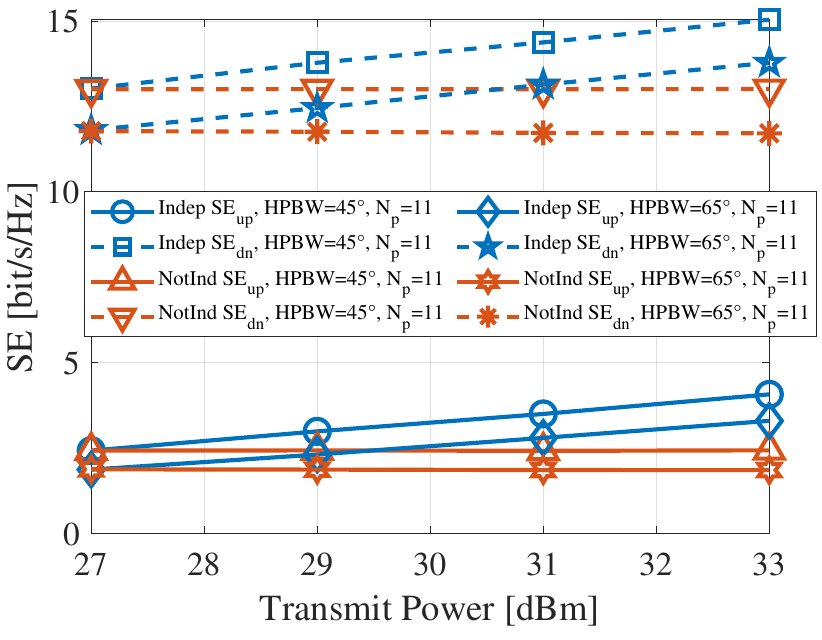}
			\label{fig_SE_IndepOrNot_HPBW_withPower}
		\end{minipage}%
	}%
	\subfigure[]{
		\begin{minipage}[t]{0.5\linewidth}
			\includegraphics[width=1.6in]{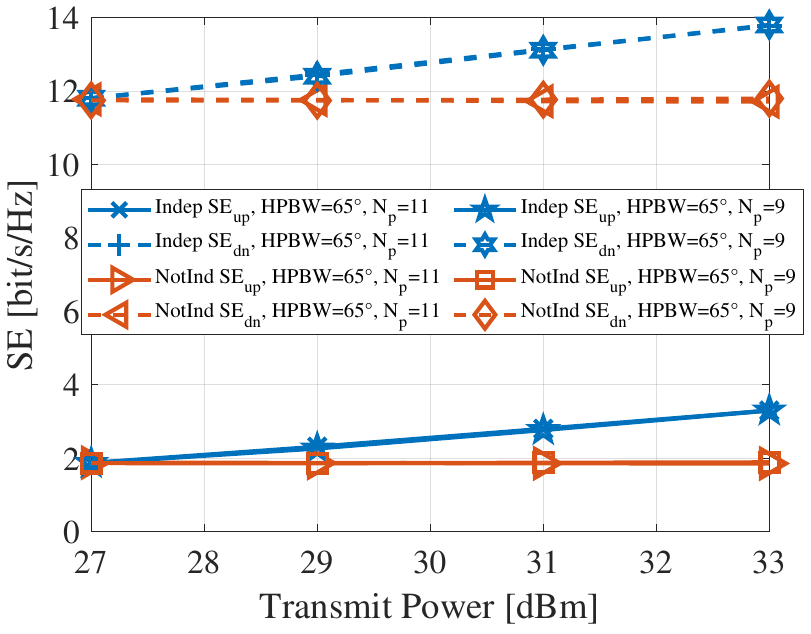}
			\label{fig_SE_IndepOrNot_Np_withPower}
		\end{minipage}%
	}
	\centering
	\caption{
			SE performance for independent (Indep.) and proposed joint (Not-Indep.) cases when $N_{\text{U}}=1$ with varying transmit power:
			(a) varying HPBW when $N_p=11$;
			(b) varying HPBW when $\text{HPBW}=25^\circ$.
	}
	\label{fig_SE_IndepOrNot_withPower}
	\vspace{-3mm}
\end{figure}

By integrating Figs.~\ref{fig_Angle_Estimation} and \ref{fig_SE}, to minimize the impact of the controller's transmit power on the estimation of the jammer's AoA, the UAV controller can remain silent, allowing the UAV to receive only the jammer's jamming signal.
Although this approach improves the estimation accuracy of the jammer's AoA by eliminating the influence of the controller's signal, it introduces an additional time overhead for jammer sensing.
Therefore, if the impact of the UAV controller on the jammer is not significant, particularly when the critical points can be identified, as shown in Fig.~\ref{fig_SE_HPBW_Up} for HPBW=45$^\circ$ and in Fig.~\ref{fig_SE_PatternNum_Up} for $N_p=7$, the system can be controlled to operate at these critical points to maximize overall system performance.

From Fig.~\ref{fig_SE_NoJammer}, we observe that even in the absence of jamming, using RPR-FAS still achieves better SE than traditional pattern-fixed antenna systems.
Moreover, because the UAV can leverage RPR-FAS's angle estimation capability to select the optimal antenna radiation direction, it achieves a higher SNR compared with pattern-fixed antenna systems, thereby extending its flight distance.

\begin{figure*}[!hb]
	\vspace{-3mm}
	\color{black}
	\subfigure[]{
		\begin{minipage}[t]{0.24\linewidth}
			\includegraphics[width=1.7in]{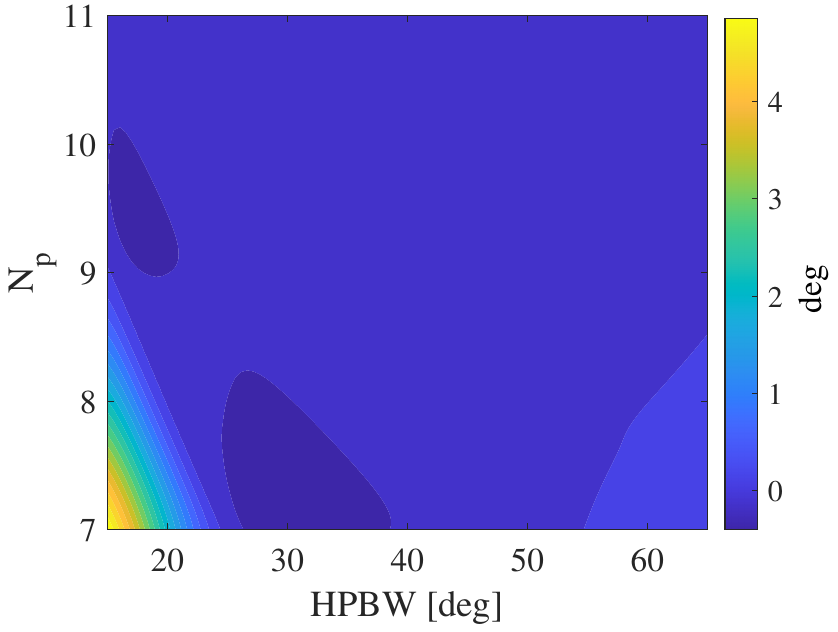}
			\label{fig_Tradeoff_AngEst_Jammer}
		\end{minipage}%
	}%
	\subfigure[]{
		\begin{minipage}[t]{0.24\linewidth}
			\includegraphics[width=1.7in]{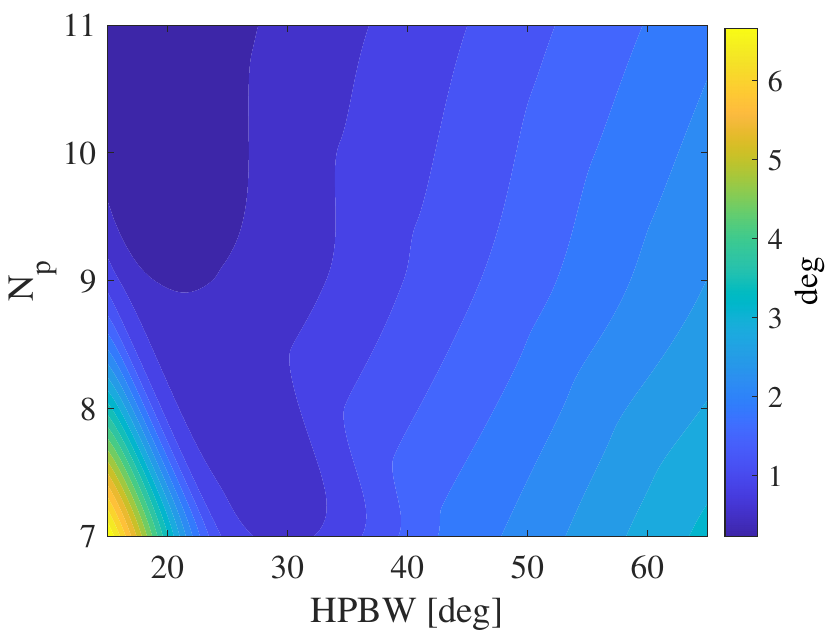}
			\label{fig_Tradeoff_AngEst_Controller}
		\end{minipage}%
	}
	\subfigure[]{
		\begin{minipage}[t]{0.24\linewidth}
			\includegraphics[width=1.7in]{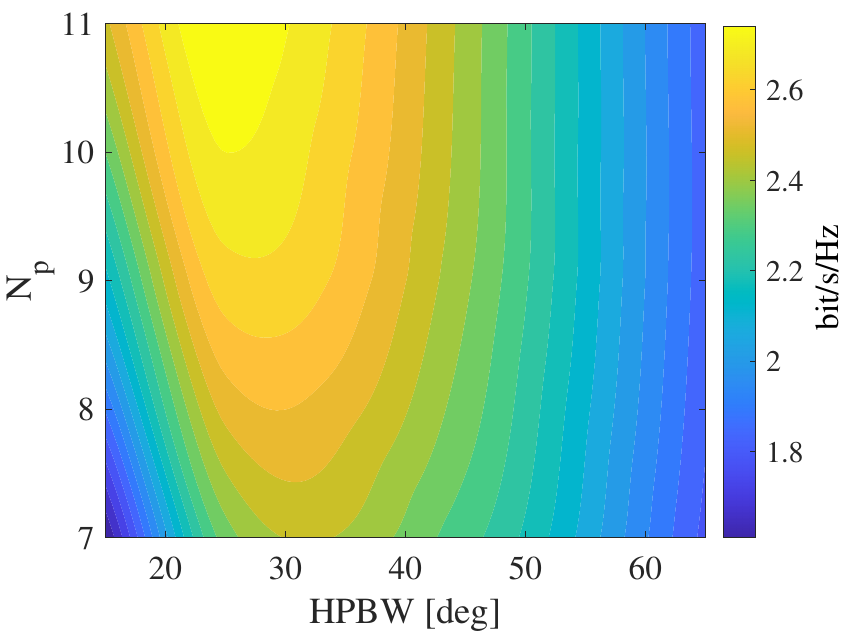}
			\label{fig_Tradeoff_SE_up}
		\end{minipage}%
	}
	\subfigure[]{
		\begin{minipage}[t]{0.24\linewidth}
			\includegraphics[width=1.7in]{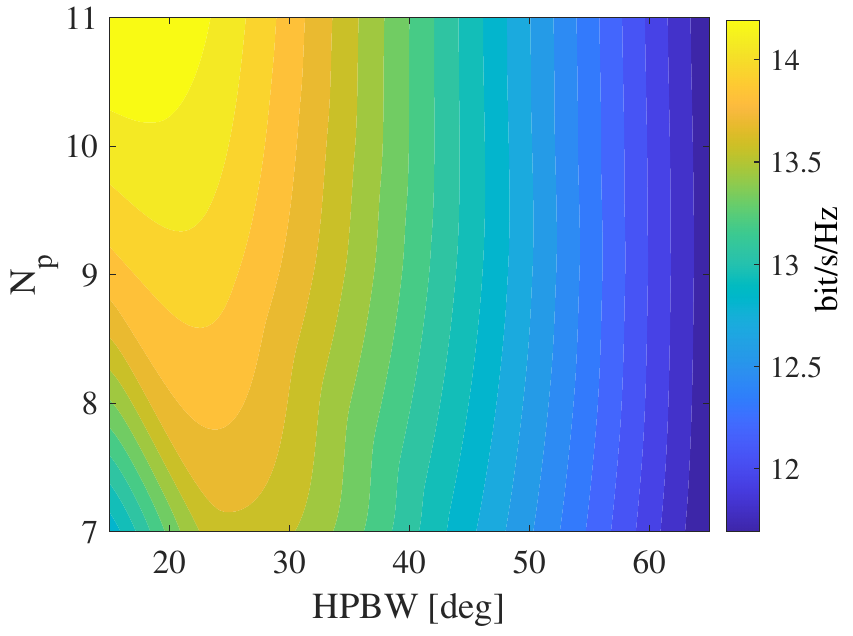}
			\label{fig_Tradeoff_SE_dn}
		\end{minipage}%
	}
	\centering
	\vspace{-2mm}	
	\caption{
			Tradeoff in choosing HPBW and $N_p$:
			(a) AoA estimation errors of the uplink signals from the jammer at the UAV;
			(b) AoA estimation errors of the uplink signals from the UAV controller at the UAV;
			(c) uplink SE;
			(d) downlink SE.
	}
	\label{fig_Tradeoff}
\end{figure*}
\begin{table*}[!hb]
	\vspace{-3mm}
	\centering
	\caption{Sensing and Communication Performance under Different System Parameter Settings with $\text{HPBW}=65^\circ$}
	\label{table_HPBW_65}
	\begin{tabular}{lllllllll}
		\noalign{\hrule height 0.5mm}
		& \begin{tabular}[c]{@{}l@{}}\textbf{Jammer}\\ \textbf{Mean}/\\ {[}$\text{deg}${]}\end{tabular} & \begin{tabular}[c]{@{}l@{}}\textbf{Jammer}\\ \textbf{Variance}/\\ {[}$\text{deg}^2${]}\end{tabular} & \begin{tabular}[c]{@{}l@{}}\textbf{Controller}\\ \textbf{Mean}/\\ {[}$\text{deg}${]}\end{tabular} & \begin{tabular}[c]{@{}l@{}}\textbf{Controller}\\ \textbf{Variance}/\\ {[}$\text{deg}^2${]}\end{tabular} & \begin{tabular}[c]{@{}l@{}}\textbf{Uplink SE}\\ \textbf{Mean}/\\ {[}$\text{bit/s/Hz}${]}\end{tabular} & \begin{tabular}[c]{@{}l@{}}\textbf{Uplink SE}\\ \textbf{Variance}/\\ {[}$\text{(bit/s/Hz)}^2${]}\end{tabular} & \begin{tabular}[c]{@{}l@{}}\textbf{Downlink SE}\\ \textbf{Mean}\\ {[}$\text{bit/s/Hz}${]}\end{tabular} & \begin{tabular}[c]{@{}l@{}}\textbf{Downlink SE}\\ \textbf{Variance}/\\ {[}$\text{(bit/s/Hz)}^2${]}\end{tabular} \\ \noalign{\hrule height 0.5mm}
		\textbf{1)}  & \textbf{0.0} & \textbf{0.0}  & 1.9  & 2.3   & 2.1  & 1.5  & 12.2  &  4.4      \\
		\textbf{2) } &  0.1 & 0.1  & 1.1   & 1.6  &14.4   & 19.9  & 13.1     & 4.9   \\
		\textbf{3)}  &  0.1 & 0.1  & \textbf{0.6}   & \textbf{0.8}  & \textbf{16.2}  & \textbf{17.6}  & \textbf{14.2}     & \textbf{4.4}   \\
		\textbf{4)}   & 86.5  & 2912.7  & 40.8   & 2046.1  & 4.3  & 48.8  & 12.7     & 12.8   \\
		\textbf{5)}  & 0.1  & 0.0  &  16.3  & 102.9  & 2.7  & 1.1  &  1.0    &  0.3  \\
		\textbf{6)}  & 0.2  &  0.2 &  11.9  & 87.8  & 6.0  & 4.0  &  1.5    & 0.5   \\
		\textbf{7)}  & 0.3  & 0.2  & 9.2   & 65.6  & 7.6  & 2.7  & 2.1     &  0.7  \\
		\textbf{8)}  & 0.1  & 0.1  &  2.1  & 2.1  & 1.9  & 1.3  &  11.7    & 5.4   \\
		\textbf{9)}  & 0.1  & 0.1  & 1.2   & 1.6  & 13.9  & 20.3  & 12.7     & 6.1   \\
		\textbf{10)}   & 0.2  & 0.2  & 0.7   & 0.9  & 15.6  & 18.5  & 13.8     & 5.5   \\
		\textbf{11)}   & 87.0  & 2923.9  & 40.4   & 2035.5  & 4.3  & 48.8  & 12.4     & 11.1   \\
		\textbf{12)}   &  0.4 & 0.3  & 9.0   & 61.6  & 7.2  & 3.0  & 1.9     & 0.7   \\ \noalign{\hrule height 0.5mm}
	\end{tabular}
\end{table*}
\begin{table*}[!t]
	\vspace{-2mm}
	\centering
	\caption{Sensing and Communication Performance under Different System Parameter Settings with $\text{HPBW}=15^\circ$}
	\label{table_HPBW_15}
	\begin{tabular}{lllllllll}
		\noalign{\hrule height 0.5mm}
		& \begin{tabular}[c]{@{}l@{}}\textbf{Jammer}\\ \textbf{Mean}/\\ {[}$\text{deg}${]}\end{tabular} & \begin{tabular}[c]{@{}l@{}}\textbf{Jammer}\\ \textbf{Variance}/\\ {[}$\text{deg}^2${]}\end{tabular} & \begin{tabular}[c]{@{}l@{}}\textbf{Controller}\\ \textbf{Mean}/\\ {[}$\text{deg}${]}\end{tabular} & \begin{tabular}[c]{@{}l@{}}\textbf{Controller}\\ \textbf{Variance}/\\ {[}$\text{deg}^2${]}\end{tabular} & \begin{tabular}[c]{@{}l@{}}\textbf{Uplink SE}\\ \textbf{Mean}/\\ {[}$\text{bit/s/Hz}${]}\end{tabular} & \begin{tabular}[c]{@{}l@{}}\textbf{Uplink SE}\\ \textbf{Variance}/\\ {[}$\text{(bit/s/Hz)}^2${]}\end{tabular} & \begin{tabular}[c]{@{}l@{}}\textbf{Downlink SE}\\ \textbf{Mean}\\ {[}$\text{bit/s/Hz}${]}\end{tabular} & \begin{tabular}[c]{@{}l@{}}\textbf{Downlink SE}\\ \textbf{Variance}/\\ {[}$\text{(bit/s/Hz)}^2${]}\end{tabular} \\ \noalign{\hrule height 0.5mm}
		\textbf{1)}  & 0.0  & 0.0  & 0.2   & 0.8  & 2.5  & 0.5  &  14.3    & 1.1   \\
		\textbf{2)}  & 0.0  & 0.0  &  0.2  & 0.6  & 15.6  & 15.1  &  15.2    & 2.1   \\
		\textbf{3)}  &  0.0 &  0.0 &  0.2  & 0.6  & \textbf{17.9}  & \textbf{10.6}  & \textbf{16.3}     & \textbf{1.2}   \\
		\textbf{4)}   & 86.7  & 2937.7  & 39.7   & 1982.3  & 4.8  & 50.5  & 11.6     & 19.1   \\
		\textbf{5)}  & 0.0  & 0.0  & 1.2   & 3.5  & 2.9  & 0.2  & 1.8     & 0.1   \\
		\textbf{6)}  & 0.0  & 0.0  & 1.0   & 3.1  & 7.5  & 1.9  & 2.3     & 0.3   \\
		\textbf{7)}  & 0.0  & 0.0  & 0.9   & 2.6  & 9.0  & 0.9  & 3.2     & 0.3   \\
		\textbf{8)}  &  0.0 &  0.0 &  0.3  & 0.8  & 2.5  & 0.5  & 14.3     & 1.1   \\
		\textbf{9)}  & 0.0  & 0.0  & 0.2   & 0.6  & 15.6  & 14.9  & 15.2     & 2.1   \\
		\textbf{10)}   & 0.0  & 0.0  & 0.2   & 0.5  & \textbf{17.8}  & \textbf{10.6}  & \textbf{16.3}     & \textbf{1.2}   \\
		\textbf{11)}   & 87.1  & 2934.4  & 40.0   & 2026.5  & 4.8  & 50.4  & 11.5     & 18.9   \\
		\textbf{12)}   & 0.0  & 0.0  & 0.9   & 2.7  & 9.0  & 0.9  & 3.2     & 0.3   \\ \noalign{\hrule height 0.5mm}
	\end{tabular}
\end{table*}

Fig.~\ref{fig_SE_IndepOrNot} shows that, at a transmit power of 27~dBm and using Algorithm~\ref{alg_AltOpt} to select the radiation pattern and combiner, the use of an additional signal dedicated to the jammer's AoA estimation makes no difference in either uplink or downlink SE. However, by varying the transmit power as in Fig.~\ref{fig_SE_IndepOrNot_withPower}, it can be seen that as the transmit power increases, the SE achieved when using an extra signal—unaffected by the UAV controller—for jammer estimation continues to improve, whereas the SE obtained with a single‐signal estimation remains constant. This confirms that the jammer's AoA estimation is influenced by the UAV controller, and that improving the accuracy of the jammer's angle estimation is important for enhancing SE.

Furthermore, comparing Figs.~\ref{fig_SE_IndepOrNot_HPBW_withPower} and \ref{fig_SE_IndepOrNot_Np_withPower} reveals that at $\text{HPBW}=65^\circ$, increasing the number of pilots $N_p$ does not improve SE, whereas at $\text{HPBW}=45^\circ$, SE shows a marked increase. This indicates that both HPBW and $N_p$ jointly affect SE, and that increasing $N_p$ is only beneficial when HPBW is small. This observation will be further illustrated in Fig.~\ref{fig_Tradeoff}.

\subsection{Overall Performance Evaluation}
From the preceding simulations and discussions, it is clear that HPBW and $N_p$ jointly determine both AoA estimation accuracy and SE. Even without cost considerations, increasing
$N_p$ is always beneficial, whereas reducing HPBW is not unconditionally better. Fig.~\ref{fig_Tradeoff} therefore establishes a clear selection criterion for HPBW.
For low $N_p$ regime, an excessively narrow HPBW yields insufficient spatial coverage, causing estimation errors—and hence reduced SE—in regions that fall outside the beam.
For high $N_p$ regime, once $N_p$ is large enough and HPBW suffices to illuminate the entire angular domain, further narrowing HPBW improves both estimation accuracy and SE.
In the extreme limit of $N_p\rightarrow \infty$ and $\text{HPBW}\rightarrow 0$, the estimated AoA converges exactly to the boresight direction of the beam in which the target resides.

The angle estimation and SE performances under various system parameter settings are presented in Table~\ref{table_HPBW_65} and Table~\ref{table_HPBW_15}.
In Table~\ref{table_HPBW_65} and Table~\ref{table_HPBW_15}, the first and second columns represent the mean and variance of the jammer's AoA estimation error. 
The third and fourth columns represent the mean and variance of the UAV controller's AoA estimation error.
The fifth and sixth columns represent the mean and variance of the uplink SE.
The seventh and eighth columns represent the mean and variance of the downlink SE.
The specific meanings of items 1)--12) are detailed in Table~\ref{table_notation}.
When generating the channel, $\mathcal{H}_{\text{C}}$ or $\mathbf{H}_{\text{C}}$ is categorized into two types: one with a LoS path and the other without a LoS path, corresponding to the first column of Table~\ref{table_notation}.
When $\mathcal{H}_{\text{C}}$ or $\mathbf{H}_{\text{C}}$ has no LoS path, $\mathbf{H}_{\text{J}}$ consists of only a single NLoS path.
The original high-storage-overhead matrix $\mathbf{C}$ and the low-storage-overhead matrix $\mathbf{C}_{\text{L}}$ are both used, aiming to demonstrate that the performance loss of using $\mathbf{C}_{\text{L}}$ instead of $\mathbf{C}$ is acceptable.
Unless otherwise specified, the parameters in Tables~\ref{table_HPBW_65} and \ref{table_HPBW_15} are set to $N_p=11$ and $P_{\text{C}}=27$ dBm.

Table~\ref{table_HPBW_65} shows the performance for HPBW = 65$^\circ$, while Table~\ref{table_HPBW_15} shows the performance for HPBW = 15$^\circ$.
The bold values in each column of these two tables indicate the best performance in that column.
From the tables, it can be observed that the MUSIC algorithm completely fails.
By comparing 1), 2), 3) with 5), 6), 7), it can be seen that the presence of a LoS path significantly improves both angle estimation performance and SE performance.
When comparing the three groups: 1), 2), 3); 5), 6), 7); and 8), 9), 10), it is evident that increasing the number of UAV antennas enhances both angle estimation and SE performance.
By comparing 1), 2), 3), 7) with 8), 9), 10), 12), it can be observed that the performance loss caused by using the low-storage-overhead reconfigurable radiation pattern gain matrix is acceptable.
Comparing Table~\ref{table_HPBW_65} and Table~\ref{table_HPBW_15}, it is evident that a smaller HPBW leads to better angle estimation and SE performance.
\vspace{-1mm}
\subsection{Insights}
Based on comprehensive simulations, we draw the following insights.
\begin{itemize}
	\item $N_p$ does not need to be large: Fig.~\ref{fig_Jammer_PatternNum}, Fig.~\ref{fig_Control_PatternNum}, and Fig.~\ref{fig_SE_PatternNum_Up}.
	
	\item When $N_p$ is large, HPBW is better if it is smaller: Fig.~\ref{fig_Jammer_HPBW}, Fig.~\ref{fig_Control_HPBW}, Fig.~\ref{fig_SE_HPBW_Up}, Fig.~\ref{fig_NoJammer_HPBW},and Fig.~\ref{fig_Tradeoff}.
	
	\item When $N_p$ is small, reducing the HPBW does not necessarily improve performance:
	Fig.~\ref{fig_Tradeoff}.
	
	\item Low-storage overhead sensing matrix achieves acceptable performance: Fig.~\ref{fig_Angle_Estimation}, Fig.~\ref{fig_SE}, Fig.~\ref{fig_SE_NoJammer}, and 1), 2), 3), 7) versus 8), 9), 10), 12) in Table~\ref{table_HPBW_65} and Table~\ref{table_HPBW_15}.
	
	\item There is a trade-off between the controller's angle estimation performance and the jammer's angle estimation performance: Fig.~\ref{fig_Jammer_Pt}, Fig.~\ref{fig_AngleEst_IndepOrNot}, and  Fig.~\ref{fig_SE}.
	
	\item The proposed RPR-FAS-empowered interference-resilient UAV communication system outperform traditional pattern-fixed antenna systems even in the absence of jamming: Fig. \ref{fig_SE_NoJammer}.
	
	\item When using the same signal for simultaneous estimation of the jammer's and controller's AoAs, simply increasing the transmit power does not necessarily improve performance; instead, measures such as adding more transmit antennas, narrowing the HPBW, or increasing $N_p$ are required:
	Fig.~\ref{fig_SE} and Fig.~\ref{fig_SE_IndepOrNot_withPower}.
\end{itemize}

\begin{table}[!t]
	\vspace{-5mm}
	\centering
	\caption{Specific System Parameter Settings in Table~\ref{table_HPBW_65} and Table~\ref{table_HPBW_15}}
	\label{table_notation}
	
	\begin{tabular}{lcccc}
		\noalign{\hrule height 0.5mm}
		& \begin{tabular}[c]{@{}l@{}}LoS Link Pre-\\ sence in $\mathcal{H}_{\text{C}}$\end{tabular} & $N_{\text{U}}$ &\begin{tabular}[c]{@{}l@{}}Sensing\\ Algorithm\end{tabular}  & $\mathbf{C}_{\text{L}}$ or $\mathbf{C}$ \\ \noalign{\hrule height 0.5mm}
		\textbf{1)}  & \ding{51}  & 1  & Algorithm \ref{alg_ParaEst}  & $\mathbf{C}$     \\
		\textbf{2)}  & \ding{51}  & 2  & Algorithm \ref{alg_ParaEst}  & $\mathbf{C}$    \\
		\textbf{3)}  & \ding{51}  & 4 &  Algorithm \ref{alg_ParaEst}  & $\mathbf{C}$  \\
		\textbf{4)}  & \ding{51}  & 4  & MUSIC   & $\mathbf{C}$  \\
		\textbf{5)}  & \ding{55}  & 1  & Algorithm \ref{alg_ParaEst}   & $\mathbf{C}$    \\
		\textbf{6)}  & \ding{55}  & 2  & Algorithm \ref{alg_ParaEst}   & $\mathbf{C}$    \\
		\textbf{7)}  & \ding{55}  & 4  & Algorithm \ref{alg_ParaEst}   & $\mathbf{C}$    \\
		\textbf{8)}  & \ding{51}  & 1 &  Algorithm \ref{alg_ParaEst}  & $\mathbf{C}_{\text{L}}$   \\
		\textbf{9)}  & \ding{51}  & 2  & Algorithm \ref{alg_ParaEst}   & $\mathbf{C}_{\text{L}}$    \\
		\textbf{10)} & \ding{51}  & 4  & Algorithm \ref{alg_ParaEst}   & $\mathbf{C}_{\text{L}}$ \\
		\textbf{11)} & \ding{51}  & 4  & MUSIC   & $\mathbf{C}_{\text{L}}$   \\
		\textbf{12)} & \ding{55}  & 4  & Algorithm \ref{alg_ParaEst}   & $\mathbf{C}_{\text{L}}$    \\ \noalign{\hrule height 0.5mm}
	\end{tabular}
	\vspace{-5mm}
\end{table}

\subsection{Practical Implementation Considerations for Deploying RPR-FAS on UAV Platforms}
\label{sec_impl_uav}

	Thanks to the single–radio frequency-chain pixel / parasitic-layer antenna prototypes reported in \cite{ref_2012_TAP_Hardware,ref_2014_TAP_Hardware,ref_2018_TAP_Hardware,ref_2022_TAP_Hardware}—all of which fit within the size, weight and power envelope surveyed for multirotor UAVs in \cite{ref_BoyuNing_Move}—the radiation-pattern reconfigurability assumed in our system model (Sec.~\ref{sec_system_model}) is already feasible. As shown in Figs.~\ref{fig_SE_PatternNum_Up} and \ref{fig_Tradeoff}, the performance already reaches a satisfactory level at $N_p = 49$.  Each pattern vector covers the $N_a\!=\!181^{2}$-point angular grid used throughout Sec.~\ref{sec_simulation}, so the raw gain matrix contains $N_pN_a\!=\!1.61\times10^{6}$ real entries. The frequency-domain compression scheme of Sec.~III-B (see Fig.~\ref{fig_LowSaveOverhead}) only requires approximate $1$ MB memory, an insignificant load for current flight-controller memory.
	Complexity analyses in Secs.~\ref{sec_controller_angle}–\ref{sec_pattern_selection} show that the overall complexity is at the level of 10 million operations, comfortably inside the real-time budget of the multi-GFLOPS embedded SoCs already deployed on small UAVs.
	Combined with the sub-millisecond switching latency measured in \cite{ref_2025_CST}, these figures substantiate that all sensing, estimation and adaptive pattern-selection steps of the proposed RPR-FAS scheme can run on-board without exceeding the size, weight and power constraints of practical UAV platforms.

\vspace{-1mm}
\section{Conclusion}\label{sec_conclusion}
This paper addresses the challenge of protecting legal UAVs from illegal interference, which poses a significant challenge for the low-altitude economy driven by the growing use of UAVs.
To tackle this challenge, we have proposed an RPR-FAS-empowered interference-resilient UAV communication scheme.
This scheme employs RPR-FAS based on the reconfigurable pixel antenna technology to enhance UAVs' angular sensing capability and SE, even in the single antenna case.
The proposed RPR-FAS-empowered interference-resilient UAV communication scheme is generalizable, requiring only the substitution of the radiation pattern gain matrix with the actual one. 
Specifically, we have proposed the LSO-OMP-MMV algorithm and a maximum likelihood estimation method based on the law of large numbers to estimate the AoAs of the UAV controller and the jammer, respectively.
By utilizing the Fourier transform to the radiation pattern gain matrix, we have reduced the LSO-OMP-MMV algorithm's storage overhead by 1--2 orders of magnitude.
Following the sensing stage, an alternating optimization approach has been utilized to obtain the optimal uplink antenna radiation pattern and combiner, while an exhaustive search combined with the water-filling algorithm has been applied to obtain the optimal downlink antenna radiation pattern and beamformer. 
Simulation results validate the effectiveness of the proposed scheme.
Moreover, simulation results show that even in the absence of interference, our approach highlights the benefits of RPR-FAS over pattern-fixed antenna systems in terms of SE.
Even though our scheme primarily targets remotely piloted UAVs jammed by illegal jammers, it can be readily extended to other scenarios, such as networked UAVs under illicit jamming or remotely piloted UAVs facing unintentional interference.
\vspace{-2mm}
\bibliographystyle{IEEEtran}
\bibliography{ref}

@ARTICLE{ref_UAV_com_1,
	author={Zhang, Chuang and Sun, Geng and Li, Jiahui and Wu, Qingqing and Wang, Jiacheng and Niyato, Dusit and Liu, Yuanwei},
	journal={IEEE Trans. Mob. Comput.}, 
	title={Multi-Objective Aerial Collaborative Secure Communication Optimization via Generative Diffusion Model-Enabled Deep Reinforcement Learning}, 
	year={2025},
	month={Apr.},
	volume={24},
	number={4},
	pages={3041-3058},
	doi={10.1109/TMC.2024.3502685}}

@ARTICLE{ref_ISAC_1,
	author={Wang, Jiacheng and Du, Hongyang and Liu, Yinqiu and Sun, Geng and Niyato, Dusit and Mao, Shiwen and In Kim, Dong and Shen, Xuemin},
	journal={IEEE Trans. Inf. Forensics Security}, 
	title={Generative {AI} Based Secure Wireless Sensing for {ISAC} Networks}, 
	year={2025},
	month={May},
	volume={20},
	number={},
	pages={5195-5210},
	doi={10.1109/TIFS.2025.3570202}}

@ARTICLE{ref_UAV_com_2,
	author={Zheng, Xiaoya and Sun, Geng and Li, Jiahui and Wang, Jiacheng and Wu, Qingqing and Niyato, Dusit and Jamalipour, Abbas},
	journal={IEEE Trans. Mob. Comput.},
	title={ {UAV} Swarm-Enabled Collaborative Post-Disaster Communications in Low Altitude Economy Via a Two-Stage Optimization Approach },
	year={2025},
	volume={},
	number={01},
	ISSN={1558-0660},
	pages={1-18}}

@ARTICLE{ref_GenAI,
	author={Sun, Geng and Xie, Wenwen and Niyato, Dusit and Mei, Fang and Kang, Jiawen and Du, Hongyang and Mao, Shiwen},
	journal={IEEE Wirel. Commun.}, 
	title={Generative {AI} for Deep Reinforcement Learning: Framework, Analysis, and Use Cases}, 
	year={2025},
	month={Jun.},
	volume={32},
	number={3},
	pages={186-195},
	doi={10.1109/MWC.001.2400176}}

@ARTICLE{ref_lhs,
	author={Liu, Hongshan and Qin, Tong and Gao, Zhen and Mao, Tianqi and Ying, Keke and Wan, Ziwei and Qiao, Li and Na, Rui and Li, Zhongxiang and Hu, Chun and et al.},
	journal={Space Sci. Technol.}, 
	title={Near-Space Communications: The Last Piece of {6G} Space--Air--Ground--Sea Integrated Network Puzzle}, 
	year={2024},
	volume={4},
	pages={0176},
	doi={10.34133/space.0176}
}

@ARTICLE{ref_lsc,
	author={Gao, Zhen and Liu, Shicong and Su, Yu and Li, Zhongxiang and Zheng, Dezhi},
	journal={IEEE J. Sel. Top. Signal Process.}, 
	title={Hybrid Knowledge-Data Driven Channel Semantic Acquisition and Beamforming for Cell-Free Massive {MIMO}}, 
	year={2023},
	month={Sept.},
	volume={17},
	number={5},
	pages={964-979},
	doi={10.1109/JSTSP.2023.3299175}}

@ARTICLE{ref_kml,
	author={Ke, Malong and Gao, Zhen and Wu, Yongpeng and Gao, Xiqi and Schober, Robert},
	journal={IEEE Trans. Signal Process.}, 
	title={Compressive Sensing-Based Adaptive Active User Detection and Channel Estimation: Massive Access Meets Massive {MIMO}}, 
	year={2020},
	month={Jan.},
	volume={68},
	number={},
	pages={764-779},
	doi={10.1109/TSP.2020.2967175}}

@ARTICLE{ref_zxy,
	author={Gao, Zhen and Zhou, Xingyu and Ning, Boyu and Su, Yu and Qin, Tong and Niyato, Dusit},
	journal={IEEE J. Sel. Areas Commun.}, 
	title={Integrated Location Sensing and Communication for Ultra-Massive {MIMO} With Hybrid-Field Beam-Squint Effect}, 
	year={2025},
	month={Apr.},
	volume={43},
	number={4},
	pages={1387-1404},
	doi={10.1109/JSAC.2025.3531551}}

@ARTICLE{refMAone,
	author={Zhu, Lipeng and Ma, Wenyan and Zhang, Rui},
	journal={IEEE Commun. Mag.}, 
	title={Movable Antennas for Wireless Communication: Opportunities and Challenges}, 
	year={2024},
	month={Jun.},
	volume={62},
	number={6},
	pages={114-120},
	doi={10.1109/MCOM.001.2300212}}

@ARTICLE{refMAtwo,
	author={Zhu, Lipeng and Ma, Wenyan and Zhang, Rui},
	journal={IEEE Trans. Wireless Commun.}, 
	title={Modeling and Performance Analysis for Movable Antenna Enabled Wireless Communications}, 
	year={2024},
	month={Jun.},
	volume={23},
	number={6},
	pages={6234-6250},
	doi={10.1109/TWC.2023.3330887}}

@ARTICLE{refMAthree,
	author={Zhu, Lipeng and Ma, Wenyan and Mei, Weidong and Zeng, Yong and Wu, Qingqing and Ning, Boyu and Xiao, Zhenyu and Shao, Xiaodan and Zhang, Jun and Zhang, Rui},
	journal={IEEE Commun. Surv. Tutor.}, 
	title={A Tutorial on Movable Antennas for Wireless Networks}, 
	year={2025},
	volume={},
	number={},
	pages={1-1},
	doi={10.1109/COMST.2025.3546373},
note={Early Access}}

@book{refRayleighQuotient,
	title={Matrix analysis},
	author={Horn, Roger A and Johnson, Charles R},
	year={2012},
	publisher={Cambridge university press}
}

@techreport{ref_38901,
  author    = {{3GPP}},
  title     = {Study on channel model for frequencies from 0.5 to 100 {GHz}},
  institution = {3rd Generation Partnership Project (3GPP)},
  number    = {TR 38.901},
  year      = {2017},
  month     = {May},
  version   = {14.0.0}
}

@ARTICLE{ref_2021_AESM_C_UAS,
  author={Wang, Jian and Liu, Yongxin and Song, Houbing},
  journal={IEEE Aerosp. Electron. Syst. Mag.}, 
  title={Counter-Unmanned Aircraft System(s) {(C-UAS)}: State of the Art, Challenges, and Future Trends}, 
  year={2021},
  month={Mar.},
  volume={36},
  number={3},
  pages={4-29},
  doi={10.1109/MAES.2020.3015537}}

@ARTICLE{ref_2024_WCM_KekeYing,
  author={Ying, Keke and Gao, Zhen and Chen, Sheng and Gao, Xinyu and Matthaiou, Michail and Zhang, Rui and Schober, Robert},
  journal={IEEE Wirel. Commun.}, 
  title={Reconfigurable Massive {MIMO}: Harnessing the Power of the Electromagnetic Domain for Enhanced Information Transfer}, 
  year={2024},
  month={Jun.},
  volume={31},
  number={3},
  pages={125-132},
  doi={10.1109/MWC.014.2200418}}

@ARTICLE{ref_IoTJ,
  author={Li, Zhuoran and Gao, Zhen and Wang, Kuiyu and Mei, Yikun and Zhu, Chunli and Chen, Lei and Wu, Xiaomei and Niyato, Dusit},
  journal={IEEE Internet Things J.}, 
  title={Unauthorized {UAV} Countermeasure for Low-Altitude Economy: Joint Communications and Jamming Based on {MIMO} Cellular Systems}, 
  year={2024},
  volume={},
  number={},
  pages={1-1},
  doi={10.1109/JIOT.2024.3491796},
  note={Early Access}}

@ARTICLE{ref_2024_TCOM_KekeYing,
  author={Ying, Keke and Gao, Zhen and Su, Yu and Qin, Tong and Matthaiou, Michail and Schober, Robert},
  journal={IEEE Trans. Commun.}, 
  title={Reconfigurable Massive {MIMO}: Precoding Design and Channel Estimation in the Electromagnetic Domain}, 
  year={2024},
  volume={},
  number={},
  pages={1-1},
  doi={10.1109/TCOMM.2024.3485587},
  note={Early Access}}

@ARTICLE{ref_LZR_JSTSP,
	author={Li, Zhuoran and Gao, Zhen and Li, Tuan},
	journal={IEEE J. Sel. Top. Signal. Process.}, 
	title={Sensing User's Channel and Location With Terahertz Extra-Large Reconfigurable Intelligent Surface Under Hybrid-Field Beam Squint Effect}, 
	year={2023},
	month={Jul.},
	volume={17},
	number={4},
	pages={893-911},
	doi={10.1109/JSTSP.2023.3278942}}

@article{ref_DroneRFa,
  author    = {N. Yu and S. Mao and C. Zhou and G. Sun and Z. Shi and J. Chen},
  title     = {{DroneRFa}: A large-scale dataset of drone radio frequency signals for detecting low-altitude drones},
  journal   = {J. Electron. Inf. Technol.},
  volume    = {45},
  pages     = {1--10},
  year      = {2023},
  month     = {Jan.}
}

@ARTICLE{ref_2021_TIM_FHS_Model,
  author={Xie, Yuelei and Jiang, Ping and Gu, Yi and Xiao, Xiao},
  journal={IEEE Trans. Instrum. Meas.}, 
  title={Dual-Source Detection and Identification System Based on {UAV} Radio Frequency Signal}, 
  year={2021},
  month={Aug.},
  volume={70},
  number={},
  pages={1-15},
  doi={10.1109/TIM.2021.3103565}}

@ARTICLE{ref_2018_CM_AntiUAV,
  author={Shi, Xiufang and Yang, Chaoqun and Xie, Weige and Liang, Chao and Shi, Zhiguo and Chen, Jiming},
  journal={IEEE Commun. Mag.}, 
  title={Anti-Drone System with Multiple Surveillance Technologies: Architecture, Implementation, and Challenges}, 
  year={2018},
  month={Apr.},
  volume={56},
  number={4},
  pages={68-74},
  doi={10.1109/MCOM.2018.1700430}}

@ARTICLE{ref_2021_Access_AntiUAV,
  author={Park, Seongjoon and Kim, Hyeong Tae and Lee, Sangmin and Joo, Hyeontae and Kim, Hwangnam},
  journal={IEEE Access}, 
  title={Survey on Anti-Drone Systems: Components, Designs, and Challenges}, 
  year={2021},
  month={Mar.},
  volume={9},
  number={},
  pages={42635-42659},
  doi={10.1109/ACCESS.2021.3065926}}

@ARTICLE{ref_2023_TWC_ContinuousOpt,
  author={Wang, Haonan and Li, Ang and Liu, Ya-Feng and Qin, Qibo and Song, Lingyang and Li, Yonghui},
  journal={IEEE Trans. Wireless Commun.}, 
  title={Achievable Rate Maximization Pattern Design for Reconfigurable {MIMO} Antenna Array}, 
  year={2023},
  month={Sept.},
  volume={22},
  number={9},
  pages={5884-5897},
  doi={10.1109/TWC.2023.3238069}}

@ARTICLE{ref_2018_TWC_ModeSelection,
  author={Hasan, Mehedi and Bahceci, Israfil and Cetiner, Bedri A.},
  journal={IEEE Trans. Wireless Commun.}, 
  title={Downlink Multi-User {MIMO} Transmission for Radiation Pattern Reconfigurable Antenna Systems}, 
  year={2018},
  month={Oct.},
  volume={17},
  number={10},
  pages={6448-6463},
  doi={10.1109/TWC.2018.2859972}}

@ARTICLE{ref_2020_AESM_CognitiveTrack,
  author={Gurbuz, Ali Cafer and Mdrafi, Robiulhossain and Cetiner, Bedri A.},
  journal={IEEE Aerosp. Electron. Syst. Mag.}, 
  title={Cognitive Radar Target Detection and Tracking With Multifunctional Reconfigurable Antennas}, 
  year={2020},
  month={Jun.},
  volume={35},
  number={6},
  pages={64-76},
  doi={10.1109/MAES.2020.2990589}}

@ARTICLE{ref_2021_TWC_ModeSelection,
  author={Zhao, Tianchi and Li, Ming and Pan, Yanjun},
  journal={IEEE Trans. Wireless Commun.}, 
  title={Online Learning-Based Reconfigurable Antenna Mode Selection Exploiting Channel Correlation}, 
  year={2021},
  month={Oct.},
  volume={20},
  number={10},
  pages={6820-6834},
  doi={10.1109/TWC.2021.3076760}}

@ARTICLE{ref_2012_TAP_Hardware,
  author={Yuan, Xiaoyan and Li, Zhouyuan and Rodrigo, Daniel and Mopidevi, Hema Swaroop and Kaynar, Oguz and Jofre, Lluís and Cetiner, Bedri A.},
  journal={IEEE Trans. Antennas Propag.}, 
  title={A Parasitic Layer-Based Reconfigurable Antenna Design by Multi-Objective Optimization}, 
  year={2012},
  month={Jun.},
  volume={60},
  number={6},
  pages={2690-2701},
  doi={10.1109/TAP.2012.2194663}}

@ARTICLE{ref_2022_TAP_Hardware,
  author={Zhang, Yujie and Han, Zixiang and Tang, Shiwen and Shen, Shanpu and Chiu, Chi-Yuk and Murch, Ross},
  journal={IEEE Trans. Antennas Propag.}, 
  title={A Highly Pattern-Reconfigurable Planar Antenna With 360° Single- and Multi-Beam Steering}, 
  year={2022},
  month={Aug.},
  volume={70},
  number={8},
  pages={6490-6504},
  doi={10.1109/TAP.2022.3161514}}

@ARTICLE{ref_2014_TAP_Hardware,
  author={Rodrigo, Daniel and Cetiner, Bedri A. and Jofre, Lluı´s},
  journal={IEEE Trans. Antennas Propag.}, 
  title={Frequency, Radiation Pattern and Polarization Reconfigurable Antenna Using a Parasitic Pixel Layer}, 
  year={2014},
  month={Jun.},
  volume={62},
  number={6},
  pages={3422-3427},
  doi={10.1109/TAP.2014.2314464}}

@ARTICLE{ref_2018_TAP_Hardware,
  author={Towfiq, Md. Asaduzzaman and Bahceci, Israfil and Blanch, Sebastián and Romeu, Jordi and Jofre, Lluís and Cetiner, Bedri A.},
  journal={IEEE Trans. Antennas Propag.}, 
  title={A Reconfigurable Antenna With Beam Steering and Beamwidth Variability for Wireless Communications}, 
  year={2018},
  month={Oct.},
  volume={66},
  number={10},
  pages={5052-5063},
  doi={10.1109/TAP.2018.2855668}}

@book{ref_Tse,
  author    = {David Tse and Pramod Viswanath},
  title     = {Fundamentals of Wireless Communication},
  publisher = {Cambridge Univ. Press},
  address   = {Cambridge, U.K.},
  year      = {2005}
}

@ARTICLE{ref_2024_TWC_Outage_Diversity,
  author={New, Wee Kiat and Wong, Kai-Kit and Xu, Hao and Tong, Kin-Fai and Chae, Chan-Byoung},
  journal={IEEE Trans. Wireless Commun.}, 
  title={Fluid Antenna System: New Insights on Outage Probability and Diversity Gain}, 
  year={2024},
  month={Jan.},
  volume={23},
  number={1},
  pages={128-140},
  doi={10.1109/TWC.2023.3276245}}

@article{ref_2024_arXiv_Maga,
  title={Fluid Antenna Systems Enabling {6G}: Principles, Applications, and Research Directions},
  author={Wu, Tuo and Zhi, Kangda and Yao, Junteng and Lai, Xiazhi and Zheng, Jianchao and Niu, Hong and Elkashlan, Maged and Wong, Kai-Kit and Chae, Chan-Byoung and Ding, Zhiguo and others},
  journal={arXiv preprint arXiv:2412.03839},
  year={2024}
}

@ARTICLE{ref_2025_CST,
  author={New, Wee Kiat and Wong, Kai-Kit and Xu, Hao and Wang, Chao and Ghadi, Farshad Rostami and Zhang, Jichen and Rao, Junhui and Murch, Ross and Ramírez-Espinosa, Pablo and Morales-Jimenez, David and Chae, Chan-Byoung and Tong, Kin-Fai},
  journal={IEEE Commun. Surv. Tutor.}, 
  title={A Tutorial on Fluid Antenna System for {6G} Networks: Encompassing Communication Theory, Optimization Methods and Hardware Designs}, 
  year={2024},
  volume={},
  number={},
  pages={1-1},
  doi={10.1109/COMST.2024.3498855},
  note={Early Access}}

@ARTICLE{ref_2024_CL_ChannelEst,
  author={Xu, Hao and Zhou, Gui and Wong, Kai-Kit and New, Wee Kiat and Wang, Chao and Chae, Chan-Byoung and Murch, Ross and Jin, Shi and Zhang, Yangyang},
  journal={IEEE Commun. Lett.}, 
  title={Channel Estimation for {FAS}-Assisted Multiuser {mmWave} Systems}, 
  year={2024},
  month={Mar.},
  volume={28},
  number={3},
  pages={632-636},
  doi={10.1109/LCOMM.2023.3347951}}

@ARTICLE{ref_2025_TWC_ChannelEst,
  author={New, Wee Kiat and Wong, Kai-Kit and Xu, Hao and Rostami Ghadi, Farshad and Murch, Ross and Chae, Chan-Byoung},
  journal={IEEE Trans. Wireless Commun.}, 
  title={Channel Estimation and Reconstruction in Fluid Antenna System: Oversampling is Essential}, 
  year={2025},
  month={Jan.},
  volume={24},
  number={1},
  pages={309-322},
  doi={10.1109/TWC.2024.3491507}}

@article{ref_2024_arXiv_ISAC,
  title={Shifting the {ISAC} Trade-Off with Fluid Antenna Systems},
  author={Zou, Jiaqi and Xu, Hao and Wang, Chao and Xu, Lvxin and Sun, Songlin and Meng, Kaitao and Masouros, Christos and Wong, Kai-Kit},
  journal={arXiv preprint arXiv:2405.05715},
  year={2024}
}

@ARTICLE{ref_2024_WCL_ISAC,
  author={Zhou, Liaoshi and Yao, Junteng and Jin, Ming and Wu, Tuo and Wong, Kai-Kit},
  journal={IEEE Wirel. Commun. Letters}, 
  title={Fluid Antenna-Assisted {ISAC} Systems}, 
  year={2024},
  month={Dec.},
  volume={13},
  number={12},
  pages={3533-3537},
  doi={10.1109/LWC.2024.3476148}}

@article{ref_2024_arXiv__ReconfigurableRadiationPattern,
  title={A Pixel-based Reconfigurable Antenna Design for Fluid Antenna Systems},
  author={Zhang, Jichen and Rao, Junhui and Ming, Zhaoyang and Li, Zan and Chiu, Chi-Yuk and Wong, Kai-Kit and Tong, Kin-Fai and Murch, Ross},
  journal={arXiv preprint arXiv:2406.05499},
  year={2024}
}

@article{ref_2025_CM_ReconfigurableRadiationPattern,
  title={Fluid Antennas: Reshaping Intrinsic Properties for Flexible Radiation Characteristics in Intelligent Wireless Networks},
  author={Lu, Wen-Jun and He, Chun-Xing and Zhu, Yongxu and Tong, Kin-Fai and Wong, Kai-Kit and Shin, Hyundong and Cui, Tie Jun},
  journal={arXiv preprint arXiv:2501.02911},
  year={2025}
}

@ARTICLE{ref_2025_OJAP_ReconfigurableRadiationPattern,
  author={Zhang, Jichen and Rao, Junhui and Li, Zan and Ming, Zhaoyang and Chiu, Chi-Yuk and Wong, Kai-Kit and Tong, Kin-Fai and Murch, Ross},
  journal={IEEE Open J. Antennas Propag. }, 
  title={A Novel Pixel-Based Reconfigurable Antenna Applied in Fluid Antenna Systems With High Switching Speed}, 
  year={2025},
  month={Feb.},
  volume={6},
  number={1},
  pages={212-228},
  doi={10.1109/OJAP.2024.3489215}}

@ARTICLE{ref_BoyuNing_Move,
	author={Ning, Boyu and Yang, Songjie and Wu, Yafei and Wang, Peilan and Mei, Weidong and Yuen, Chau and Björnson, Emil},
	journal={IEEE Wireless Commun.},
	title={Movable Antenna-Enhanced Wireless Communications: General Architectures and Implementation Methods},
	year={2025},
	month={Oct.},
	volume={32},
	number={5},
	pages={108-116}}

@ARTICLE{ref_BoyuNing_OJCS,
	author={Ning, Boyu and Tian, Zhongbao and Mei, Weidong and Chen, Zhi and Han, Chong and Li, Shaoqian and Yuan, Jinhong and Zhang, Rui},
	journal={IEEE Open J. Commun. Soc.}, 
	title={Beamforming Technologies for Ultra-Massive {MIMO} in Terahertz Communications}, 
	year={2023},
	month={Feb.},
	volume={4},
	number={},
	pages={614-658},
	doi={10.1109/OJCOMS.2023.3245669}}

@ARTICLE{ref_UAV_1,
	author={Mei, Weidong and Zhang, Rui},
	journal={IEEE Wirel. Commun.}, 
	title={Aerial-Ground Interference Mitigation for Cellular-Connected {UAV}}, 
	year={2021},
	month={Feb.},
	volume={28},
	number={1},
	pages={167-173},
	doi={10.1109/MWC.001.2000173}}

@ARTICLE{ref_UAV_2,
	author={Liu, Liang and Zhang, Shuowen and Zhang, Rui},
	journal={IEEE Trans. Wireless Commun.}, 
	title={Multi-Beam {UAV} Communication in Cellular Uplink: Cooperative Interference Cancellation and Sum-Rate Maximization}, 
	year={2019},
	month={Oct.},
	volume={18},
	number={10},
	pages={4679-4691},
	doi={10.1109/TWC.2019.2926981}}

@ARTICLE{ref_UAV_3,
	author={Zhou, Lingyun and Chen, Xihan and Hong, Mingyi and Jin, Shi and Shi, Qingjiang},
	journal={IEEE Trans. Veh. Technol.}, 
	title={Efficient Resource Allocation for Multi-{UAV} Communication Against Adjacent and Co-Channel Interference}, 
	year={2021},
	month={Oct.},
	volume={70},
	number={10},
	pages={10222-10235},
	doi={10.1109/TVT.2021.3104279}}

@ARTICLE{ref_UAV_4,
	author={Shakhatreh, Hazimf and Sawalmeh, Ahmad and Hayajneh, Khaled F. and Abdel-Razeq, Sharief and Malkawi, Wa’Ed and Al-Fuqaha, Ala},
	journal={IEEE Open J. Commun. Soc.}, 
	title={A Systematic Review of Interference Mitigation Techniques in Current and Future {UAV}-Assisted Wireless Networks}, 
	year={2024},
	month={Apr.},
	volume={5},
	number={},
	pages={2815-2846},
	doi={10.1109/OJCOMS.2024.3392623}}

@ARTICLE{ref_UAV_5,
	author={Su, Yueyue and Qi, Nan and Huang, Zanqi and Yao, Rugui and Jia, Luliang},
	journal={China Commun.}, 
	title={Cooperative anti-jamming and interference mitigation for {UAV} networks: A local altruistic game approach}, 
	year={2024},
	month={Feb.},
	volume={21},
	number={2},
	pages={183-196},
	doi={10.23919/JCC.fa.2021-0759.202402}}


\vfill

\end{document}